\documentclass[12pt]{spieman}  
\usepackage{amsmath,amsfonts,amssymb}
\usepackage{graphicx}
\usepackage{setspace}
\usepackage{tocloft}
\usepackage{rotating}
\usepackage{pdflscape}
\usepackage{afterpage}
\usepackage{siunitx}

\DeclareSIUnit{\nothing}{\relax}

\title{Complex spectral line profiles resulting from cryogenic deformation of the SINFONI/SPIFFI diffraction gratings}

\author[a,b,*]{E. M. George}
\author[a,c]{D. Gr\"{a}ff}
\author[a]{M. Hartl}
\author[a]{H. Huber}
\author[a]{F. Eisenhauer}
\author[a]{H. Feuchtgruber}
\affil[a]{Max Planck Institut f\"{u}r Extraterrestrische Physik, Gie{\ss}enbachstr. 1, 85748 Garching, Germany}
\affil[b]{European Southern Observatory, Karl-Schwarzchild-Str. 2, 85748 Garching, Germany}
\affil[c]{ETH Z\"{u}rich Laboratory of Hydraulics, Hydrology and Glaciology, H\"{o}nggerbergring 26, 8093 Z\"{u}rich, Switzerland}

\cftpagenumbersoff{figure}
\cftpagenumbersoff{table} 
\begin{document} 
\maketitle


\begin{abstract}
The integral field spectrograph, SPIFFI, has complex line profile shapes that vary with wavelength and pixel scale, the origins of which have been sought since the instrument construction. SPIFFI is currently operational as part of SINFONI at the VLT, and will be upgraded and incorporated into the new VLT instrument ERIS. We conducted an investigation of the line profiles based on measurements we could take with the instrument calibration unit, as well as laboratory measurements of spare SPIFFI optical components. Cryogenic measurements of a spare SPIFFI diffraction grating showed significant periodic deformation. These measurements match the cryogenic deformation expected from bimetallic bending stress based on a finite element analysis of the lightweighted grating blank. The periodic deformation of the grating surface gives rise to satellite peaks in the diffraction pattern of the grating. An optical simulation including the cryogenic grating deformation reproduces the behavior of the SPIFFI line profiles with both wavelength and pixel scale as measured with the instrument calibration unit. The conclusion is that cryogenic deformation of the diffraction gratings is responsible for the non-ideal line profiles, and that the diffraction gratings should be replaced during the upgrade for optimal instrument performance.
\end{abstract}


{\noindent \footnotesize\textbf{*}E. M. George,  \linkable{egeorge@eso.org} }


\section{Introduction}
\label{sec:intro}  

SINFONI (Spectrograph for INtegral Field Observations in the Near Infrared)\cite{eisenhauer03}  has been operational as a facility instrument for the ESO Very Large Telescope (VLT) since 2005. It is made up of the adaptive optics module MACAO (Multi-Application Curvature Adaptive Optics)\cite{bonnet03} and the integral field spectrograph (IFS) SPIFFI (SPectrometer for Infrared Faint Field Imaging)\cite{iserlohe04}. The instrument has been scientifically productive in several areas over its 11-year lifetime, and is currently in use for several high-profile scientific programs. An upgraded version of the instrument will be included in the new VLT Adaptive Optics (AO) instrument ERIS\cite{amico12, kuntschner14} (Enhanced Resolution Imager and Spectrometer) as the Integral Field Unit (IFU) subsystem SPIFFIER (SPectrometer for Infrared Faint Field Imaging Enhanced Resolution)\cite{george16}.  

SPIFFI operates at wavelengths from 1.0-2.5 \SI{}{\micro\meter} in selectable J, H, K, or H+K bands with a choice of pixel scale of 25, 100, or 250 mas/px. The spectroscopic line profile shapes of SPIFFI are complex and vary with wavelength and pixel scale. These complex line profiles result in approximately a factor of two degradation of the instrument resolution in the short wavelengths of J-band, that smoothly rises to approximately the design value of the resolution by the long wavelengths of K-band.\cite{graeff16, george16} The origins of these line profiles have been investigated since the construction of the instrument\cite{iserlohe04}, and effort has gone into correcting for them in atmospheric subtraction.\cite{thatte12} The source of the line profile variation needed to be found so that it could be fixed in the upcoming instrument upgrade for ERIS.

 As reported in George et al. (2016)\cite{george16}, after replacing the potentially troublesome spectrometer collimator mirrors during an upgrade of the instrument in January 2016, we were able to track down the likely origin of the line profile degradation to cryogenic bimetallic bending stress in the diffraction grating blanks that resulted in quilting of the grating surface. However, before procuring new diffraction gratings, we needed to prove that the gratings were the source of the line profile degradations. Because SPIFFI is currently in use at the VLT as part of SINFONI, we were limited in what measurements we were able to take--in particular, we were limited to only measurements that could be taken with the instrument calibration unit, as well as laboratory experiments on spare optical components. This investigation is presented in this paper, and our approach is as follows: 
\begin{enumerate}
\item{Take super-sampled line profiles in all wavelengths and pixel scales using the instrument calibration unit. (Section \ref{sec:lineProfiles}) }
\item{Evaluate the potential cryogenic deformation from the grating blank design using Finite Element Analysis. (Section \ref{sec:FEA})}
\item{Verify calculated grating blank deformations by cryogenic interferometric wavefront measurements of one of the spare SPIFFI diffraction gratings. (Section \ref{sec:cryo})}
\item{Insert grating deformation into an optical simulation to create simulated line profiles including the effect of the grating deformation. (Section \ref{sec:opticalsimulation})}
\item{Compare optical simulation results to measurements from the instrument calibration unit. (Section \ref{sec:simmatch})}
\end{enumerate}
Finally, we discuss our plans for the instrument upgrade along with the practical implications for diffraction grating design, and conclude in section \ref{sec:discussionconclusion}. 

\section{SPIFFI Line Profiles}
\label{sec:lineProfiles}  

\begin{figure}[htbp!]
	\begin{center}
		\resizebox{1.0\textwidth}{!}{
			\includegraphics[height=7cm, trim={0.4cm 0.1cm 0.9cm 0.3cm}, clip]{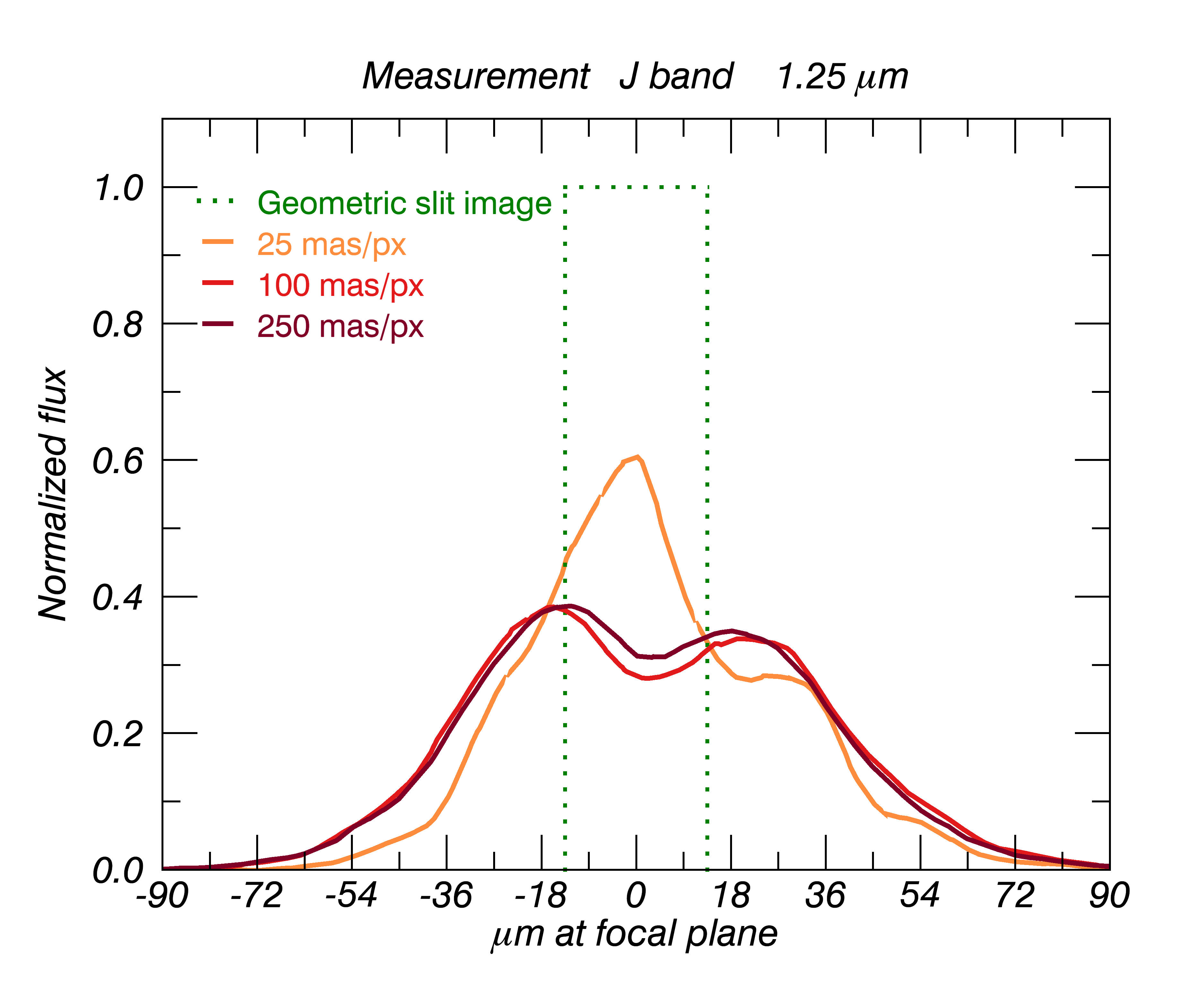}
			\includegraphics[height=7cm, trim={1.0cm 0.1cm 0.9cm 0.3cm}, clip]{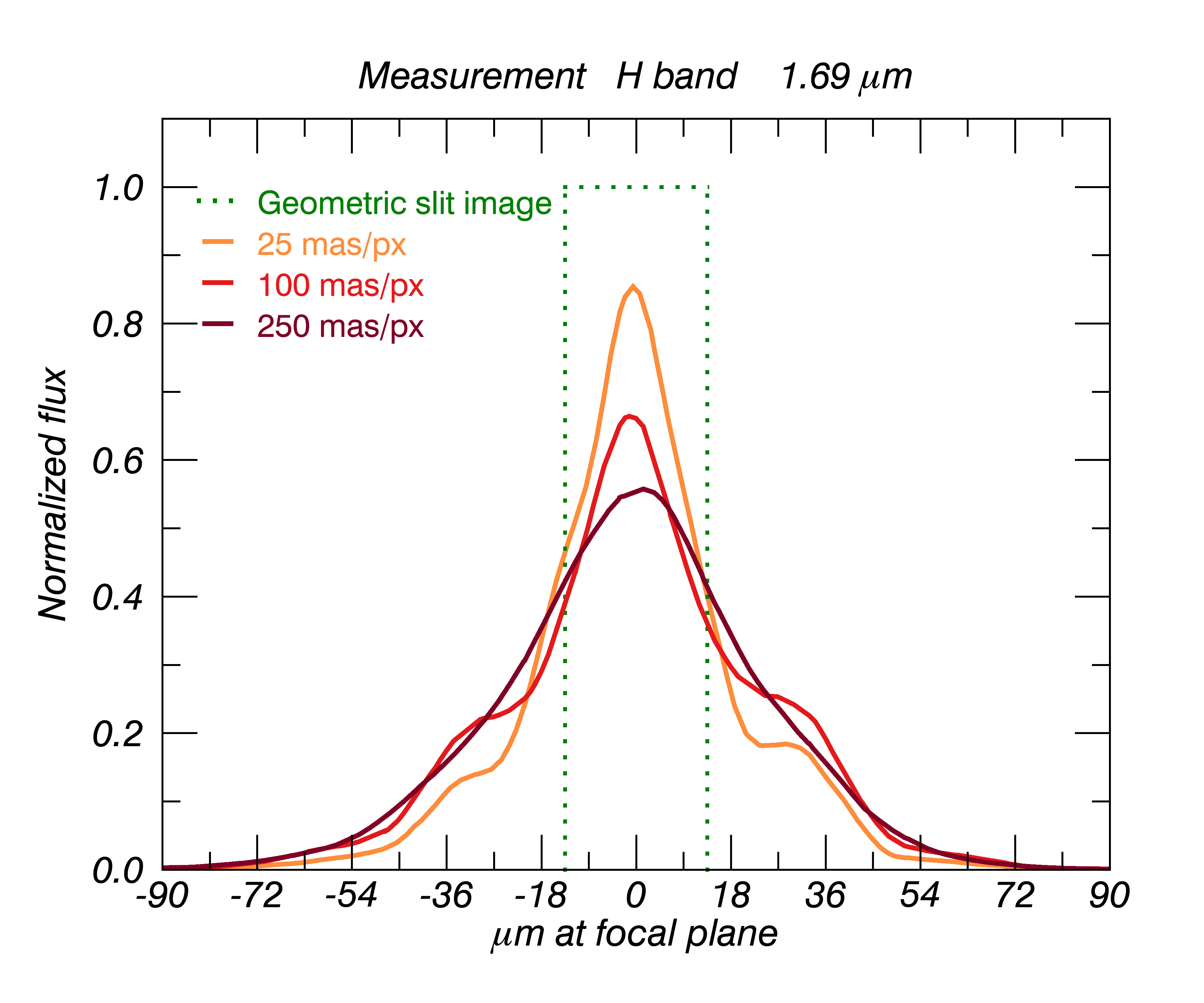}
			\includegraphics[height=7cm, trim={1.0cm 0.1cm 0.9cm 0.3cm}, clip]{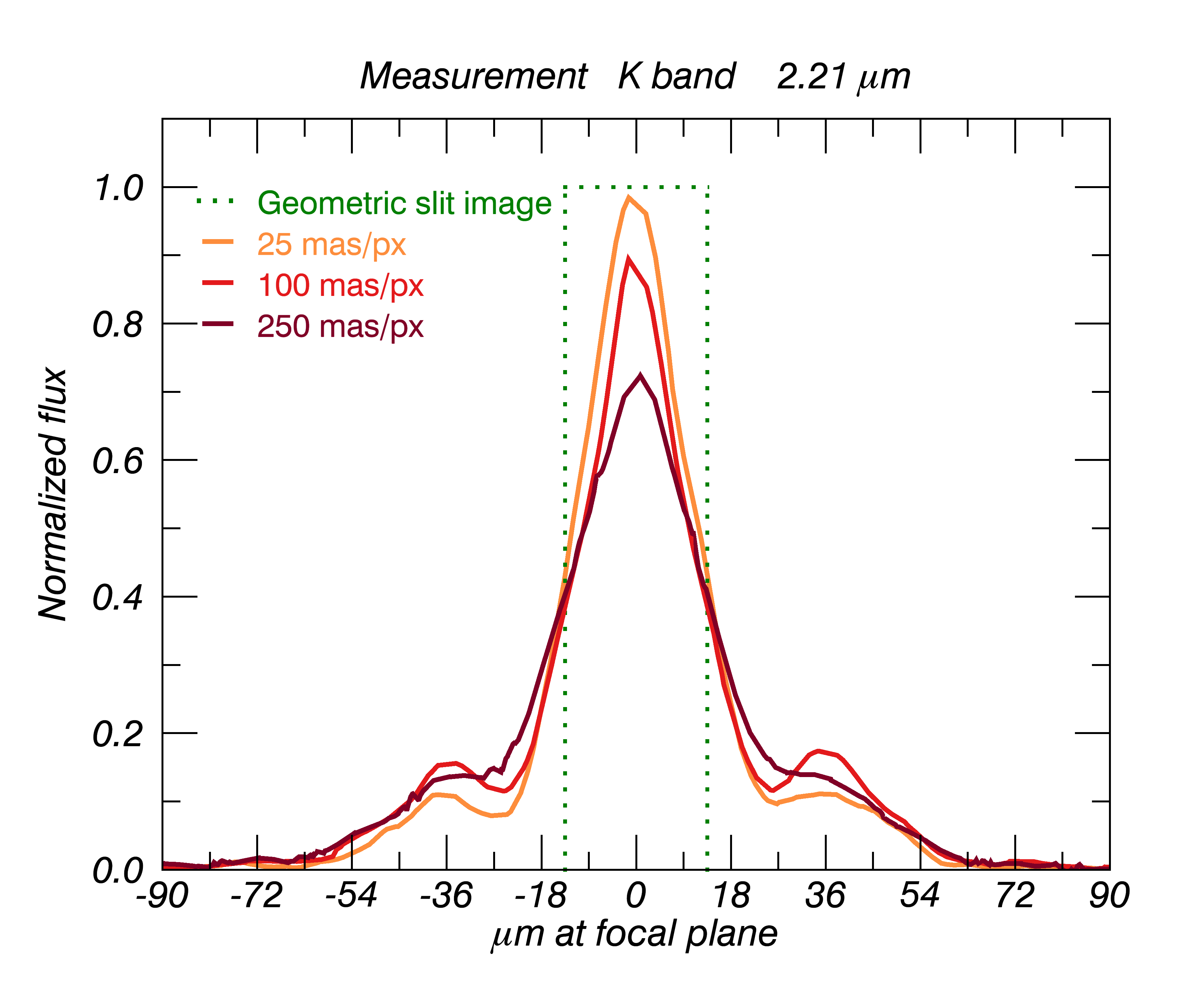}
		}
		\caption[Spectral line profiles in J-, H-, and K- band]{Spectral line profiles for all pixel scales (25 mas/px, 100 mas/px and 250 mas/px) in J-band (left), H-band (center) and K-Band (right) normalized to the integrated flux.}
		\label{fig:measured_lsf}
	\end{center}
\end{figure}

The spectral line profiles we expect to measure are a cross section of an image of the slit in the focal plane. However, the measured profiles are asymmetric with shoulders and vary in shape with wavelength, pixel scale, and spatial detector position along the pseudo-slit. SPIFFI is designed to approximately Nyquist sample the slit image, so super-sampled line profiles are required to measure the shape of the line profiles in detail. To obtain super-sampled line profiles, Thatte et al. (2012) \cite{thatte12} describes a method using OH lines. This method was slightly adapted and used to measure the line profiles of the instrument using the spectral lines of the arc lamps in the calibration unit of SINFONI. In this way, we obtained a catalog of line profiles that covers the entire wavelength range of the instrument in all pixel scales and at all pseudo-slit locations. Detailed plots of the spectral line profiles of SPIFFI and their variability with wavelength, over single slitets, and along the pseudo-slit are available in Gr\"{a}ff (2016)\cite{graeff16} for all bandpasses and pixel scales, and the measured line profiles presented here are produced using the methodology therein. In this paper, all plots, simulations, and analyses are done using measurements with the new collimator as installed in 2016. Figure \ref{fig:measured_lsf} shows for the three single bandpasses of SPIFFI representative spectral line profiles from a single detector column (column 940 in slitlet 16 close to the center of the detector) in all three pixel scales.

In K-band, distinct shoulders at a distance of 36 \SI{}{\micro\meter} in the focal plane (equivalent to 2 detector pixels) from the center of the line are present, and appear to be strongest in the 100 mas pixel scale. In H-band these shoulders are nearly not seen in the largest pixel scale, rather the peak just appears broadened, however, the shoulders appear in the two smaller pixel scales. In J-band, the line profiles look different. In the 25 mas pixel scale a peak with asymmetric shoulders is present; the shoulder on the right side is more distinct than on the left side.  The two larger pixel scales in J-band differ from the other bands even more dramatically -- a double peak behavior can be seen. Here the spacing of the double peak is between 36-45 \SI{}{\micro\meter} (2.0-2.5 pixels) for the 100 and 250 mas pixel scales. 

We additionally took data at single wavelengths on separate diffraction gratings using variable diffraction orders (see appendix \ref{sec:appendixa}). In these measurements the double peak behaviour at 1.25 \SI{}{\micro\meter} appears most strongly on the J-band grating, and is nearly not visible with 1.25 \SI{}{\micro\meter} light on the H- and K-band gratings. This indicates that the J-band grating must be somehow different from the H- and K-band gratings, and was our first clue that the gratings must be affecting the line profiles, as otherwise the light path through the optics is identical for the measurements. 

The line profiles in the 25mas pixel scale vary strongly over the length of each single slitlet, which is most noticeable in J-band but only weakly in H- and K-bands.\cite{thatte12, graeff16}. The upgrade of the instrument did not affect this. The effect of varying line profiles also exists in the 100 and 250 mas pixel scales, though not as strongly as in the 25 mas pixel scale. The variation of the line profiles is discussed more in section \ref{sec:pupilposLSF}.

\section{SPIFFI Lightweighted Diffraction Gratings}
\label{sec:gratings}  

As telescopes and instruments become larger, their cryogenic optical elements become larger too, and it can be necessary to lightweight the optical components to keep both the inertial and thermal masses low. Lightweighting techniques have also long been developed for space applications where keeping the mass low is a priority. However, lightweighted structures can result in quilting of the optical surface\cite{matthews14, kendrick01, kroedel10}, and cryogenic optics deform under differential thermal stresses, such as those due to a Coefficient of Thermal Expansion (CTE) mismatch in bimetallic structures.\cite{vukobratovich97, rohloff10, kinast14} The SPIFFI diffraction gratings have both a lightweighted design and a CTE mismatch between the blank material and polishing layer, which results in significant deformation at their operating temperature of 80K.

\subsection{Design}
\label{sec:gratingDesign}  

\begin{figure}[htbp!]
	\begin{center}
		\includegraphics[width=0.5\textwidth]{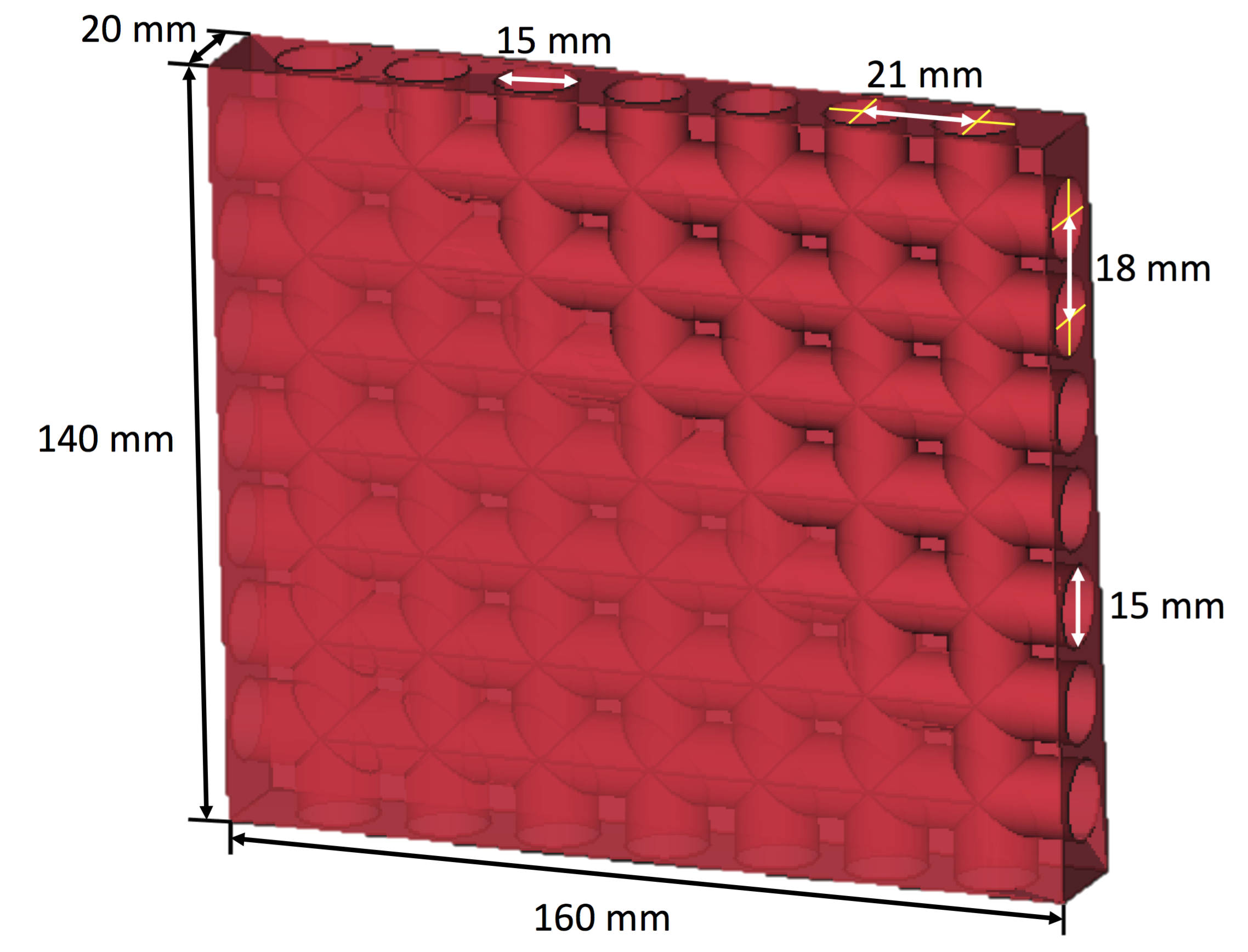}
		\caption[CAD model of the grating blank]{CAD model of the aluminum grating blank with the lightweighting holes used for the simulations. The real blank has additional small notches and holes for fixation screws in the corners.}
		\label{fig:grating_blank}
	\end{center}
\end{figure}

SPIFFI is equipped with four diffraction gratings. The gratings for J-, H-, and K-band are operated in 2\textsuperscript{nd} order and the grating for the combined H+K-band in 1\textsuperscript{st} order. All four diffraction gratings are directly ruled on identical blanks made of 6061 aluminum alloy. The dimensions of the grating blanks are 160 x 140 x 20 mm. Seven lightweighting holes with 15 mm diameter are drilled through the blank in the X- and Y- directions (the Z- direction is normal to the grating surface). Figure \ref{fig:grating_blank} shows a CAD model of the grating blank that was used for the simulations in this paper. The Aluminum blank was electroless Nickel-phosphorus (NiP) plated with a layer thickness between 100-200 \SI{}{\micro\meter} on all surfaces. The two large faces of the NiP-plated blank were then polished symmetrically to be flat. The remaining NiP layer on these surfaces is between 50-90\% of the original layer thickness. Finally a 2-3 \SI{}{\micro\meter} thick gold layer was applied on the front surface before the gratings were ruled. 

\subsection{Finite Element Analysis}
\label{sec:FEA}  

\begin{figure}[htbp!]
	\begin{center}
		\resizebox{1.0\textwidth}{!}{
			\includegraphics[height=7cm, trim={0.2cm 0.1cm 0.85cm 0.1cm}, clip]{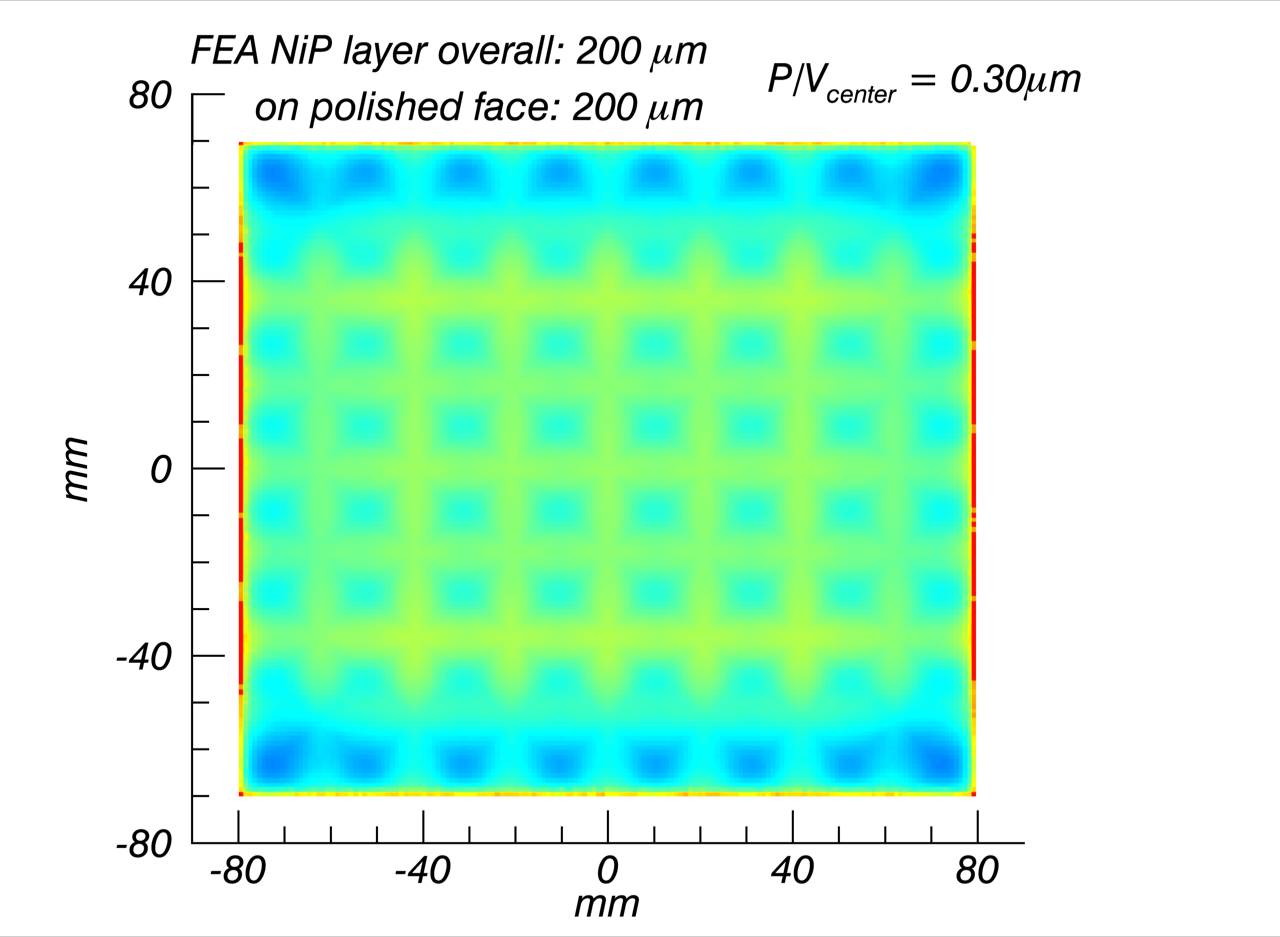}
			\includegraphics[height=7cm, trim={0.5cm 0.1cm 0.85cm 0.1cm}, clip]{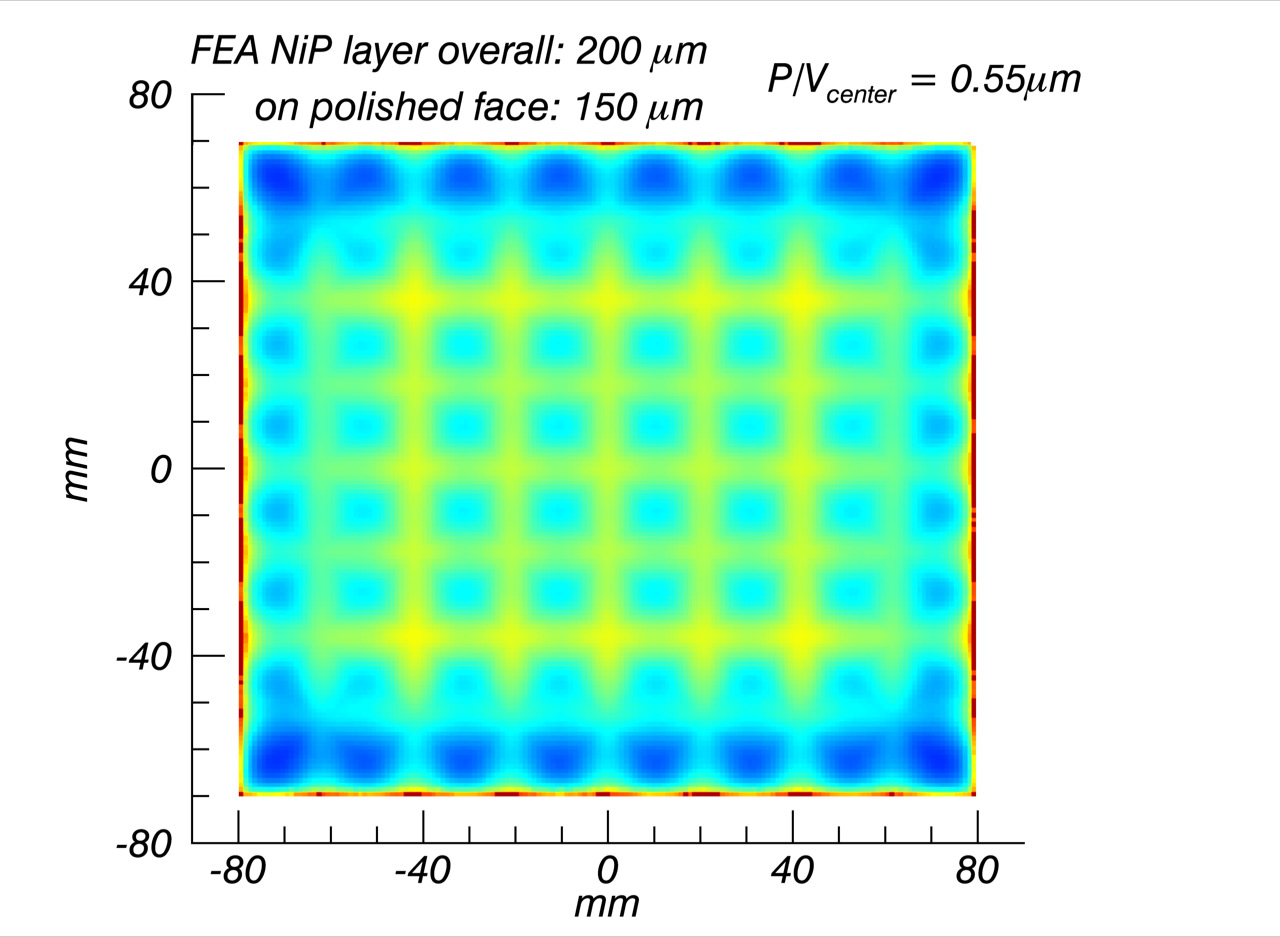}
			\includegraphics[height=7cm, trim={1.5cm 0.2cm 0 0.2cm}, clip]{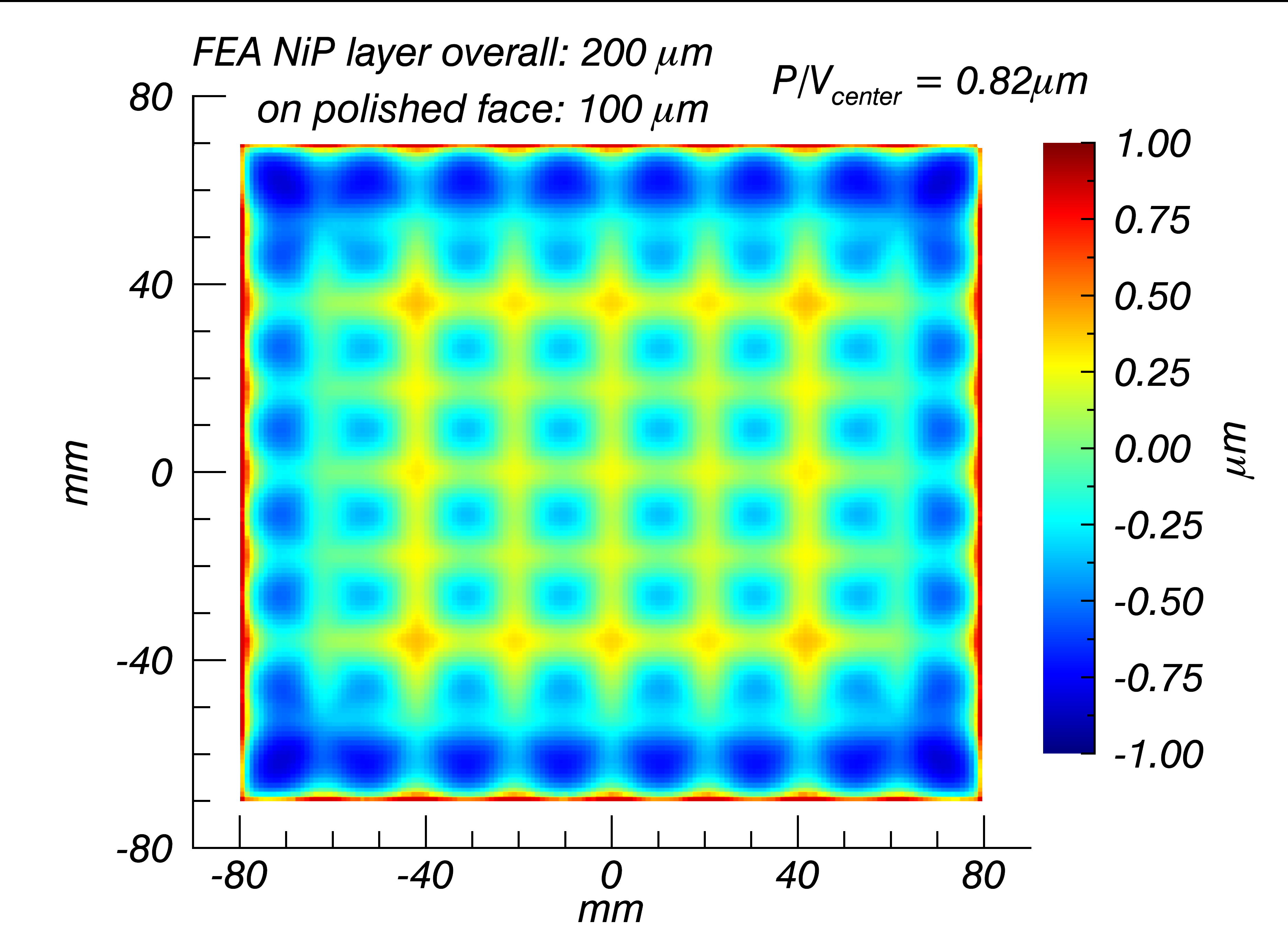}
		}
		\caption[FEA simulation of surface deformation]{FEA simulation of the grating surface deformation when cooling a grating blank with an original NiP layer thickness of 200 \SI{}{\micro\meter} from room temperature to 80K. From left to right, the remaining layer thickness after polishing are 100\%, 75\%, and 50\%. The peak-to-valley in the centre region (P/V$_{center}$ values) are calculated for rectangle of 82 x 72 mm in the center of the grating corresponding to a grid of four lightweighting holes. The deviations at the edges are larger. All images are on the same color scale.}
		\label{fig:deformation}
	\end{center}
\end{figure}

To quantify bimetallic stresses and the resulting deformations of the grating surface when the grating is cooled down from room temperature to the operating temperature of SPIFFI at 80K, we implemented a Finite Element Analysis (FEA) of the Aluminum grating blank with the NiP coating. Unfortunately, we only know roughly the original layer thickness and a range of possible values for the thickness post-polishing. We therefore simulated three layer thicknesses (100, 150, and 200 \SI{}{\micro\meter}) that span the range provided by the manufacturer, as well as polishing steps that left 50\%, 75\%, or 100\% of the original layer thickness. Figure \ref{fig:deformation} shows the deformation resulting from the three polishing values for an original layer thickness of 200 \SI{}{\micro\meter}. 

The bimetallic stresses deform the surface resulting in a regular grid structure. We define the X-direction to be the dispersion direction and the Y-direction to be the ruling direction, therefore, the period of the grid structure is $D_x$ = 21 mm and $D_y$ = 18 mm. Of particular interest is the deformation in the central parts of the grating, since the outer parts and especially the corners of the grating are vignetted by a pupil mask. The pupil mask also covers the areas that may be affected by the small notches and holes for fixation screws in the real grating blank. The peak-to-valley (P/V) deformation in the central region of 82 mm x 72 mm (orientated on the lightweighting holes) is listed in table \ref{tab:FEAdef} for the nine gratings simulated. We chose to report the P/V in this region as a proxy for the amplitude of the periodic deformation, as the deformation in this region is the most regular and is not affected by edge effects. 

\begin{table}[htb]
\begin{center}
\begin{tabular}{c|ccc}
\hline
 &Polished to &Polished to &Polished to \\
 Original Layer & 100\% thickness & 75\% thickness & 50\% thickness \\
 thickness [\SI{}{\micro\meter}] & P/V$_{center}$ [\SI{}{\micro\meter}] & P/V$_{center}$ [\SI{}{\micro\meter}] & P/V$_{center}$ [\SI{}{\micro\meter}]  \\
\hline
\hline
200 & 0.30 & 0.55 & 0.82 \\
150 & 0.23 & 0.43 & 0.63 \\
100 & 0.16 & 0.30 & 0.44\\
 \hline
\end{tabular}
\vspace{0.05in}
\caption[Table of FEA grating deformations]{FEA simulated P/V deformation of the grating in the central region of 82 mm x 72 mm (orientated on the lightweighting holes) for different layer thicknesses and polishing depths. The P/V deformation at the edges is larger by up to a factor of two due to edge effects.}
\label{tab:FEAdef}
\end{center}
\end{table}

In general, thicker original layers result in more deformation. Additionally, for a given original layer thickness, more surface polishing results in more deformation. This indicates that mis-matched layer thicknesses between the inside and outside of the lightweighting structure results in more deformation, however, given the specific geometry of our grating blank, we still found deformation even with matched layer thicknesses. Simulations with thicker NiP on the surface than inside of the lightweighting holes are not presented here, as this is not a realistic scenario given the manufacturing process of the grating. However, the edges cases (NiP only inside the holes, or only on the surface) are presented in appendix \ref{sec:appendixB}, and show that the deformations induced by these two layers oppose each other.

Given the large range in P/V deformation values for relatively small changes in the amount of polishing done to the grating blanks in the FEA, and our unknown layer thicknesses, a cryogenic measurement had to be performed to get an idea of the actual deformation of the SPIFFI diffraction gratings.

\subsection{Cryogenic Measurements}
\label{sec:cryo} 
To measure cryogenic bimetallic bending effects of the grating and verify our FEA analysis, we set up the SPIFFI J-band spare grating on a rotatable stage in a cryogenically cooled vacuum chamber at MPE. The grating was installed on a stress-free mount with heat straps  from the cold plate attached to the four corners of the grating using the small holes for fixation screws. A temperature sensor was installed directly on top of the grating near one corner using another of these fixation screw holes.

\begin{figure}[htbp!]
	\begin{center}
		\resizebox{0.6\textwidth}{!}{
			\includegraphics[width=1.0\textwidth]{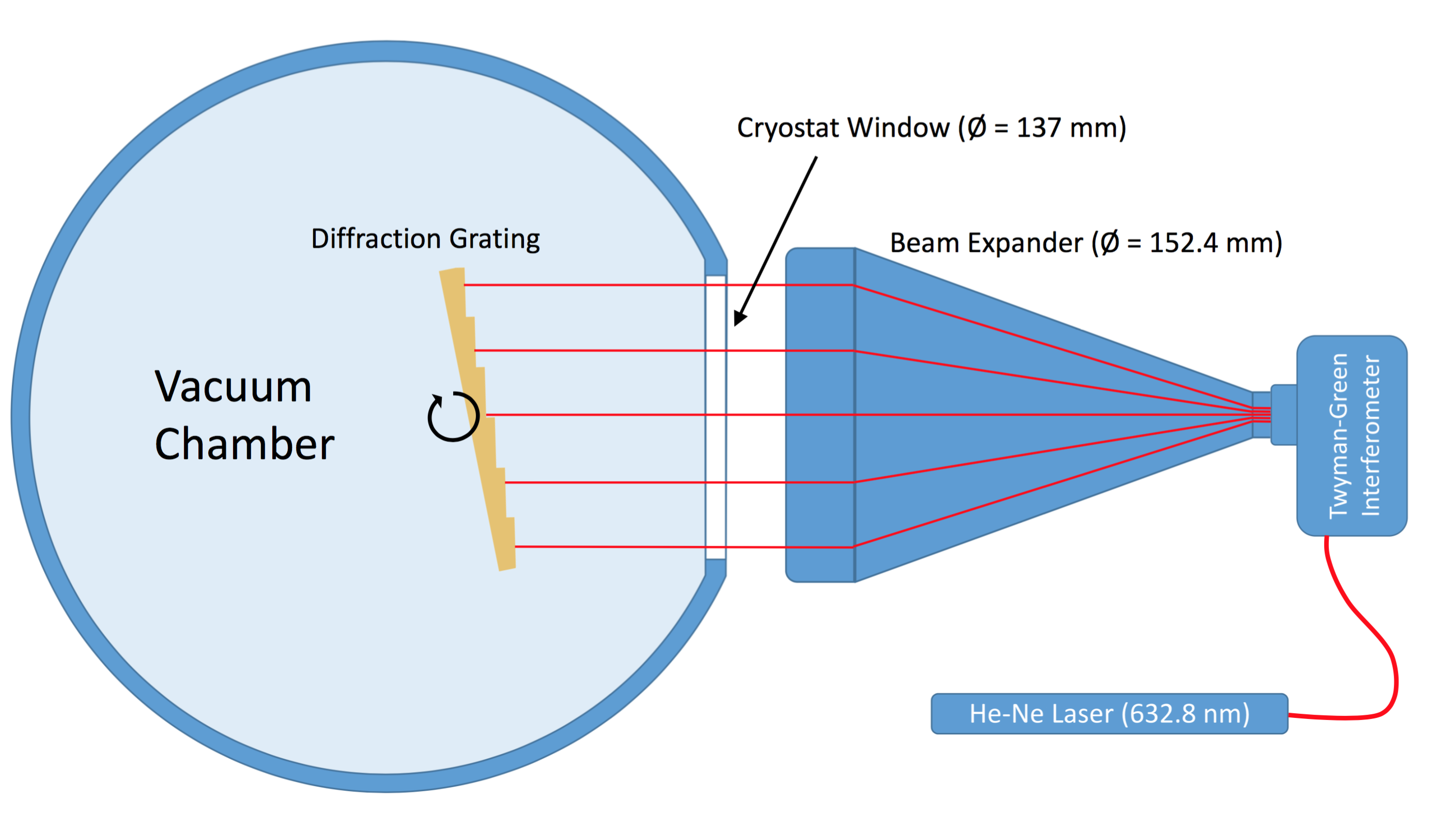}
		}
		\caption[Layout of the cryogenic grating surface measurement]{Layout of the grating surface measurement at cryogenic temperatures. The SPIFFI J-band spare grating is mounted on a rotatable stage in a test cryostat. Through an optical window (137mm diameter) the grating surface deformation is measured interferometrically in Littrow configuration via a static Fourier Fringe Analysis (FFA).}
		\label{fig:cryo_scheme}
	\end{center}
\end{figure}

For the measurements of the grating surface we used a FISBA \SI{}{\micro\nothing}Phase$\textsuperscript{\textregistered}$ 500 interferometer together with its beam expander lens providing an 152.4 mm diameter output beam. The interferometer is a Twyman-Green phase-shifting type, operated with a stabilized He-Ne laser at a wavelength of 632.8 nm. The interferometer and beam expander lens were placed outside of the cryostat and we measured the grating surface in Littrow configuration (see figure \ref{fig:cryo_scheme}) through an optical quality cryostat window. The diameter of the cryostat window is 137 mm, and thus is also the limiting size for the interferogram of the grating surface. 

The test cryostat at MPE could not be shielded easily against vibrations, and thus it was not possible to measure the grating surface in the phase scanning mode of the interferometer. We therefore used a static fringe analysis. Therefore the phase value of a certain pixel cannot be measured independently, and thus the phase sign gets lost, and the results are less accurate than a phase-shifting measurement. In exchange, only one image is needed to complete the measurement, so it is fast enough for us to obtain interferograms in unstable environments. We used the Fourier Fringe Analysis (FFA) \cite{macy83} which is very effective if the surface deformations are small ($< \lambda/5$). For this kind of static fringe analysis, a basic tilt has to be applied to the grating surface resulting in a carrier frequency in the interferogram. These fringes are afterwards removed in the Fourier domain to obtain a surface deviation map. The filtering used in the Fourier domain affects the final P/V value measured for higher-frequency deformations, and thus care must be taken to ensure an accurate measurement. 

\begin{figure}[htbp!]
	\begin{center}
		\resizebox{0.5\textwidth}{!}{
			\includegraphics[height=7cm, trim={0 0.1cm 0 0.1cm}, clip]{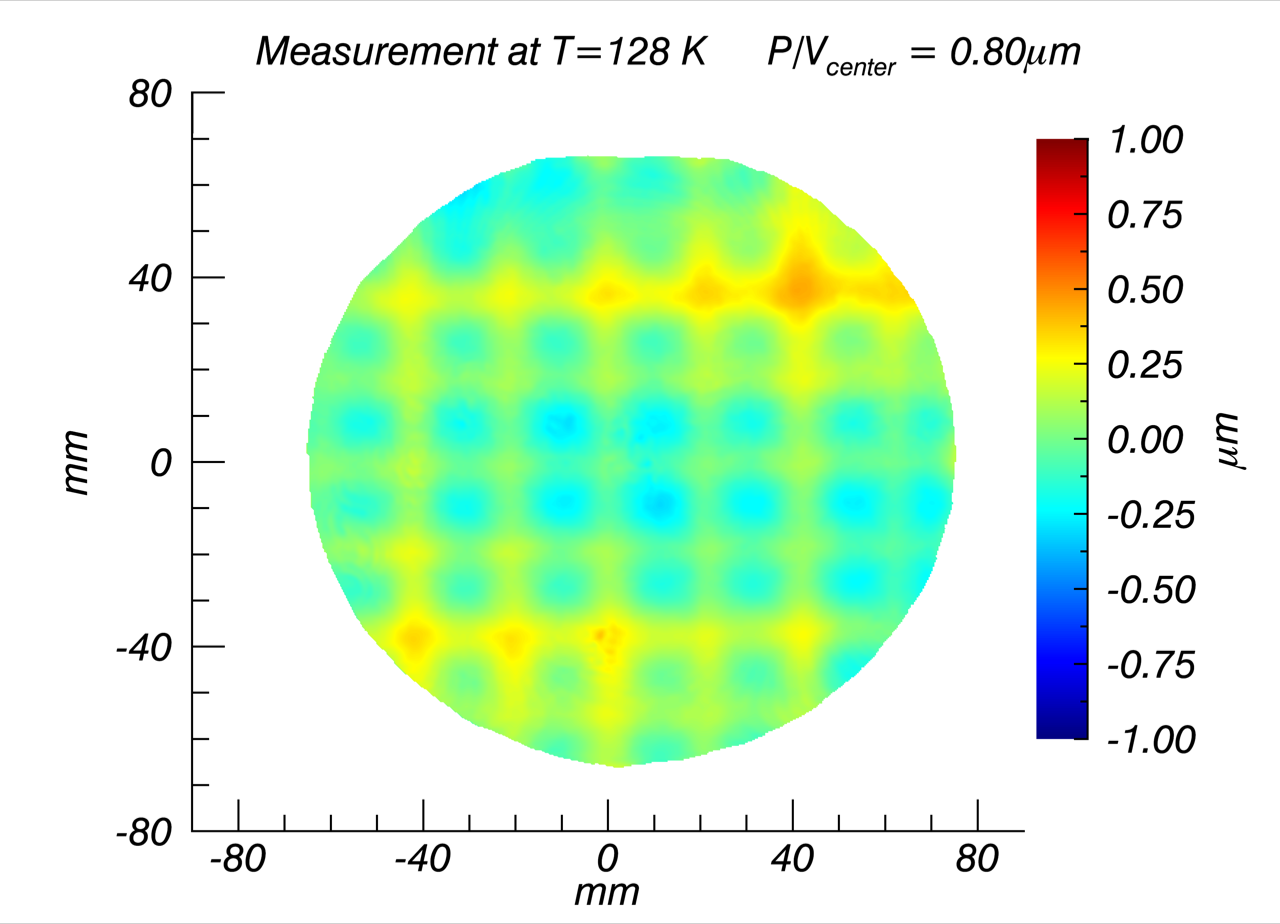}
		}
		\caption[Grating surface deformation from the cryogenic measurement]{Image of the surface deformation from cooling the SPIFFI spare J-band grating from room temperature to 128K. The measurement shown is (Cold surface) - (Warm surface) to show only the effects of cryogenic deformation. The aperture of the measurement was limited by the size of the cryostat window. The P/V$_{\mathrm{center}}$ of 0.80 \SI{}{\micro\meter} is calculated over the same area as the FEA. The color scale on this measurement is the same as in figure \ref{fig:deformation} for easy comparison.}
		\label{fig:deformationmeas}
	\end{center}
\end{figure}

The result of the FFA measurement is shown in figure \ref{fig:deformationmeas}. The surface deviation shown is the difference between the cold measurement at 128 K and the warm measurement at room temperature, in order to show only the deformation due to cryogenic bimetallic stresses. The P/V$_{\mathrm{center}}$ deformation is 0.80 \SI{}{\micro\meter}, however, this value is also affected by the oblique astigmatism that can be recognized in the interferogram. Without this astigmatism, the P/V value is approximately 0.60 \SI{}{\micro\meter}. This lies comfortably in the range of values simulated for NiP layer and polishing thicknesses provided by the manufacturer. 

In our measurement campaign, the lowest temperature reached at the grating was 103 K. The grating is heated by thermal radiation through the cryostat window, preventing the grating from reaching a lower temperature. The surface measurement taken at a temperature of 103 K was found to be identical to the one at 128 K to within the measurement uncertainty, though the quality of the 103K measurement was not as good as the measurement at 128K shown in figure \ref{fig:deformationmeas}. It should also be noted that this was a measurement of a single spare grating--each grating blank was polished separately, so it is likely that the NiP layer thicknesses are different on the different gratings.

\section{Optical Simulation}
\label{sec:opticalsimulation}  

Diffraction theory shows that a regular periodic structure, such as the measured deformation on our diffraction gratings, will result in a diffraction pattern with peaks located at $\mathrm{sin{\theta} = \lambda/D}$.\cite{bornwolf99} Since we have a periodic structure in both the X- and Y- directions, we expect that the PSF of the instrument will be diffracted in a grid pattern in the spatial {\textit{and}} spectral directions. We use CODE V optical simulation software to model the effect of the grating surface deformation on the line profiles of the instrument. 

\subsection{Setup and Parameters}
\label{sec:simSetup}  

In order to simplify the model, we adopt an approximate setup for simulating the spectrograph. We assume a point source at the location of the image slicer. The light is collimated by a CODE V lens module with a focal length of 2873 mm (identical to the instrument collimator focal length, though in the instrument the collimator is a Three Mirror Anastigmat (TMA)). We do not simulate the full TMA, as we have measured the TMA wavefront from the location of the spectrograph pupil and include it in the simulations as an interferogram applied to the CODE V collimator lens module. The diffraction gratings for the J, H, K, and H+K bands are setup as listed in table \ref{tab:gratparams}. The diffracted light is then imaged onto the focal plane by a CODE V lens module with a focal length of 342 mm, identical to the instrument camera. The setup allows us to place an interferogram of the surface deformation of the grating on the CODE V grating surface. Due to the anamorphic distortion of the diffraction gratings, the slit width geometrically imaged onto the detector is 27 \SI{}{\micro\meter} using the J, H, and K band gratings and 29 \SI{}{\micro\meter} using the H+K band grating.

\begin{figure}[htbp!]
\begin{center}
\resizebox{0.8\textwidth}{!}{
\includegraphics[width=1.0\textwidth,trim={0 0 0 0},clip=true]{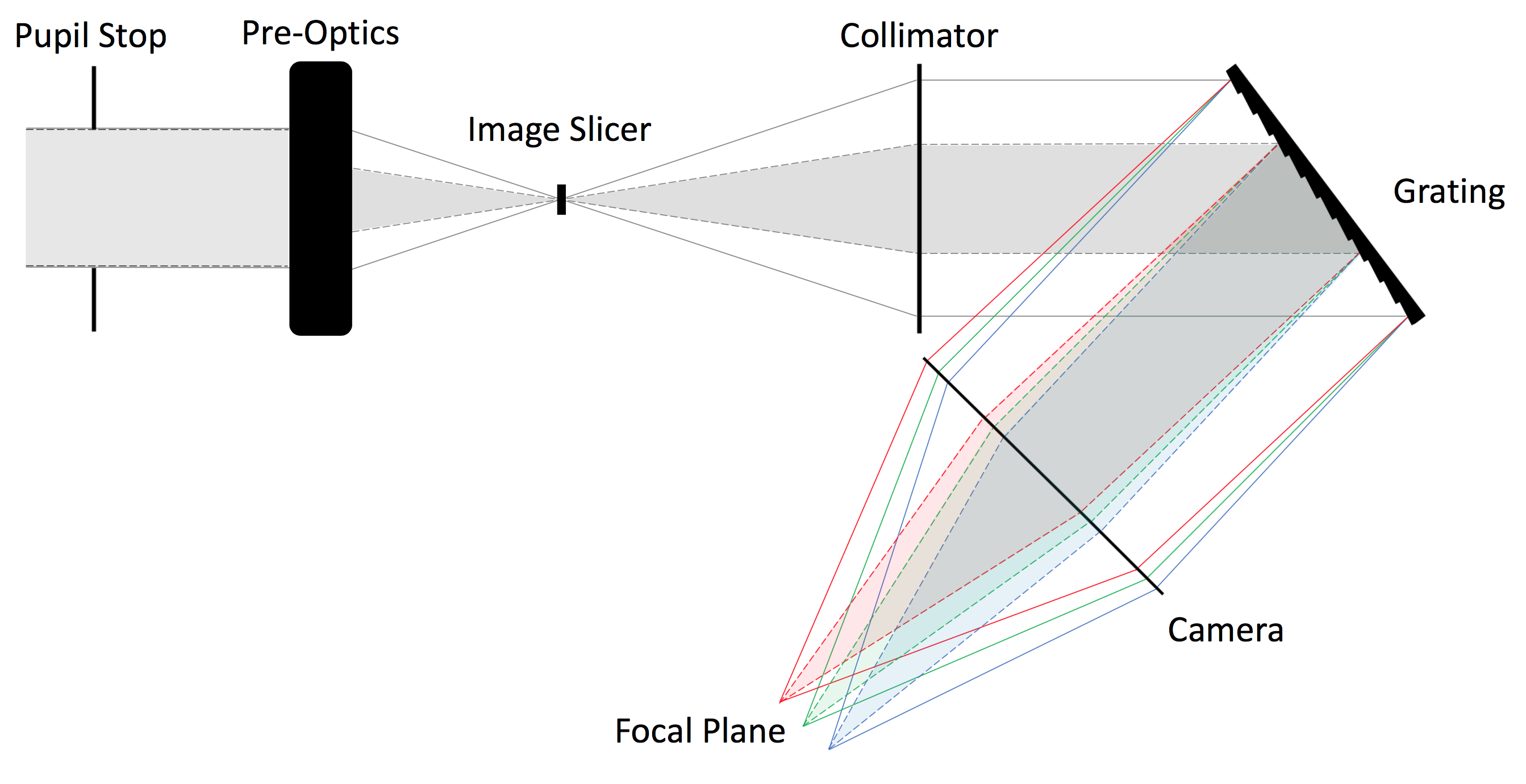}
}
\vspace{0.1cm}
\caption[Layout of the optical simulation in CODE V for the J-band grating]{Layout of the optical simulation in CODE V for the J-band grating (not to scale). All other gratings were setup in the same way. The wavelengths shown in blue, green, and red are 1.10, 1.25, and 1.40 \SI{}{\micro\meter}, coving the J-band. The pre-optics change the NA of the light entering the image slicer. The shaded area shows the geometric light path for the 100 mas pixel scale, while the solid (outer) lines show the light path for the 250 mas pixel scale. The 25 mas pixel scale is not shown, but would have an even smaller light cone than the 100 mas pixel scale.}
\label{fig:CodeVlayout}
\end{center}
\end{figure}

Figure \ref{fig:CodeVlayout} shows the layout of the simulation for the case of the J-band grating. The figure also shows a ``black box" pre-optics and the instrument cold stop. The pre-optics act as a scale changer for the incoming light, and for a fully illuminated slit (the case for our spectral line profile measurements) the effect is to change the Numerical Aperture (NA) of the light entering the spectrograph slit for the different pixel scales. We account for this effect by using the CODE V 2D Partial Phase Coherence function, which allows the user to specify the relative NA of the incoming light beam and an effective slit size. In figure \ref{fig:CodeVlayout}, the shaded area shows the geometric light path for the 100 mas pixel scale, while the solid (outer) lines show the light path for the 250 mas pixel scale.

\begin{table}[htb]
\begin{center}
\begin{tabular}{ccccc}
\hline
Band  & Grating spacing & Wavelengths & Diffraction Order& Incidence Angle\\
 & [\SI{}{\micro\meter}] & [\SI{}{\micro\meter}] & & [Degrees]\\
\hline
\hline
J & 5.758 & (1.10, 1.25, 1.40) & -2 & -36.0908\\
H & 7.747 & (1.45, 1.65, 1.85) & -2 & -35.8277\\
K & 9.684 & (1.95, 2.20, 2.45) &-2 & -36.7344\\
H+K & 9.961 & (1.45, 1.95, 2.45) &-1 & -28.5817\\
 \hline
\end{tabular}
\vspace{0.05in}
\caption[Table of grating parameters for the SPIFFI instrument used in our CODE V simulation]{Table of grating parameters for the SPIFFI instrument used in our CODE V simulation. For the gratings, (Reflection Angle - Incidence Angle) = 45 Degrees. All parameters in the simulation are for the 80K operating temperature.}
\label{tab:gratparams}
\end{center}
\end{table}

Unless otherwise noted, throughout this section the default grating deformation used in the simulation is the one calculated by the FEA with an original layer thickness of 200 \SI{}{\micro\meter} and 75\% of the layer thickness (150 \SI{}{\micro\meter}) remaining after polishing. Additionally, throughout this section we show results only for the J- and K- bands, as the behaviour of H-band is intermediate between the two. H-band results are discussed in section \ref{sec:simmatch}.

\subsection{Point Spread Function}
\label{sec:PSF}  

\begin{figure}[htbp!]
\begin{center}
\resizebox{1.0\textwidth}{!}{
\includegraphics[height=7cm, trim = {0 0 3.5cm 0.1cm}, clip]{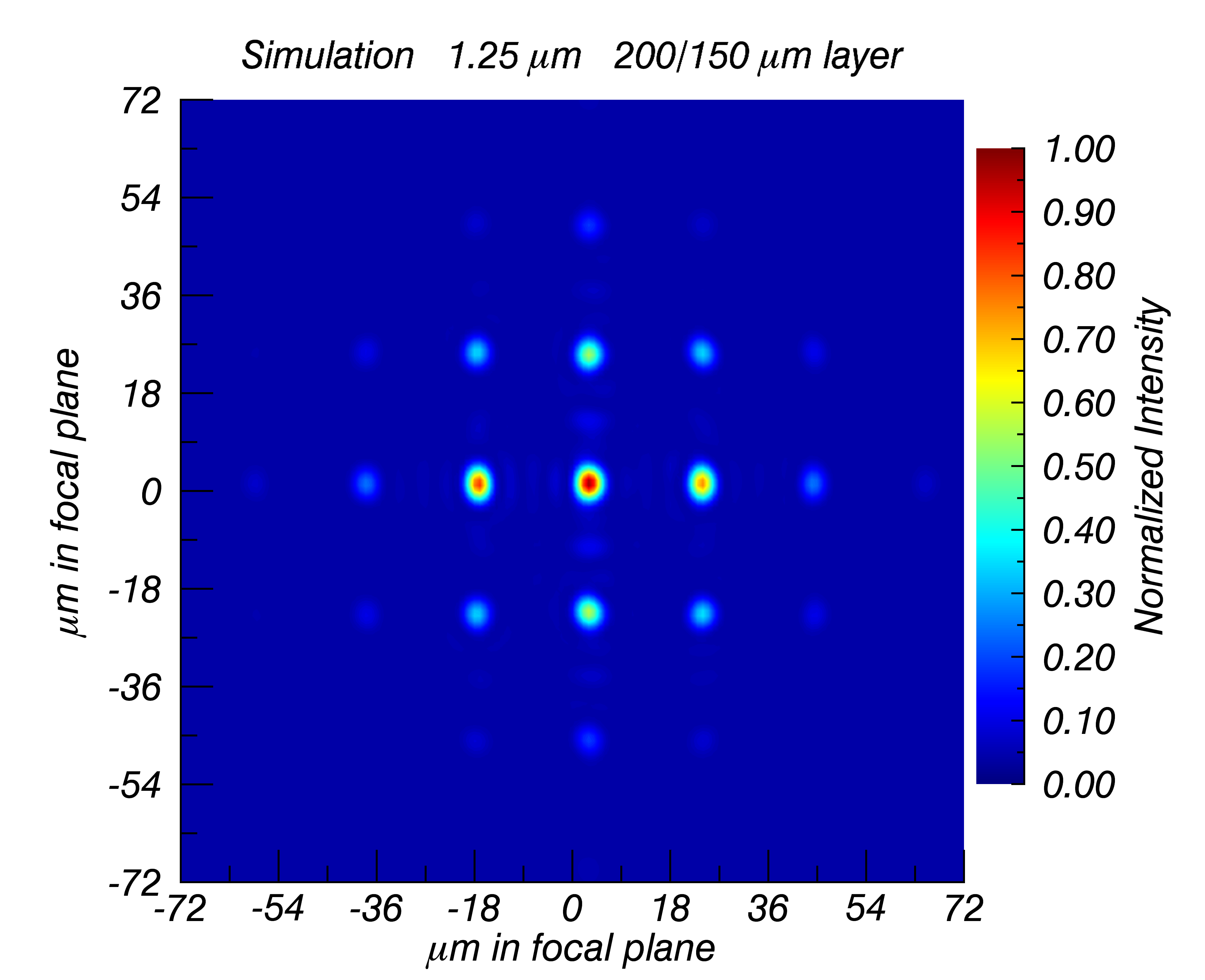}
\includegraphics[height=7cm, trim = {3.0cm 0 0 0.1cm}, clip]{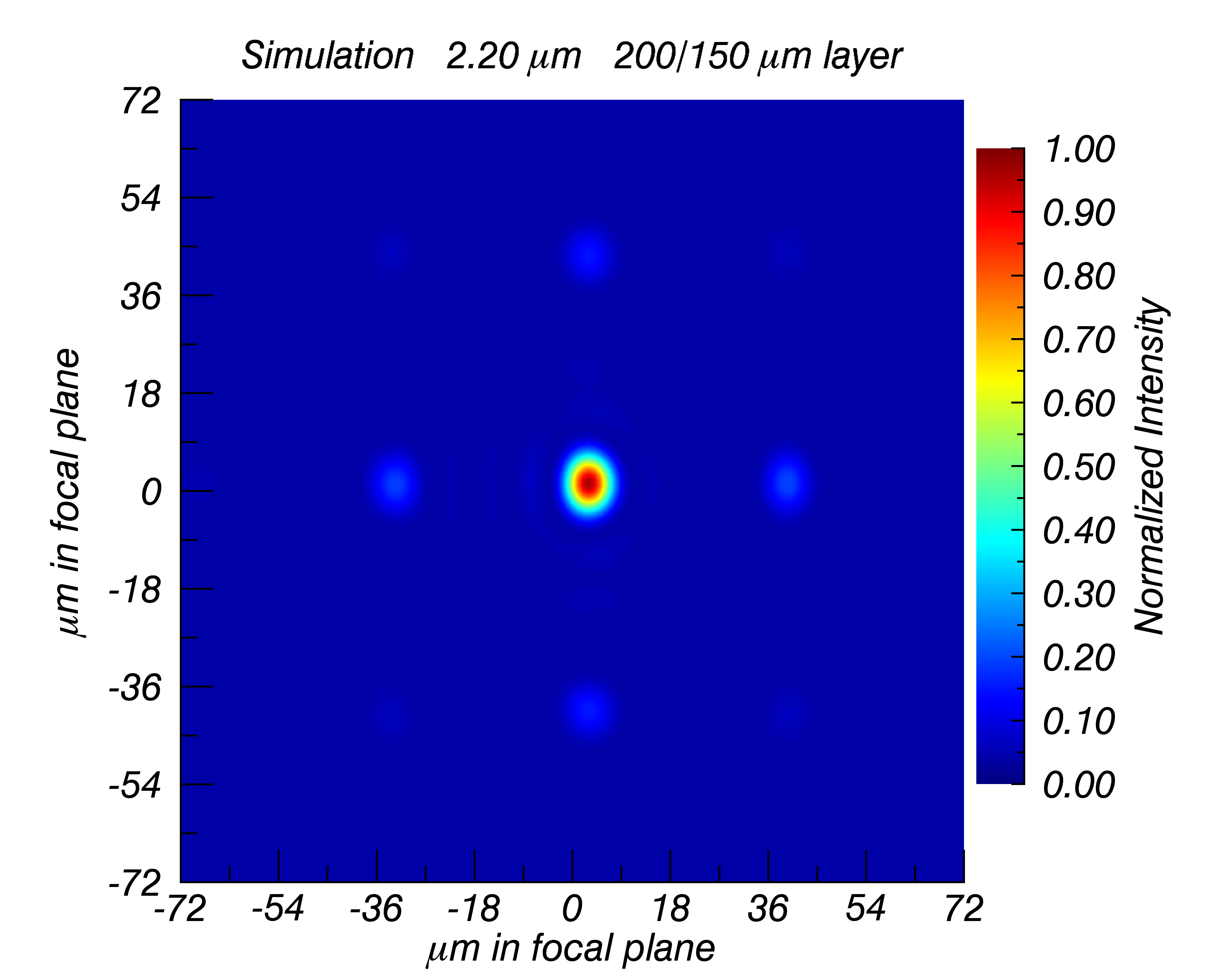}
}

\caption[Examples of the PSFs produced in the optical simulation]{Examples of the PSFs produced in the optical simulation for FEA calculated grating surface for wavelengths in the center of J-band (1.25 \SI{}{\micro\meter}, {\textbf{left}}) and K-band (2.2 \SI{}{\micro\meter}, {\textbf{right}}). The spectral direction is the X-direction, while the slit extends in the Y-direction.}
\label{fig:CodeVPSF}
\end{center}
\end{figure}

To see the diffraction effects of the grating surface deformation, we begin by calculating a PSF for a point source. On the grating, we place an interferogram of the grating surface deformation. Figure \ref{fig:CodeVPSF} shows the resulting PSFs for the deformation calculated by the FEA for the J- and K-band setups. Note that the grid structure of the PSF has a spacing that is a function of wavelength, as expected from diffraction theory. In J-band the spacing of the peaks is approximately one detector pixel (18 \SI{}{\micro\meter}) in the focal plane, while in K-band the spacing is approximately 2 detector pixels (36 \SI{}{\micro\meter}) in the focal plane.

At shorter wavelengths, diffraction orders farther from the center grow in strength relative to the central order. This is a general result expected for sinusoidal deformations as the wavelength approaches the deformation amplitude.\cite{bornwolf99} Additionally, for deformation amplitudes comparable to $\sim\lambda/2$ or larger, the central order(s) can be suppressed.\cite{bornwolf99, bonzel68} For the largest deformation amplitudes tested, we see central order suppression especially in J-band. This is explored in the next section.

\subsection{Line Spread Function}
\label{sec:LSF}  

The spectral line profile is the measured line spread function of the instrument. This section shows how the cryogenic deformation of the diffraction gratings affects the spectral line profile of the instrument.

\subsubsection{Basic LSF}
\label{sec:basicLSF}

\begin{figure}[htbp!]
\begin{center}
\resizebox{1.0\textwidth}{!}{
\includegraphics[width=1.0\textwidth]{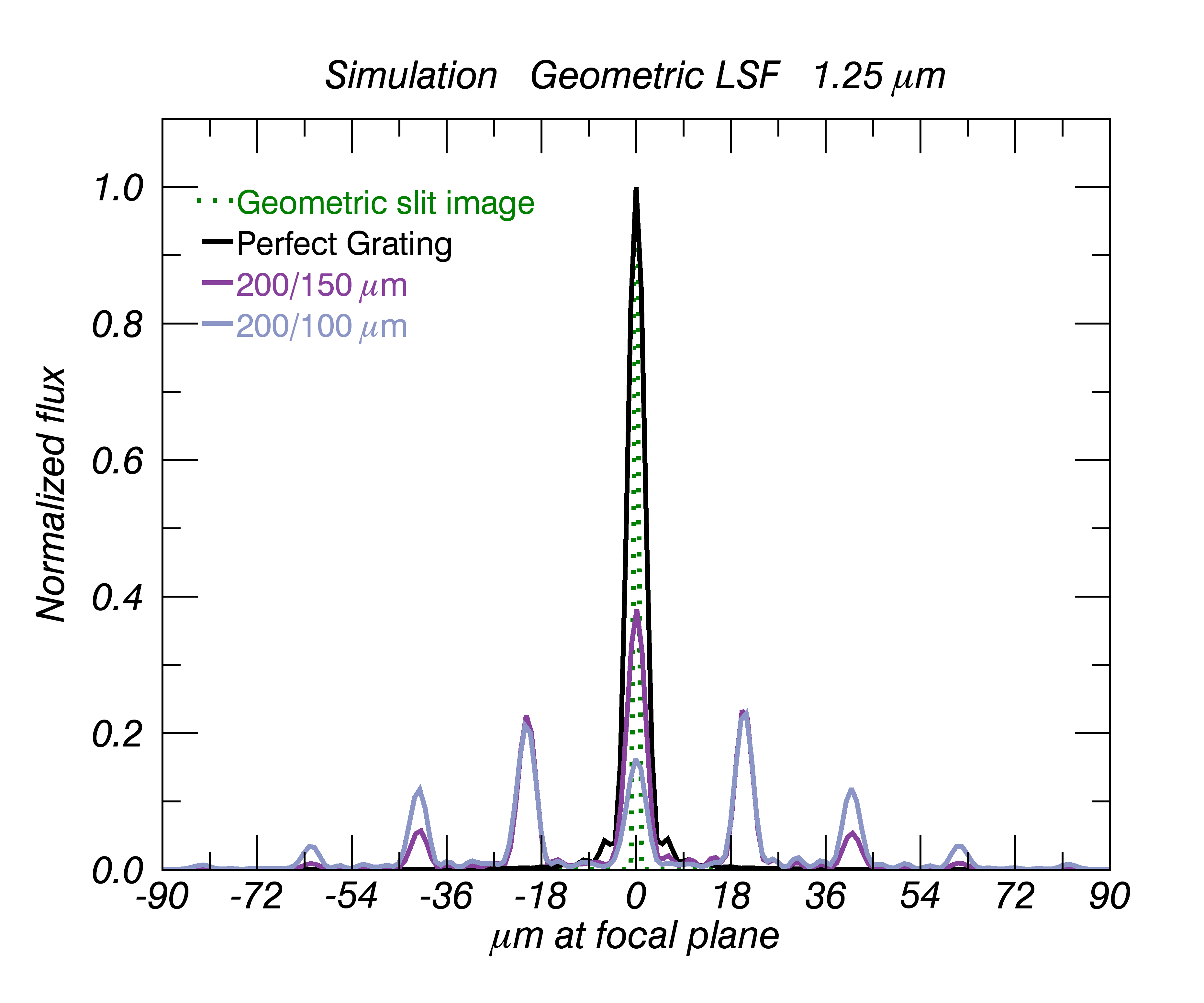}
\includegraphics[width=1.0\textwidth]{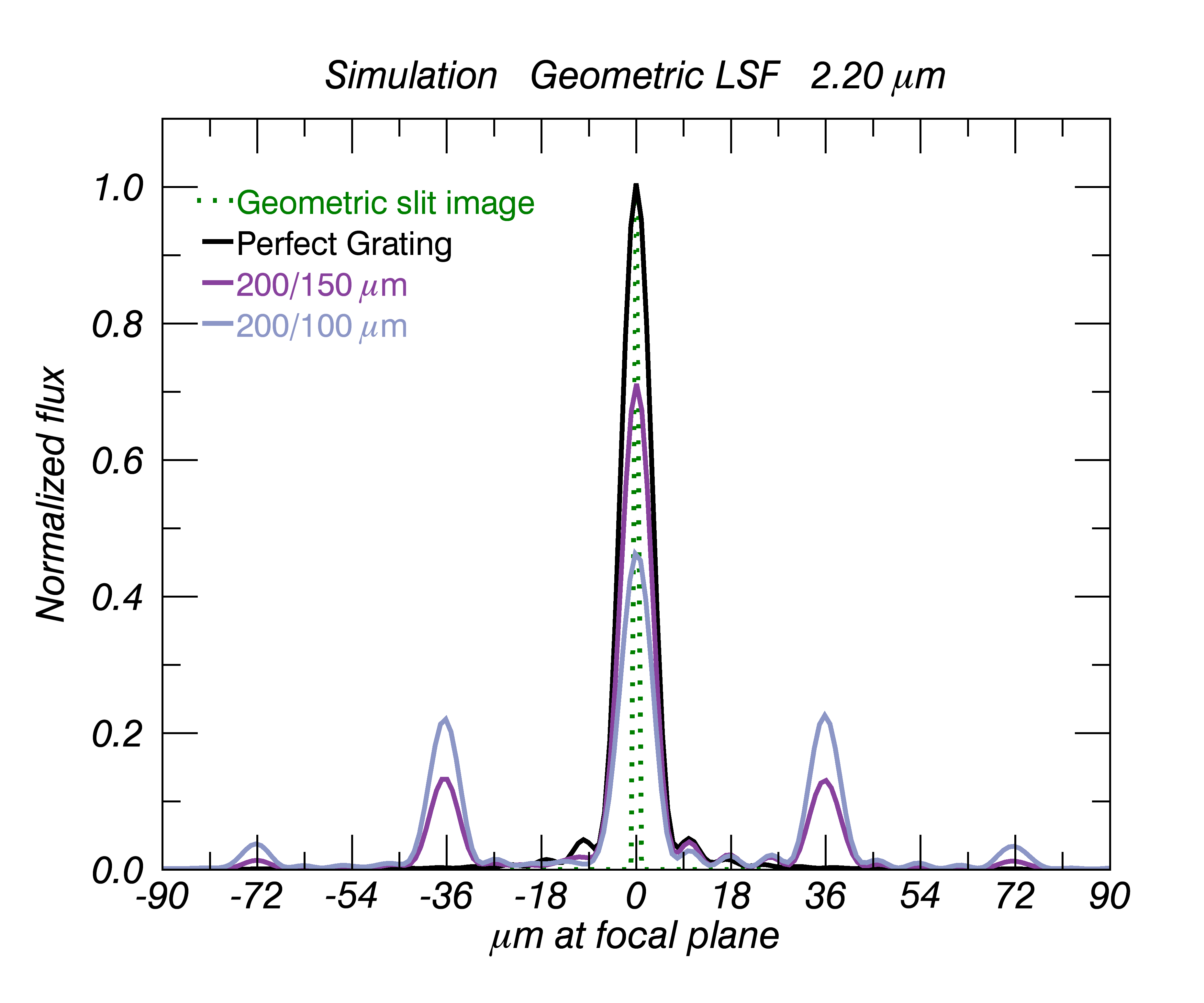}
}
\resizebox{1.0\textwidth}{!}{
\includegraphics[width=1.0\textwidth]{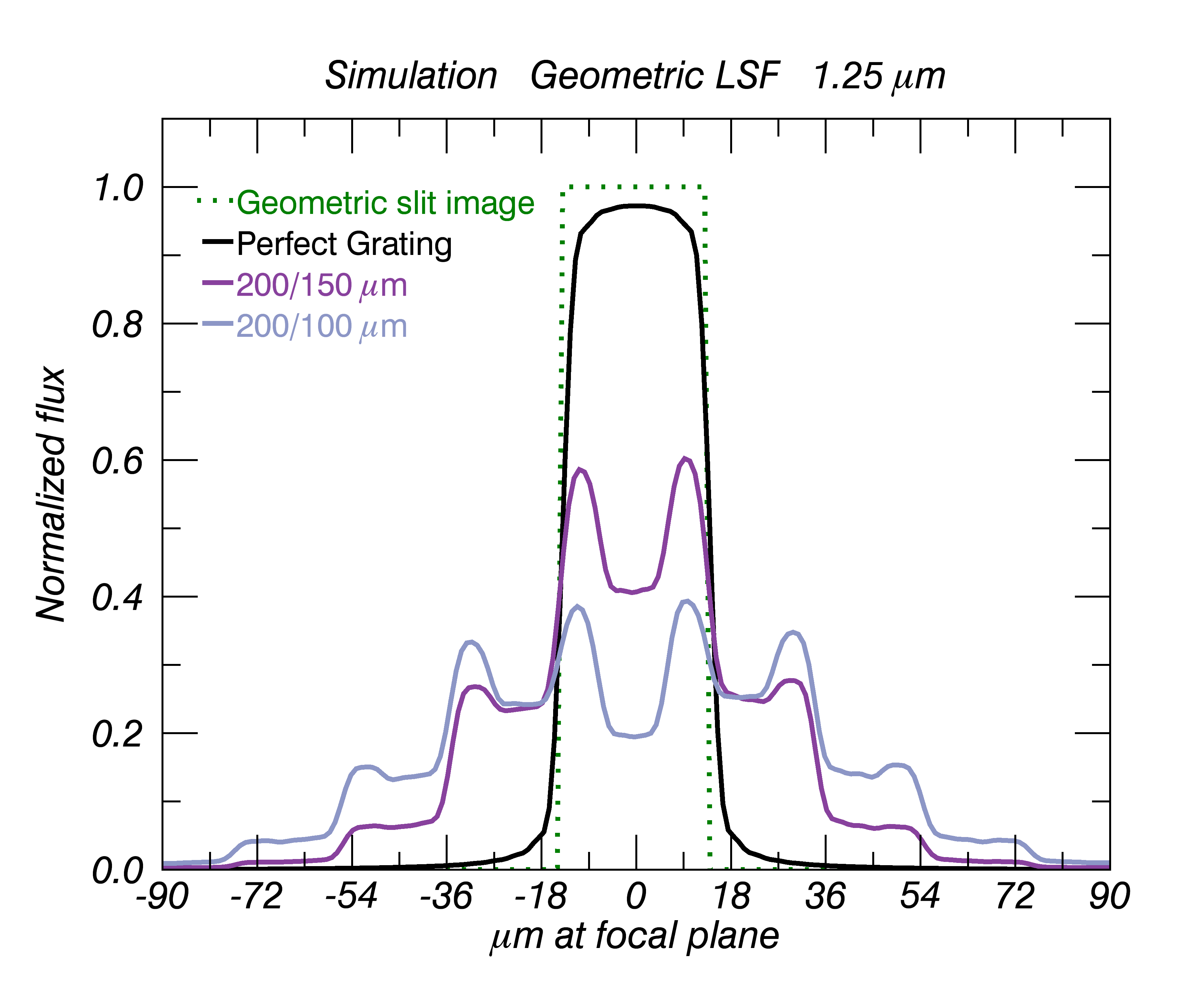}
\includegraphics[width=1.0\textwidth]{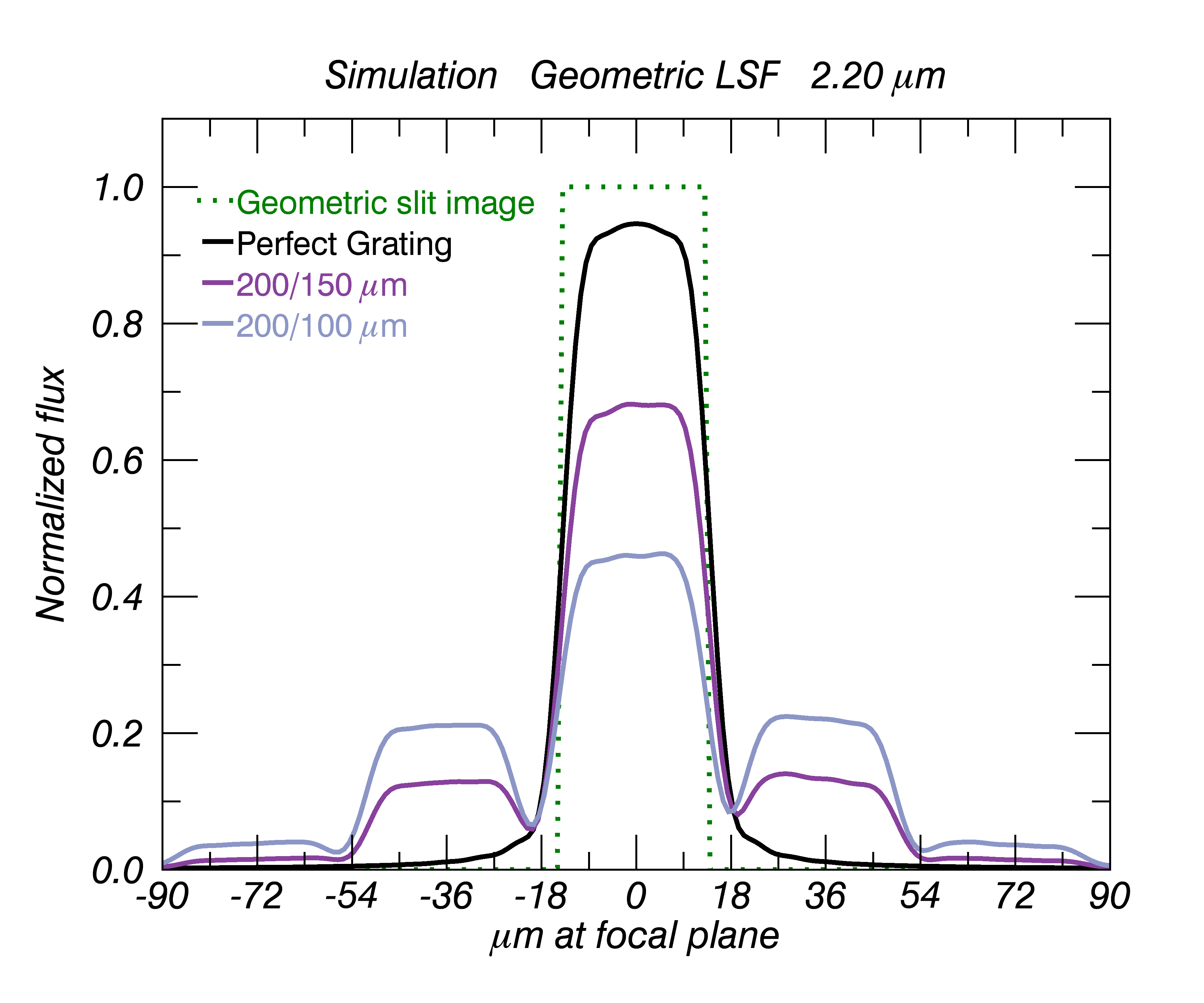}
}
\caption[Examples of basic LSFs produced by convolving the slit image with the PSFs produced by the optical simulation]{Examples of basic LSFs produced by convolving the slit image with the PSFs produced by the optical simulation for two different amplitudes of FEA calculated distortion as well as a ``perfect grating.'' The {\textbf{top row}} shows the result for an infinitely thin slit, while the {\textbf{bottom row}} shows the result for the SPIFFI slit width.  LSFs are shown for wavelengths at the center of J-band (1.25 \SI{}{\micro\meter}, {\textbf{left}}) and K-band (2.2 \SI{}{\micro\meter}, {\textbf{right}}). In general, as the amplitude of the grating deformation increases, power shifts from the central peak to the shoulders.}
\label{fig:CodeV_basicLSF}
\end{center}
\end{figure}

In its most basic form, the line spread function of the instrument is simply a cross section of the convolution of the PSF with the image of the slit in the focal plane. In SPIFFI, the slit image is 27 \SI{}{\micro\meter} wide in the focal plane. To build intuition about how the amplitude of the grating deformation affects the line profile of the instrument, in figure \ref{fig:CodeV_basicLSF} we plot this ``basic LSF" for both an infinitely thin slit (equivalent to collapsing the PSF plotted in figure \ref{fig:CodeVPSF} along the slit direction), and the actual SPIFFI slit size for two different amplitudes of the grating deformation.  In general, as the amplitude of the deformation increases, more power shifts from the central peak to the shoulders, and for large amplitudes the central diffraction order is suppressed. 

By construction, the LSF produced in this way does not include any diffraction effects from the slit. The ``horns" that appear in the J-band LSF in the lower left plot in figure \ref{fig:CodeV_basicLSF} are a result of the fact that the spacing of the diffraction orders is smaller than the slit width in J-band, and they apprear purely from the convolution of the PSF shown in figure \ref{fig:CodeVPSF} with the slit image. In the next section, we will include diffraction effects from the slit.

\subsubsection{Partial Phase Coherence LSF}
\label{sec:PPCLSF}

As discussed in section \ref{sec:lineProfiles}, the line profiles vary depending on the pixel scale selected (25, 100, or 250 mas/px). Since our line profile measurements are taken with the slit fully illuminated by a flat-field source, the only difference between the three pixel scales is the numerical aperture (NA) of the incoming light illuminating the slit. This has two consequences. The first effect can be seen in figure \ref{fig:CodeVlayout}, where for smaller pixel scales the geometric light path through the instrument results in smaller beam footprints on the spectrograph optics. The second is that the incoming light at the image slicer will diffract differently at the slit edges depending on the illumination NA. We can model both effects in CODE V using the 2D partial phase coherence (PPC) function, which includes the full diffraction effects of partially coherent light propagating through the optical system. 

In the setup of the 2D PPC, we use a rectangular object that has an equivalent slit width of 27 \SI{}{\micro\meter} and length of 400 \SI{}{\micro\meter}\footnote{The actual SPIFFI slit length is 1152 \SI{}{\micro\meter} in the focal plane, but the 400 \SI{}{\micro\meter} length reduces simulation times and is sufficiently long to avoid edge effects.} in the focal plane. For the three pixel scales [250, 100, 25] mas/px, we use a relative NA of the incoming light of [1.0, 0.4, 0.1] with respect to the 250 mas pixel scale. The left column of figures \ref{fig:CodeV_2DPPC_LSF_nograt} and \ref{fig:CodeV_2DPPC_LSF} shows the LSF in the center of the slit that results from the CODE V partial phase coherence calculation in the three pixel scales for the case of a perfect grating (figure \ref{fig:CodeV_2DPPC_LSF_nograt}) and a single amplitude of the grating deformation (figure \ref{fig:CodeV_2DPPC_LSF}).

In general, diffraction at the slit edges is largest in the smallest pixel scale due to the smaller NA of the illumination. This appears as ``ringing" at the slit edges. The ringing is larger at smaller pixel scales (smaller NA), and at longer wavelengths. The behaviour with NA and wavelength are both expected from diffraction theory.\cite{bornwolf99} Note that in the J-band plots of figure \ref{fig:CodeV_2DPPC_LSF}, the "ringing" is confused with the ``horns" (which are a result of the diffraction order spacing). An additional effect that appears in the 2D PPC simulation is that the size of the shoulders is significantly smaller in the 25 mas pixel than the larger pixel scales in both J and K band. This is due to the smaller beam footprint on the diffraction grating, minimizing the effect of the periodic deformation of the grating surface.

\afterpage{
\clearpage 
\begin{landscape}
\begin{figure}[htbp!]
\begin{center}
\resizebox{1.4\textwidth}{!}{
\includegraphics[width=1.0\textwidth]{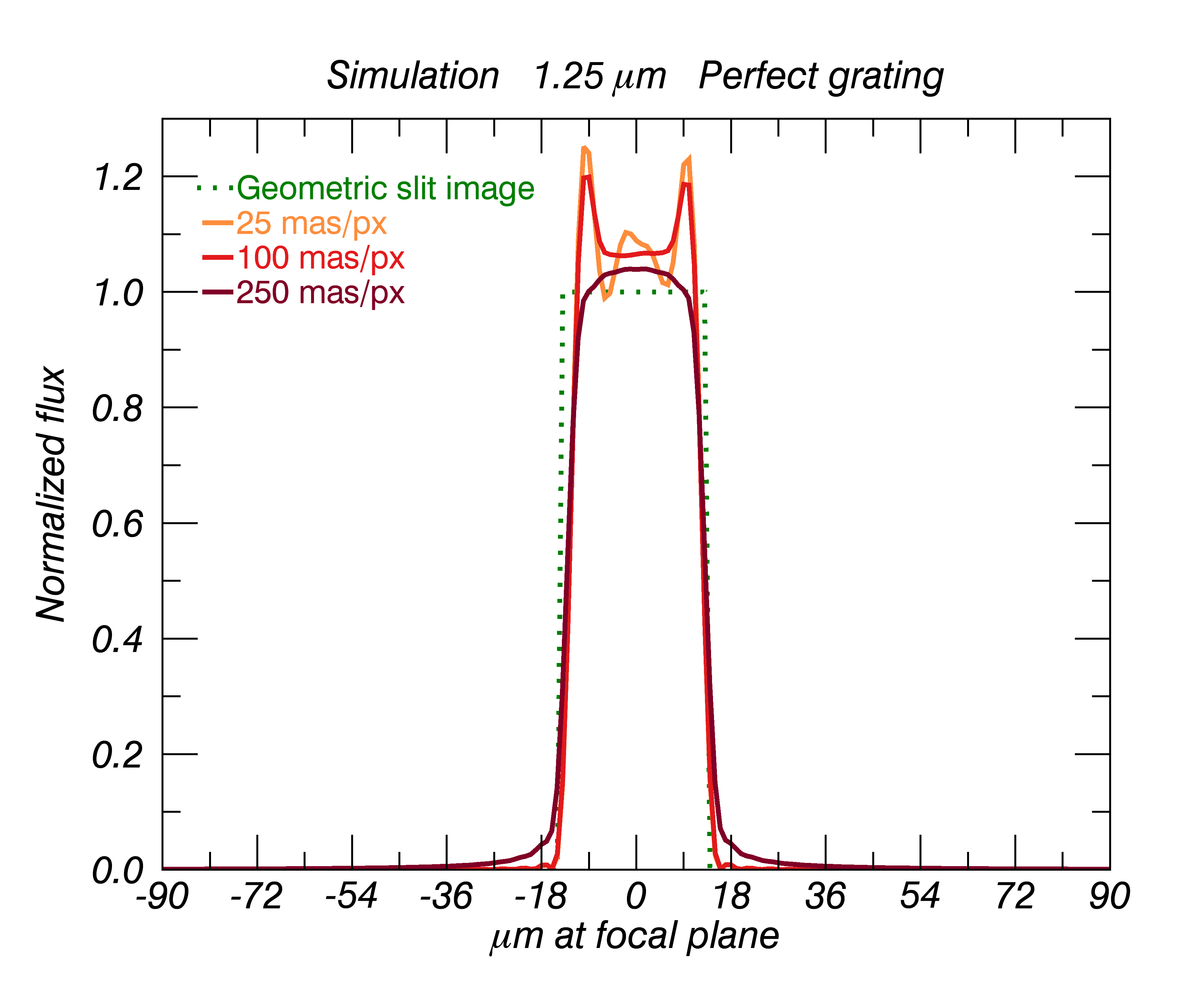}
\includegraphics[width=1.0\textwidth]{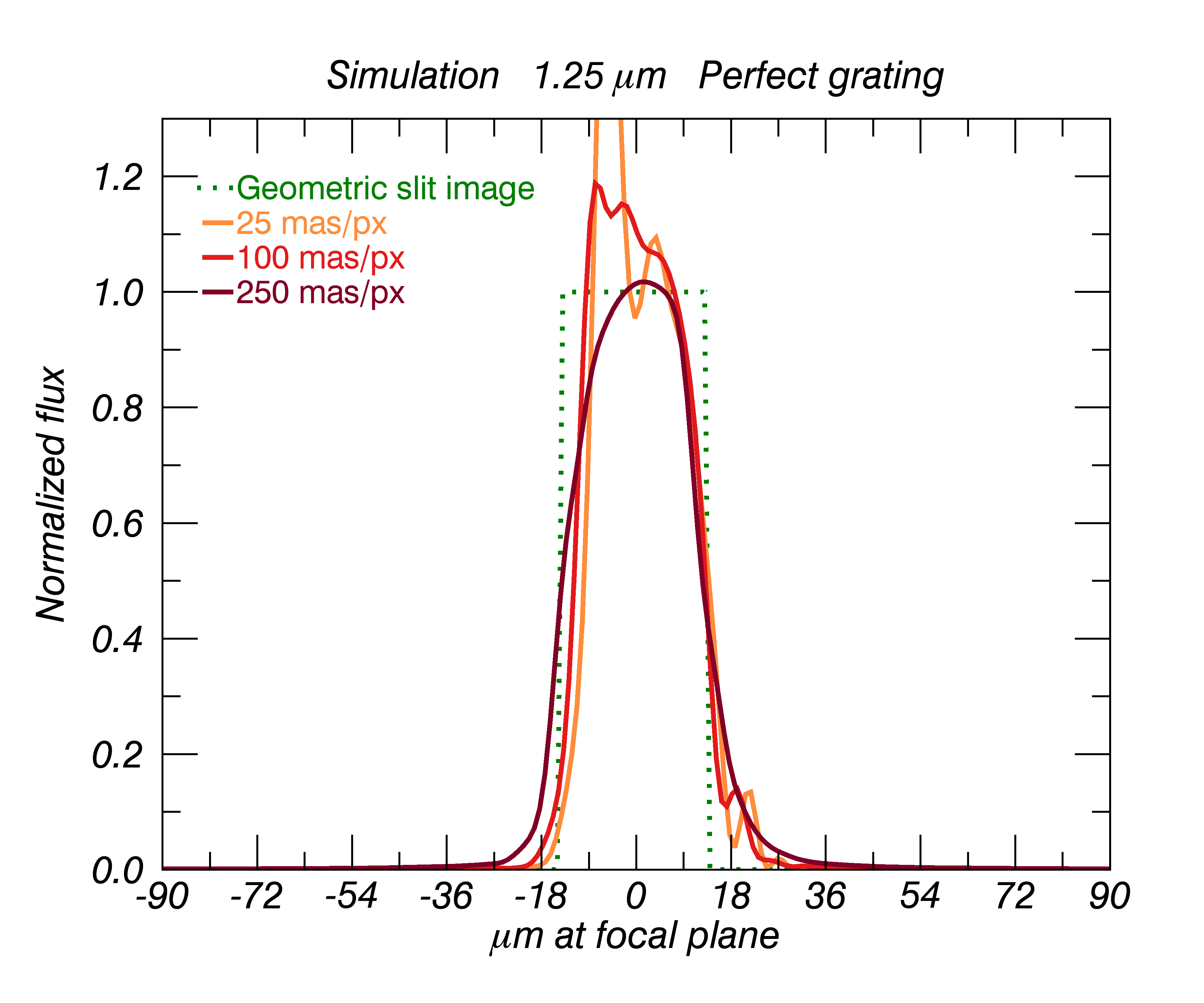}
\includegraphics[width=1.0\textwidth]{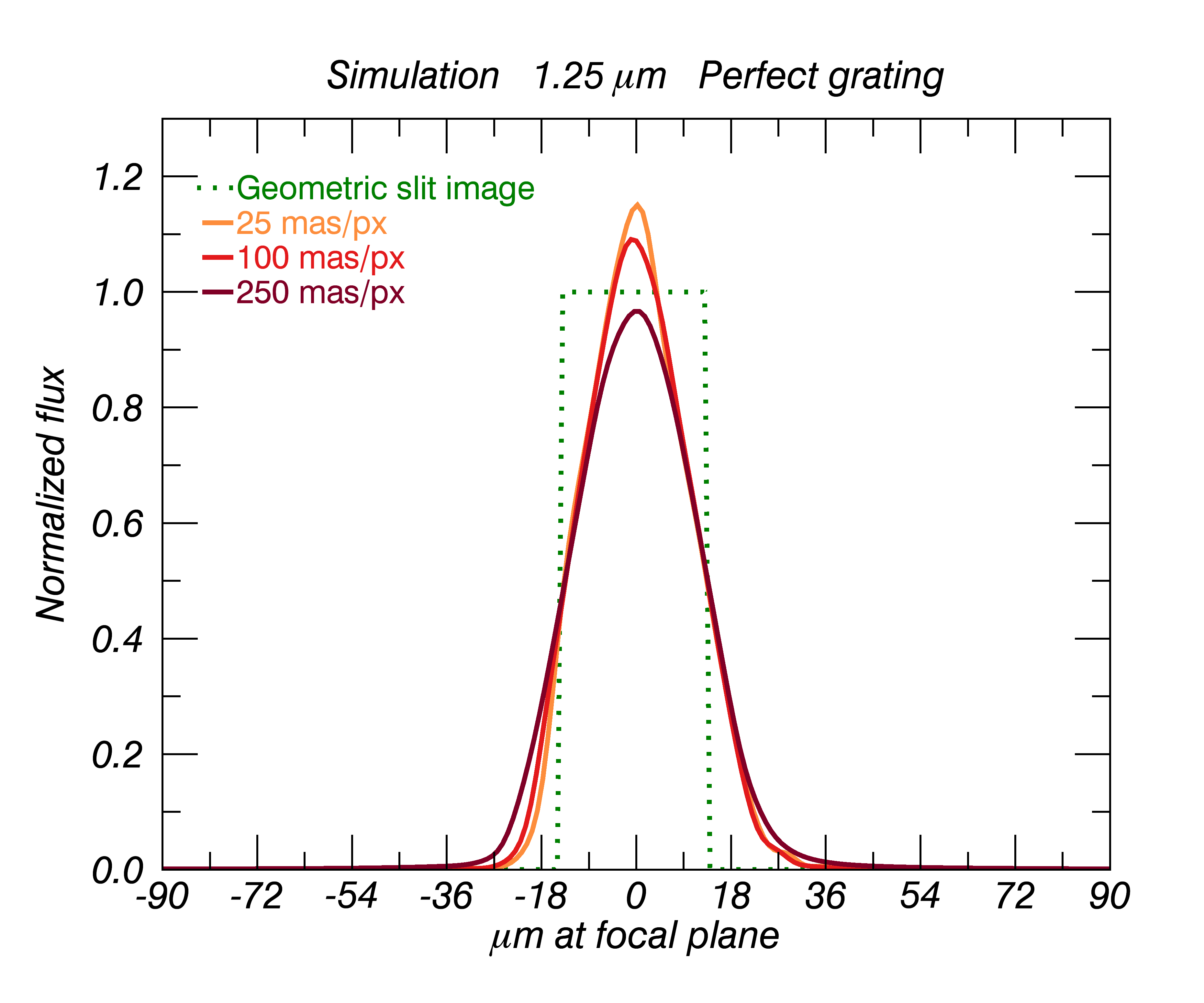}
}
\resizebox{1.4\textwidth}{!}{
\includegraphics[width=1.0\textwidth]{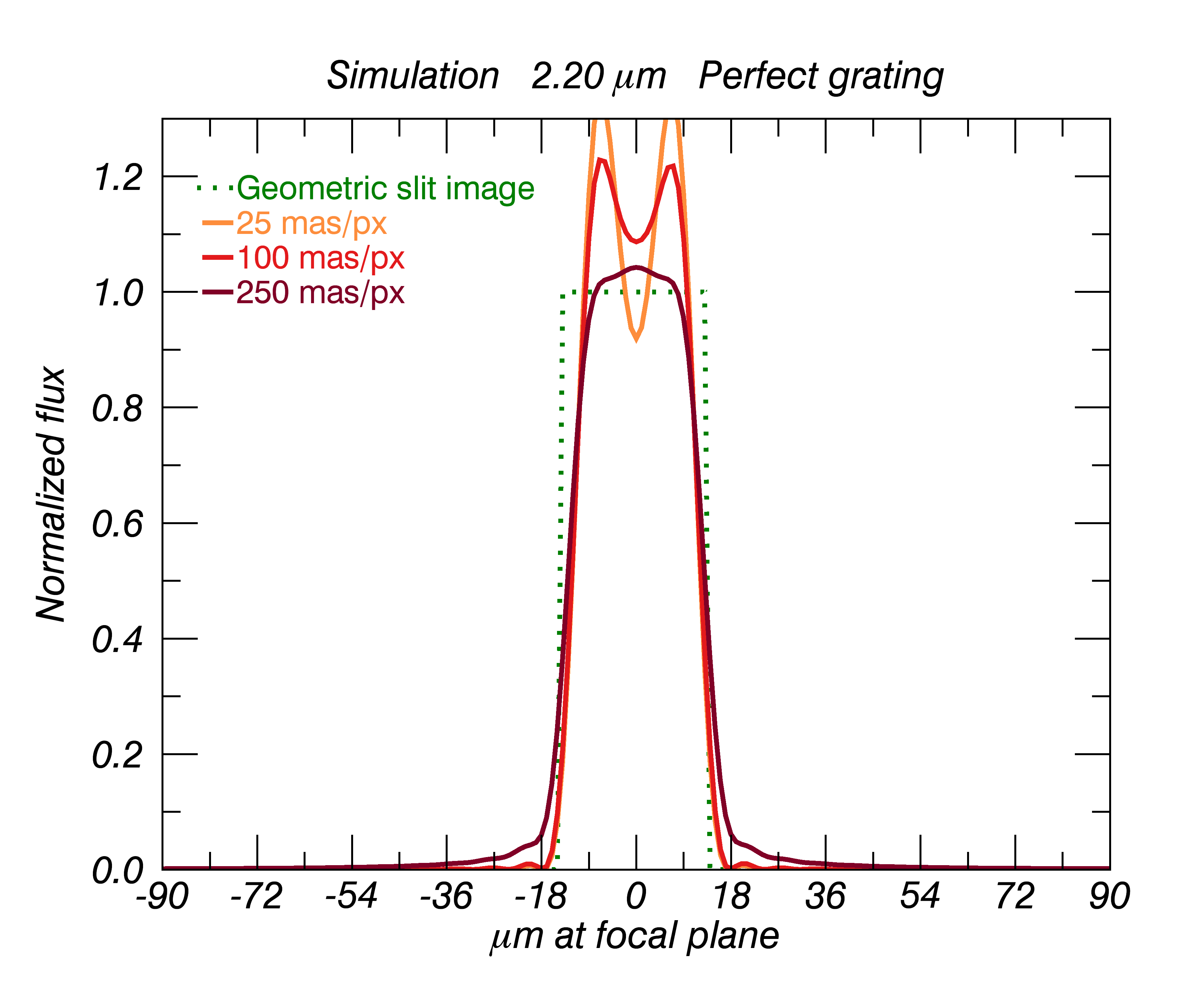}
\includegraphics[width=1.0\textwidth]{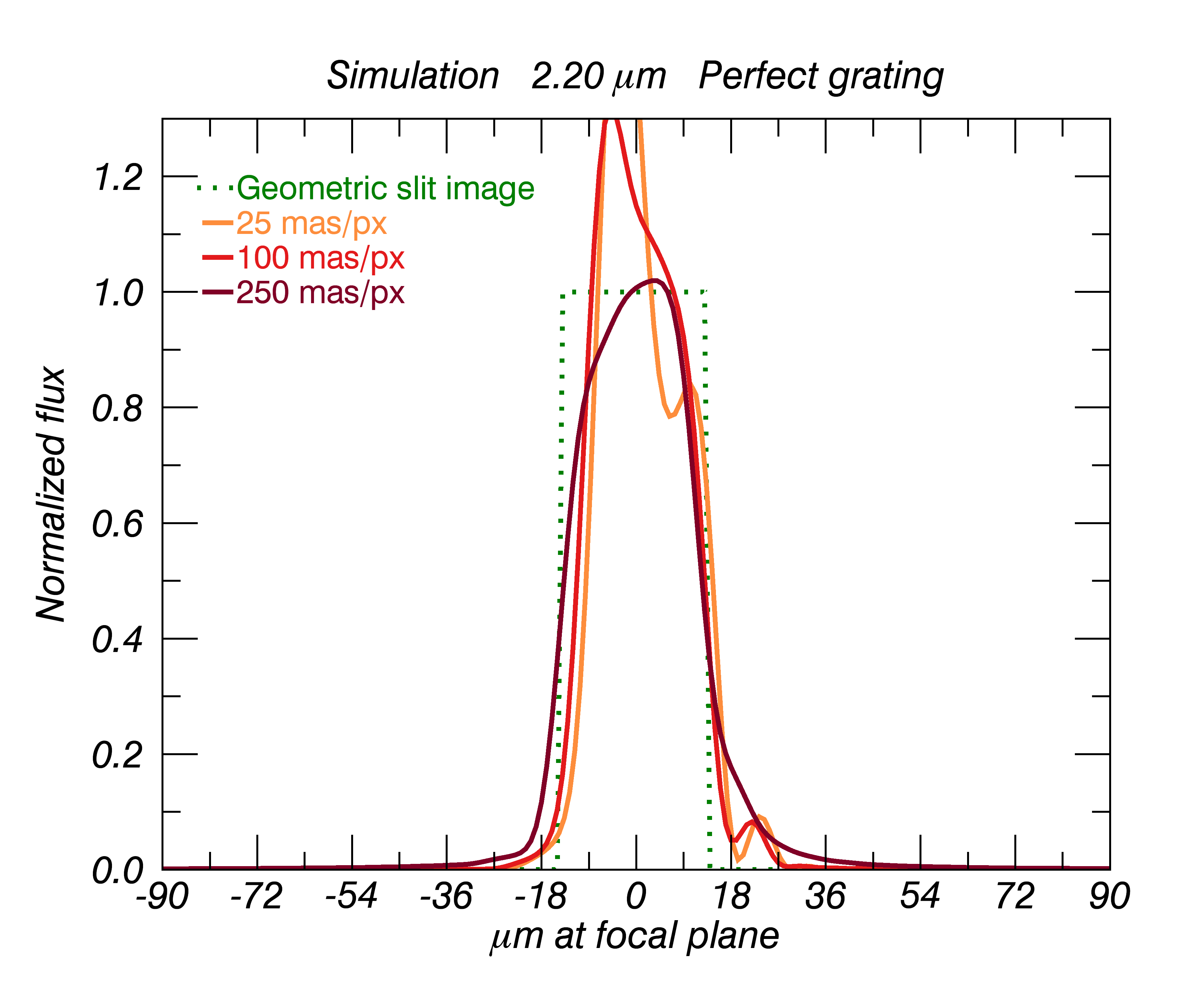}
\includegraphics[width=1.0\textwidth]{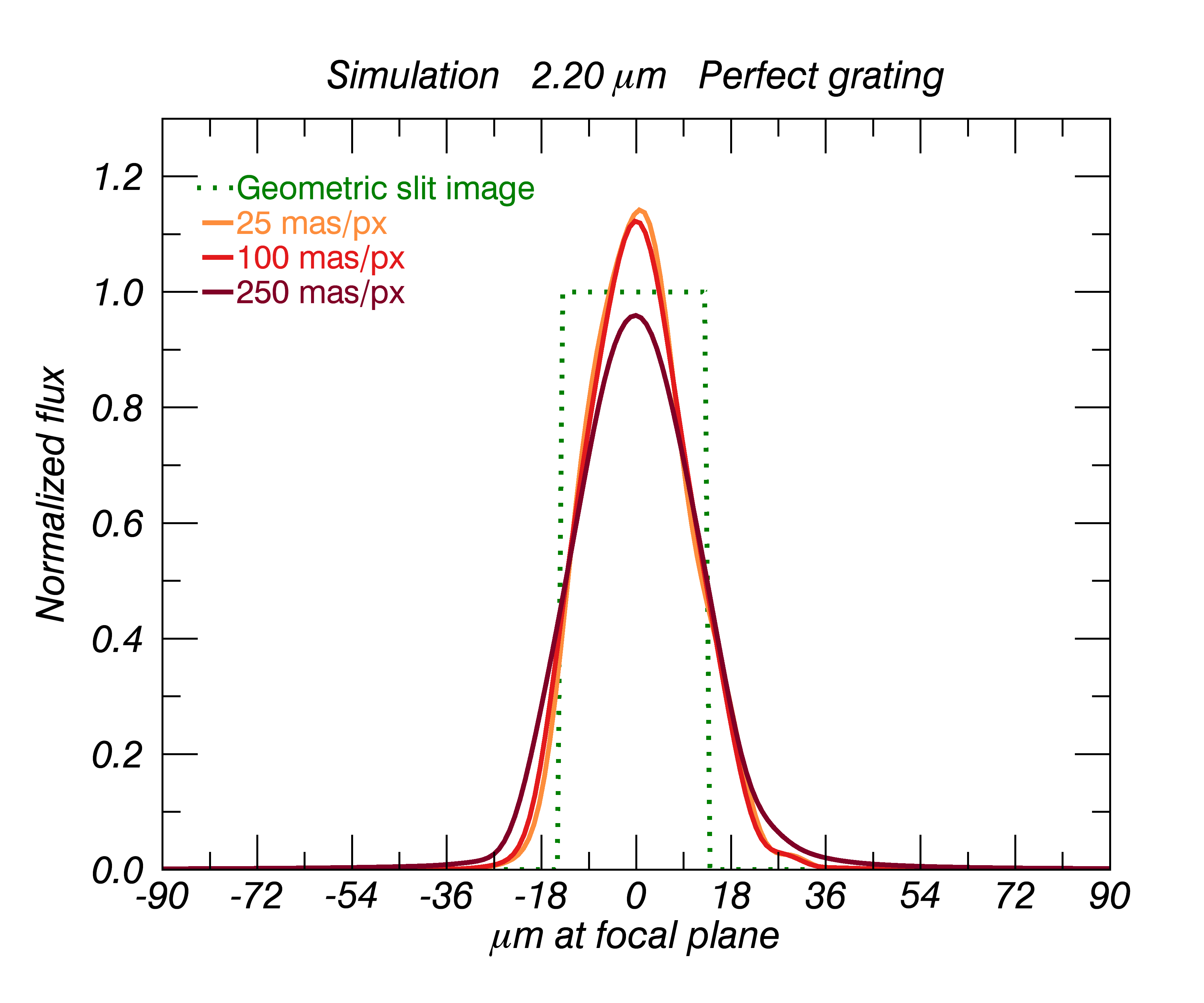}
}
\caption[Examples of 2D Partial Phase Coherence LSFs in all three pixel scales for a perfect grating]{Examples of 2D Partial Phase Coherence LSFs in all three pixel scales for a perfect grating. LSFs are shown in the {\textbf{top row}} for wavelengths at the center of J-band (1.25 \SI{}{\micro\meter}) and in the {\textbf{bottom row}} for the center of K-band (2.2 \SI{}{\micro\meter}). The {\textbf{left column}} includes the effects of Partial Phase Coherence, the {\textbf{middle column}} additionally includes the collimator wavefront error (measured on the mirrors installed in January 2016), and the {\textbf{right column}} additionally includes the effects of the detector pixel function, and thus depicts the simulated detected line profiles. The slit diffraction effects are largest in the 25 mas pixel scale in both J and K band, and the collimator wavefront error shifts power to the left. The detector pixel function smooths out the detected line profile, such that not much difference is seen between J and K band simulated detected line profiles for the case of a perfect grating. }
\label{fig:CodeV_2DPPC_LSF_nograt}
\end{center}
\end{figure}
\end{landscape}
}

\afterpage{
\clearpage 
\begin{landscape}
\begin{figure}[htbp!]
\begin{center}
\resizebox{1.4\textwidth}{!}{
\includegraphics[width=1.0\textwidth]{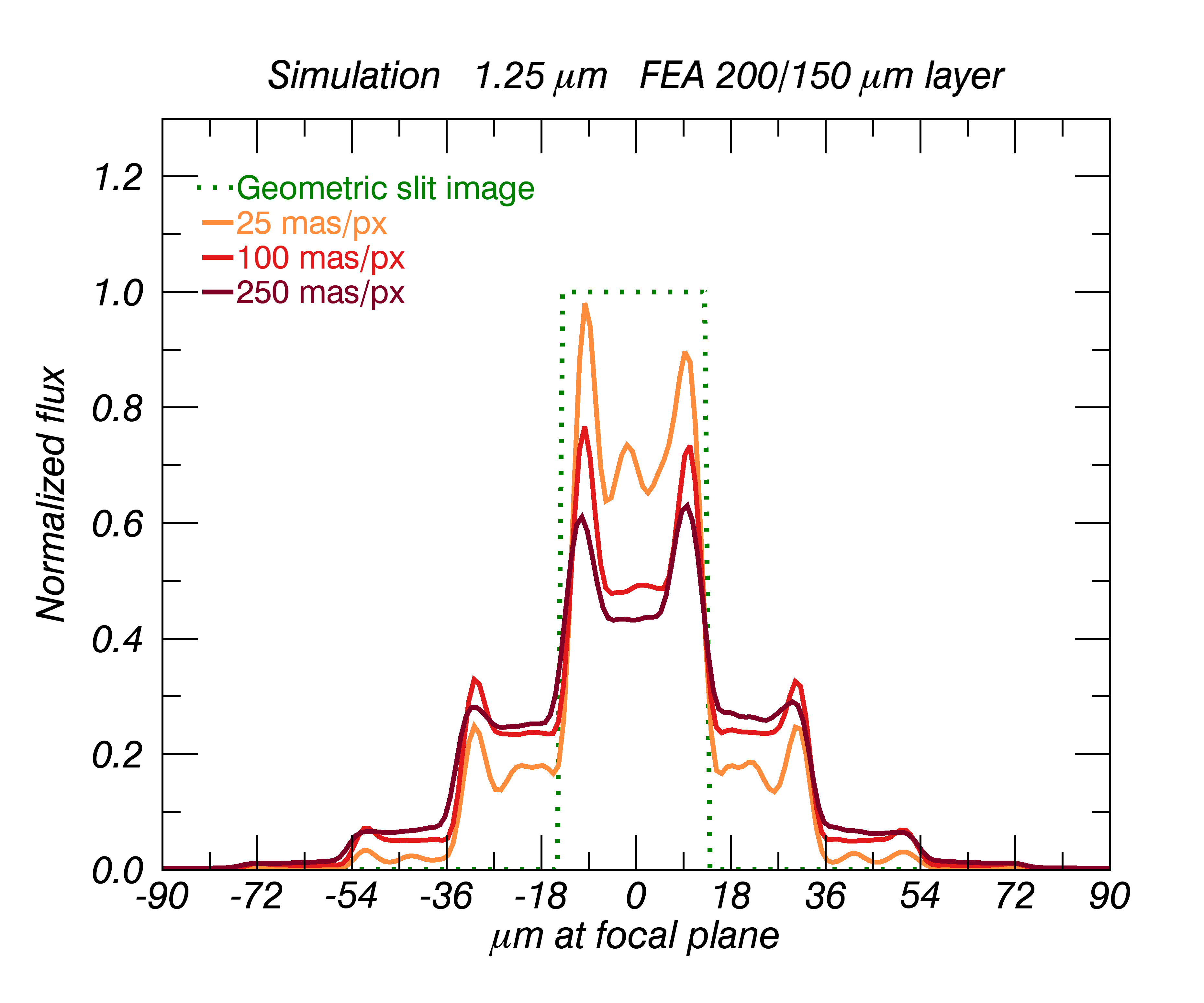}
\includegraphics[width=1.0\textwidth]{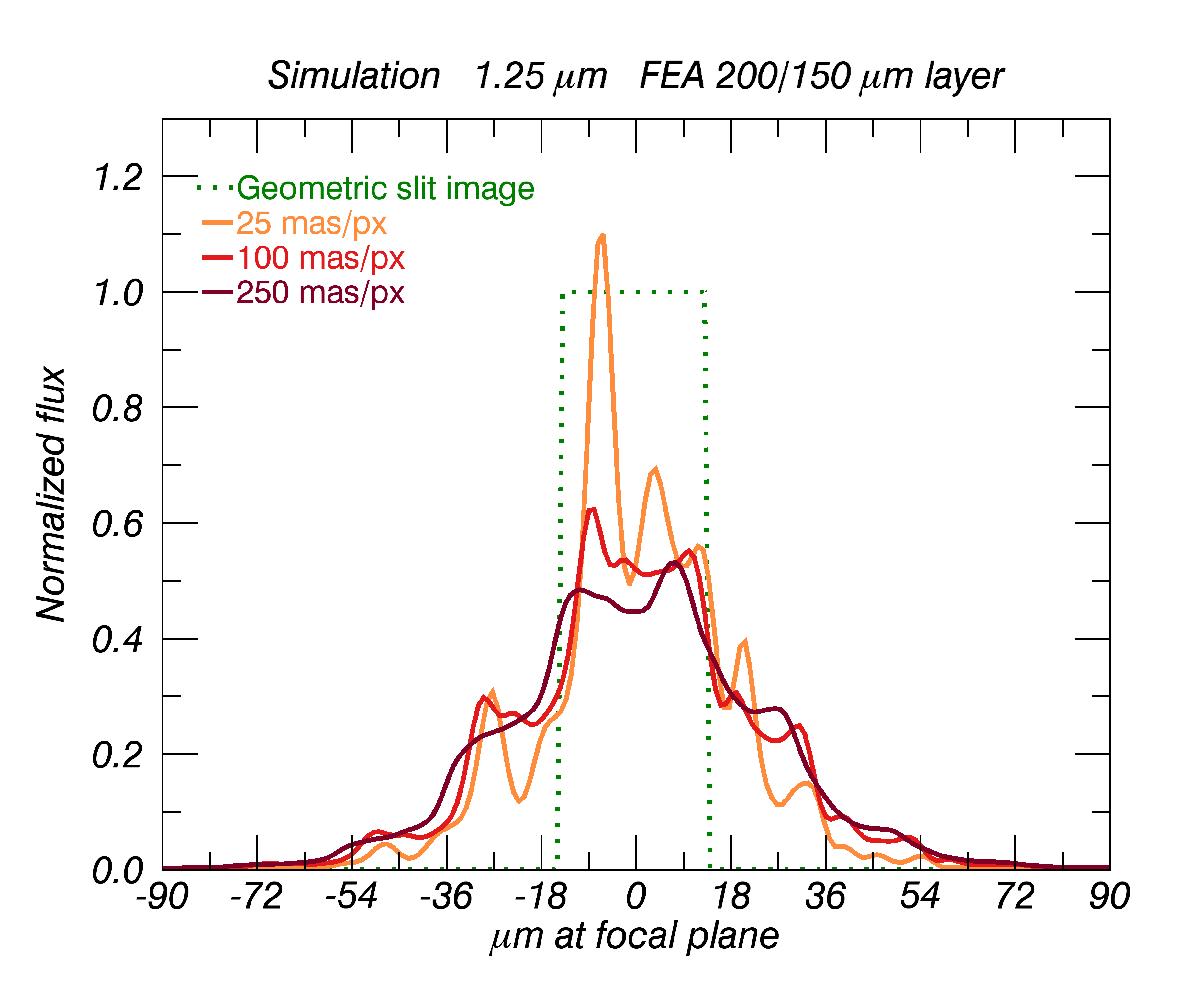}
\includegraphics[width=1.0\textwidth]{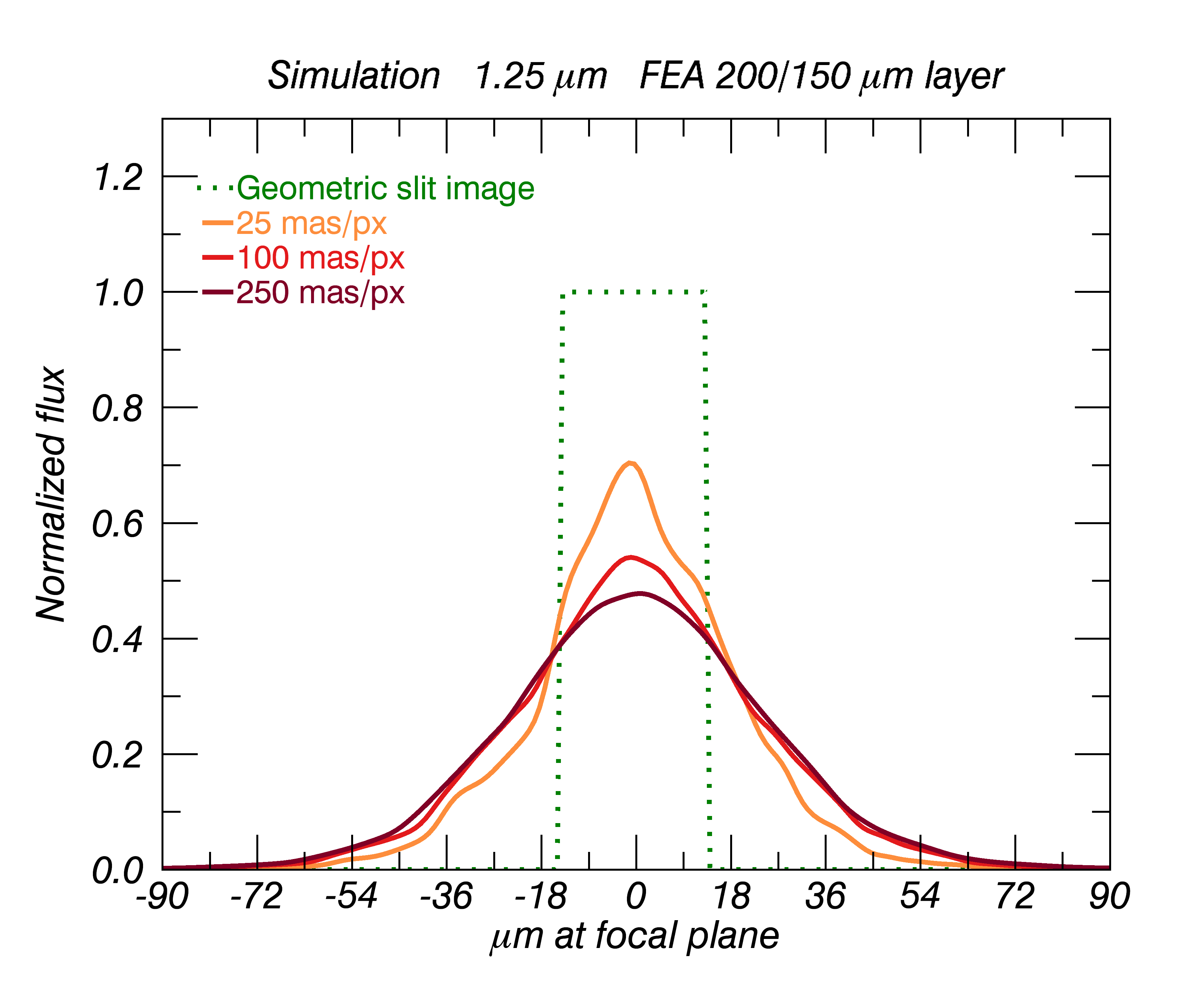}
}
\resizebox{1.4\textwidth}{!}{
\includegraphics[width=1.0\textwidth]{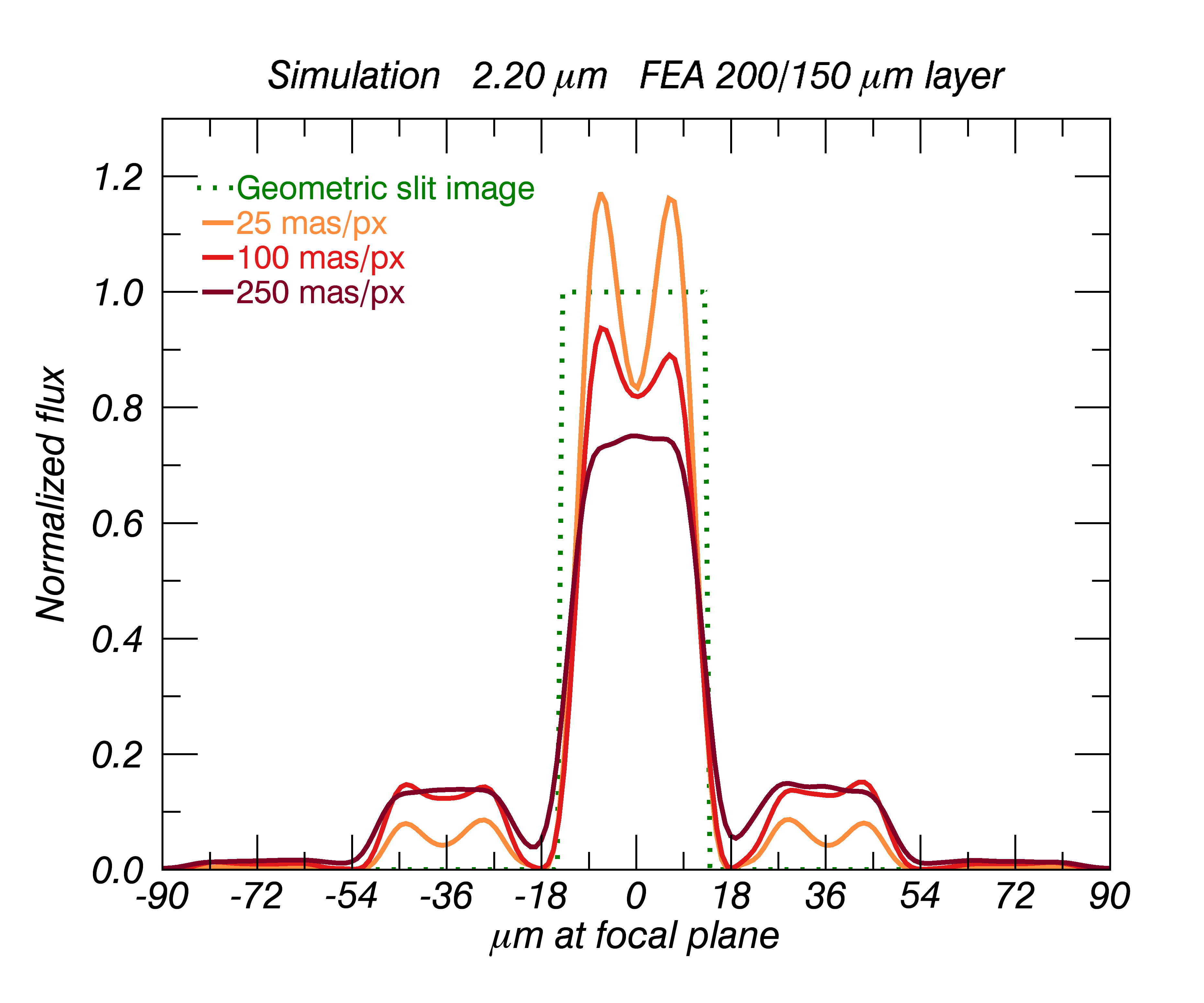}
\includegraphics[width=1.0\textwidth]{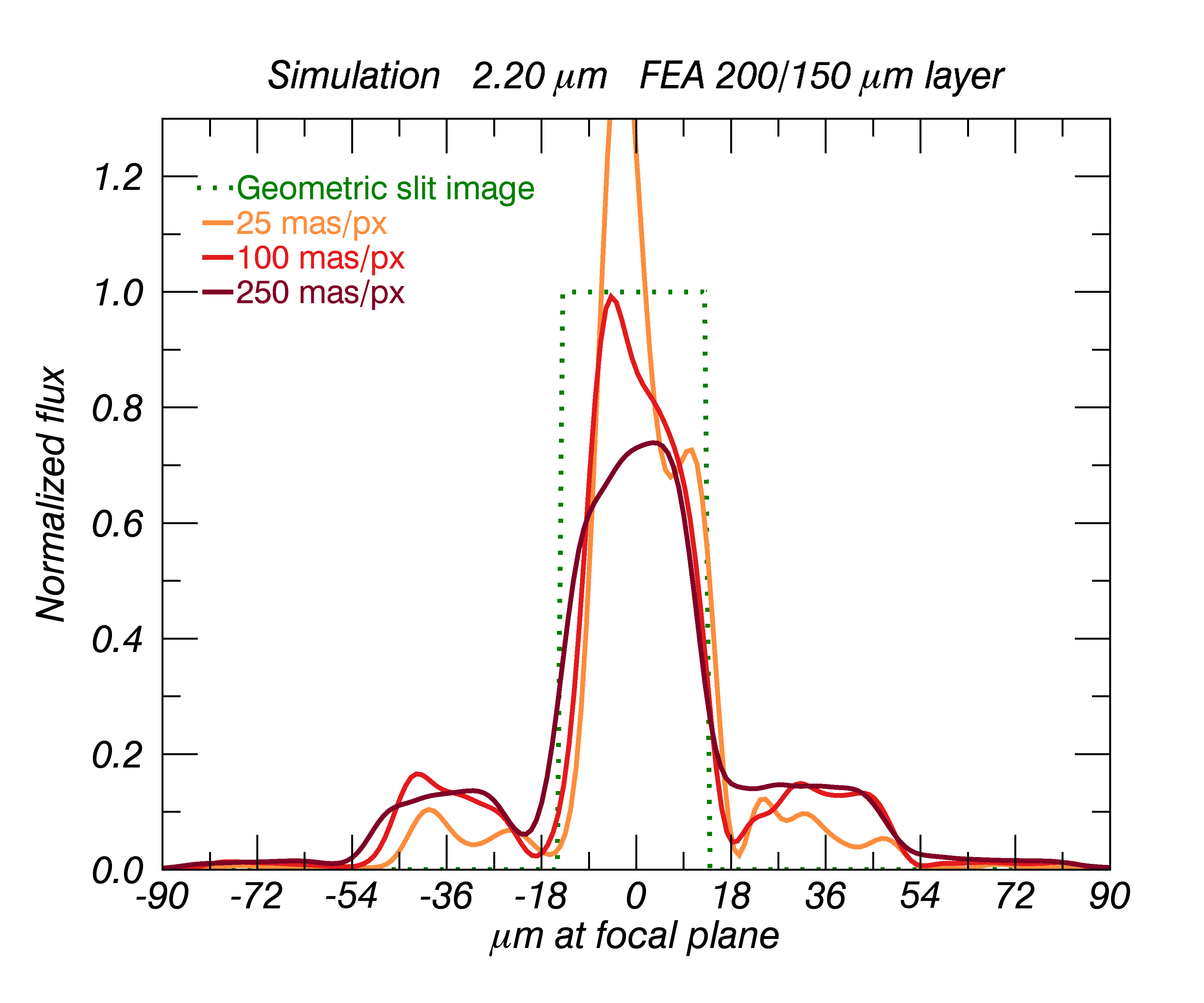}
\includegraphics[width=1.0\textwidth]{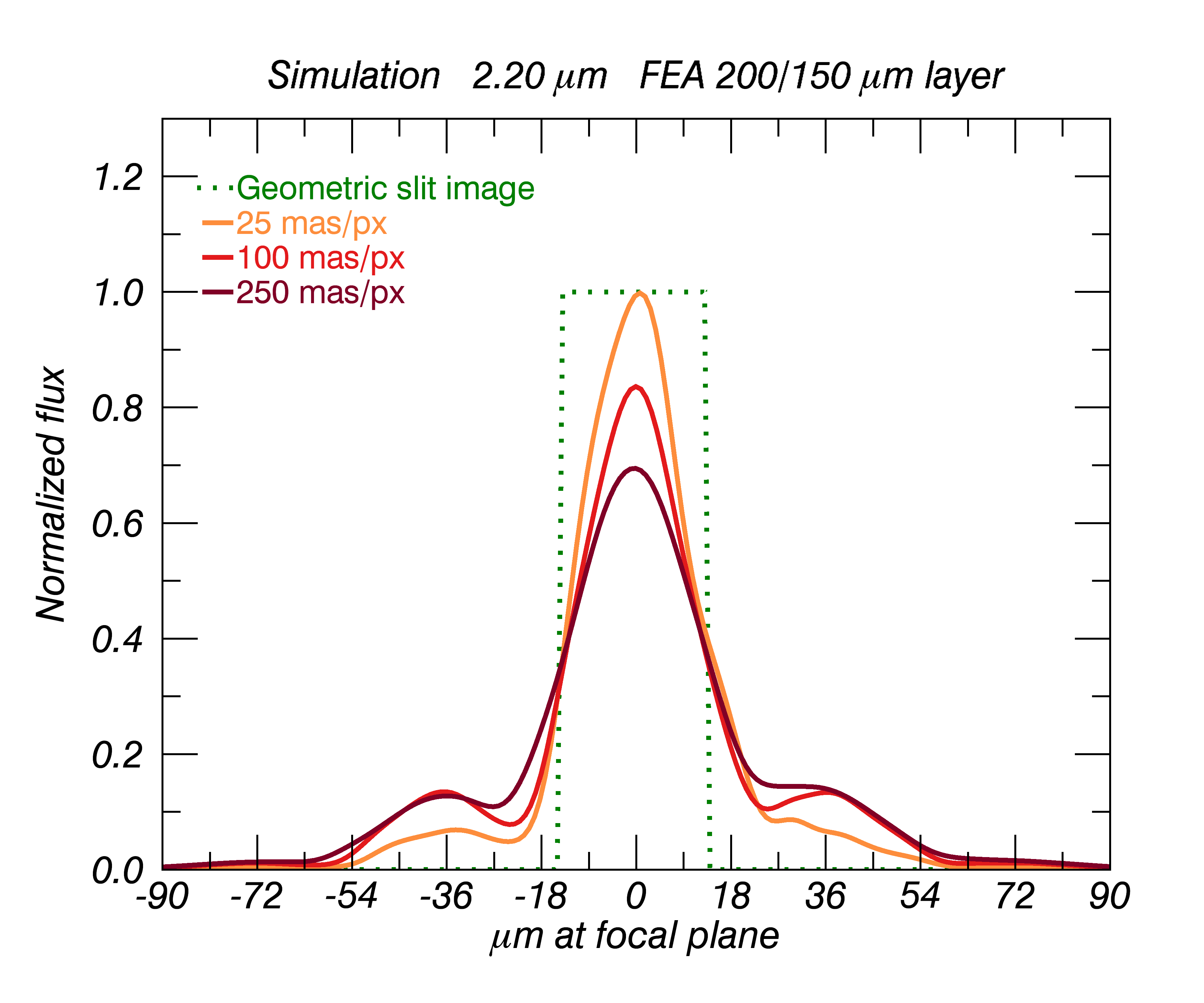}
}
\caption[Examples of 2D Partial Phase Coherence LSFs in all three pixel scales]{Examples of 2D Partial Phase Coherence LSFs in all three pixel scales with a grating deformation from the 200 \SI{}{\micro\meter} thick NiP polished to 75\% thickness (150 \SI{}{\micro\meter}). LSFs are shown in the {\textbf{top row}} for wavelengths at the center of J-band (1.25 \SI{}{\micro\meter}, left) and in the {\textbf{bottom row}} for the center of K-band (2.2 \SI{}{\micro\meter}, right). The {\textbf{left column}} includes the effects of Partial Phase Coherence, the {\textbf{middle column}} additionally includes the collimator wavefront error (measured on the mirrors installed in January 2016), and the {\textbf{right column}} additionally includes the effects of the detector pixel function, and thus depicts the simulated detected line profiles. The size of the shoulders is significantly smaller in the 25 mas pixel scale in both J and K band, and the collimator wavefront error shifts power into the left shoulder. The detector pixel function smooths out the detected line profile. }
\label{fig:CodeV_2DPPC_LSF}
\end{center}
\end{figure}
\end{landscape}
}

\subsubsection{The Effect of Other Wavefront Errors}
\label{sec:otherwavefronts}

So far, we have been discussing the effects of only the cryogenic distortion of the grating blank on the line profiles of the instrument. However, it is known that the rest of the spectrograph optics also have wavefront errors. In order to get the most realistic possible line profile out of our simulations, we must also include these additional wavefront errors. We measured the wavefront error of the spectrometer collimator mirrors installed in January 2016 from the grating position, and apply the measured wavefront of the new collimator mirrors to the collimator in the optical model. The cryogenic measurements of the spare grating showed an additional astigmatism, despite the stress-free mounting. Since we only have measurements of a single spare grating (and not the actual gratings inside of the instrument), we cannot know the exact additional wavefront errors, however, we can add additional wavefront errors of appropriate size to the grating surface using the Zernike polynomials. 

We tested several different additions of wavefront errors into the optical model, and found that in general, the addition of wavefront errors on the grating, collimator, or camera results in asymmetric power in the shoulders. Since we only have an exact measurement of the new collimator mirror wavefront errors, we show the results of including the wavefront error of the collimator only in the model. This is primarily a horizontal coma term and a 45 degree astigmatism term, with a total wavefront P/V of 1.31 \SI{}{\micro\meter} and an RMS of 0.31 \SI{}{\micro\meter}, as seen in the rightmost image of Figure 4 in George et al. (2016)\cite{george16}. This wavefront is applied to the perfect collimator in the optics model. The middle columns of figures \ref{fig:CodeV_2DPPC_LSF_nograt} and \ref{fig:CodeV_2DPPC_LSF} includes the collimator wavefront error. The primary effect is that power becomes distributed asymmetrically in the various diffraction peaks and shoulders.

\subsubsection{Simulated Detected Line Profiles}
\label{sec:realisticLSF}

The simulation results allow us to probe the processes that affect the line profiles of SPIFFI. However, to compare these simulations to the measured line profiles of the instrument, we must process the simulation output to match what we measure using our super-sampled line profile technique. The detector pixel pitch is 18 \SI{}{\micro\meter}, therefore the detected super-sampled line profile is a cross section of the 2D convolution of an 18 \SI{}{\micro\meter} x 18 \SI{}{\micro\meter} wide square with the slit image at the detector. The right column of figures \ref{fig:CodeV_2DPPC_LSF_nograt} and \ref{fig:CodeV_2DPPC_LSF}  shows the line profiles after simulation output has been convolved with the detector pixel function. The main effect of this convolution is a smoothing out of higher-frequency effects such as the diffraction spikes caused by the slit edges. 

\subsubsection{The Effect of Pupil Position on the  LSF}
\label{sec:pupilposLSF}

\begin{figure}[htbp!]
\begin{center}
\resizebox{1.0\textwidth}{!}{
\includegraphics[trim={0.5cm 0 2.9cm 0}, clip]{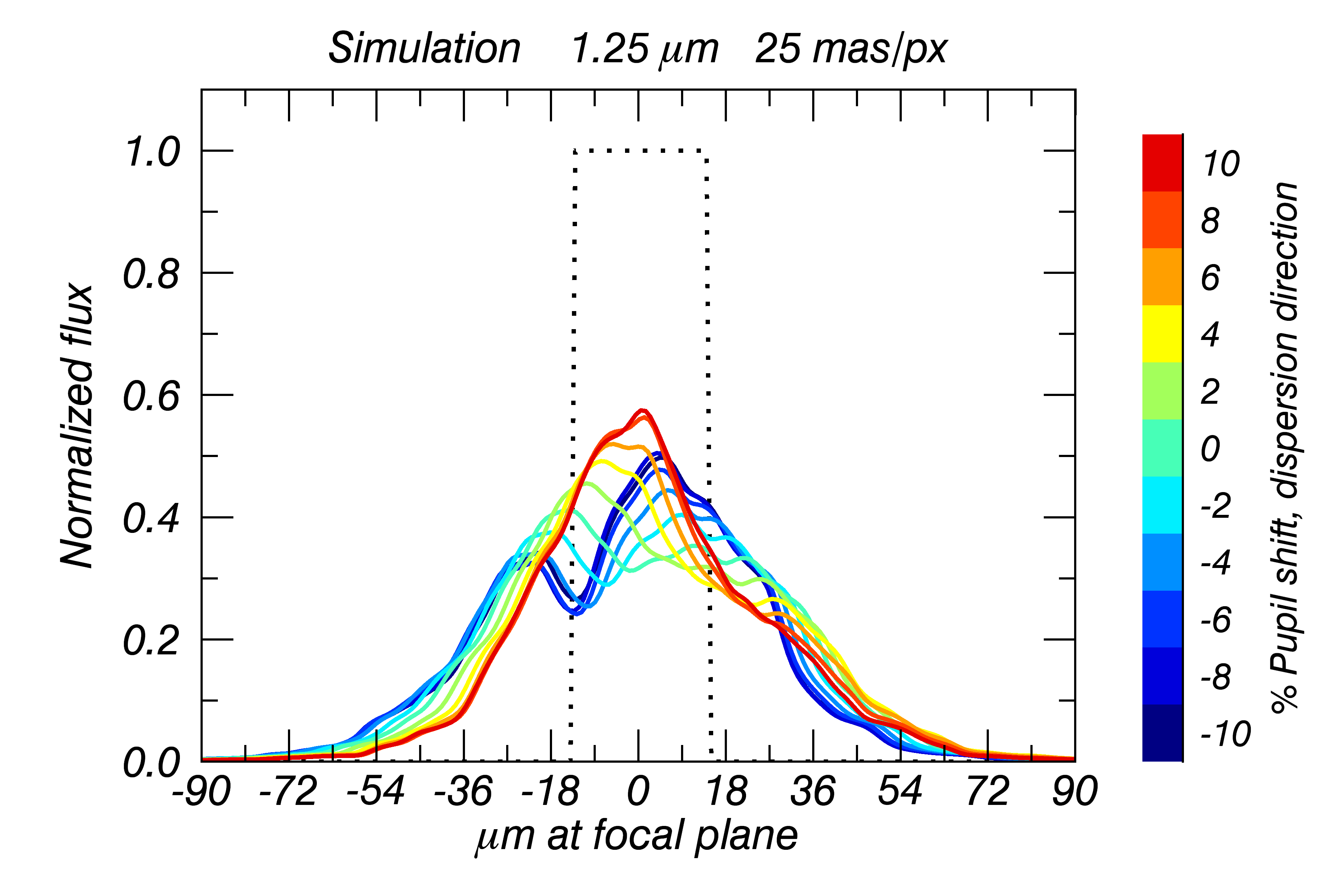}
\includegraphics[]{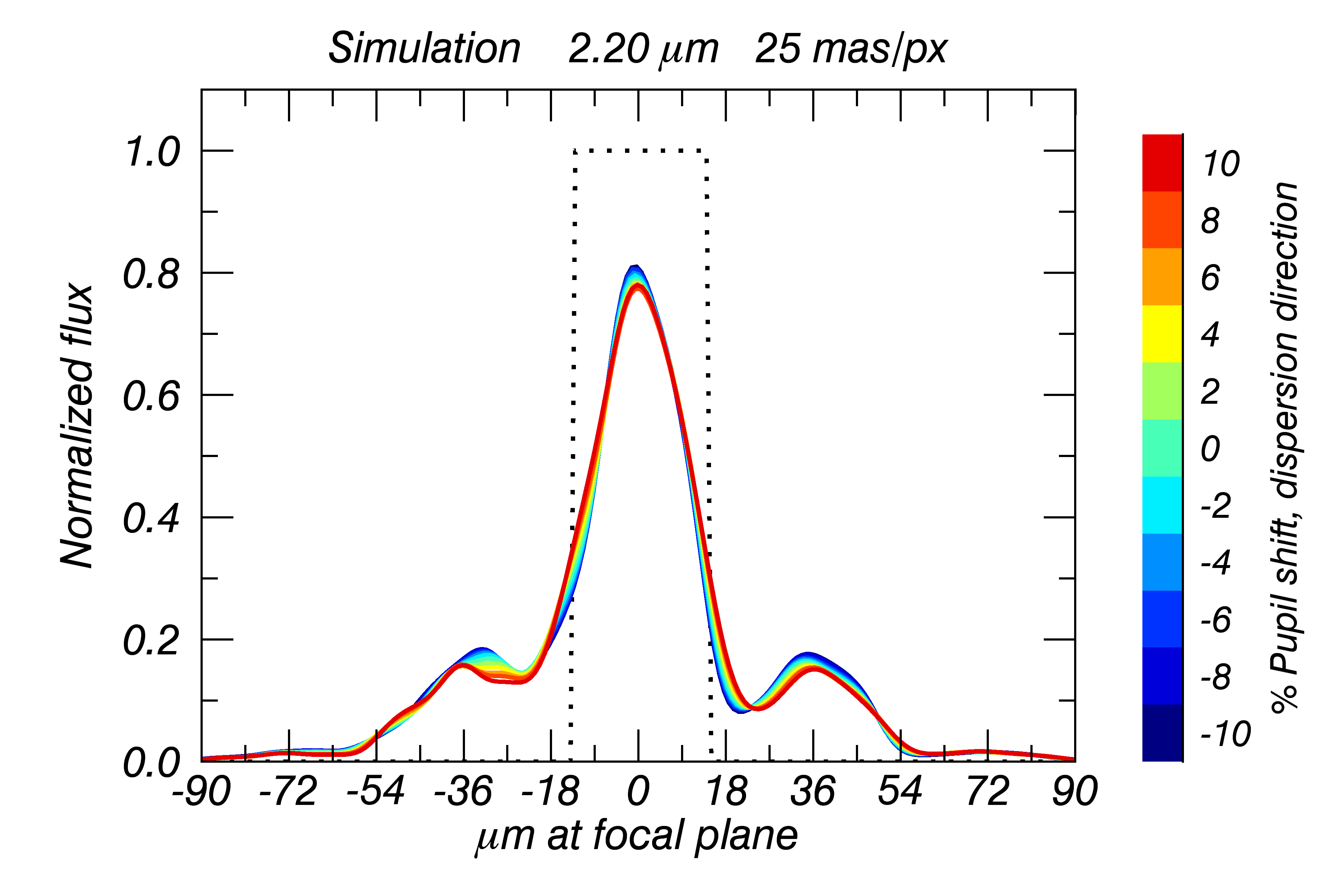}
}
\caption[Effect of varying pupil position]{Line profiles in the J- ({\textbf{left column}}) and K- ({\textbf{right column}}) bands for a range of pupil positions in the 25 mas pixel scale for the grating deformation from the 200 \SI{}{\micro\meter} thick NiP layer polished to 50\% (100 \SI{}{\micro\meter}) thickness. The geometric slit image is shown as a dotted line. The 100 and 250 mas scales are not shown, since there is essentially zero variation with pupil position in these pixel scales.}
\label{fig:pupilshift}
\end{center}
\end{figure}

During the upgrade in 2016, we moved the pupil position on the grating slightly in the X-direction (the dispersion direction). Additionally, the pupil position on each individual grating is slightly different. We explored the effect of the pupil position on the line profiles by shifting the pupil and repeating the simulation. We shifted the pupil by $\pm$10\%, which for our entrance pupil diameter of 110 mm corresponds to a shift of $\pm$11 mm on the grating surface. The periodic structure in the y-direction has a period of 21 mm, so this pupil shift magnitude covers all phases of the periodic structure. Figure \ref{fig:pupilshift} shows the line profiles for a range of pupil positions in the 25 mas pixel scale for the 200 \SI{}{\micro\meter} thick NiP layer polished to 50\% (100 \SI{}{\micro\meter}) thickness. The change in line profile with pupil position is smaller for the grating deformations with smaller amplitudes, so we show here the largest amplitude grating deformation to highlight the effect. The 100 and 250 mas pixel scales are not shown, since there is essentially zero variation of the line profiles with pupil position in these pixel scales. 

The results of the pupil shift simulation show that the amplitude and shape of the 25 mas pixel scale (mas/px) line profiles depends on the exact pupil position on the grating, while exact pupil position has very little effect on the 100 mas/px lines, and almost no effect on the 250 mas/px lines. This can be understood when one considers that due to the low NA of the light entering the image slicer in the smallest pixel scale, only a small portion of the grating is illuminated in the 25 mas/px scale, on the order of a single period of the grating deformation (neglecting the diffraction effects of the slicer on the illumination). Figure \ref{fig:CodeVlayout} shows schematically the effect of smaller NA illumination on the beam footprint on the grating. Thus for the 25 mas/px scale the shape of the wavefront after the grating depends more strongly on the pupil position. Since the grating shows periodic surface deformations on scales smaller than the illuminated pupil size in the 100 mas/px and 250 mas/px scale, for these pixel scales the effect of a pupil shift on the line profiles is much smaller, because the Fourier transform of the surface deformation is nearly identical for small relative shifts on a periodic surface. Additionally, the effect is larger at smaller wavelengths--in our simulation, only the J-band showed very strong line profile variation with pupil position.

\section{Matching Simulations to Data}
\label{sec:simmatch}

In this section we examine each band to determine how well each band's simulations match the data.

\subsection{K-band}
\label{sec:Kband}

\begin{figure}[htbp!]
\begin{center}
\resizebox{1.0\textwidth}{!}{
\includegraphics[width=1.0\textwidth]{k_variation_pxscale_smoothnew.png}
\includegraphics[width=1.0\textwidth]{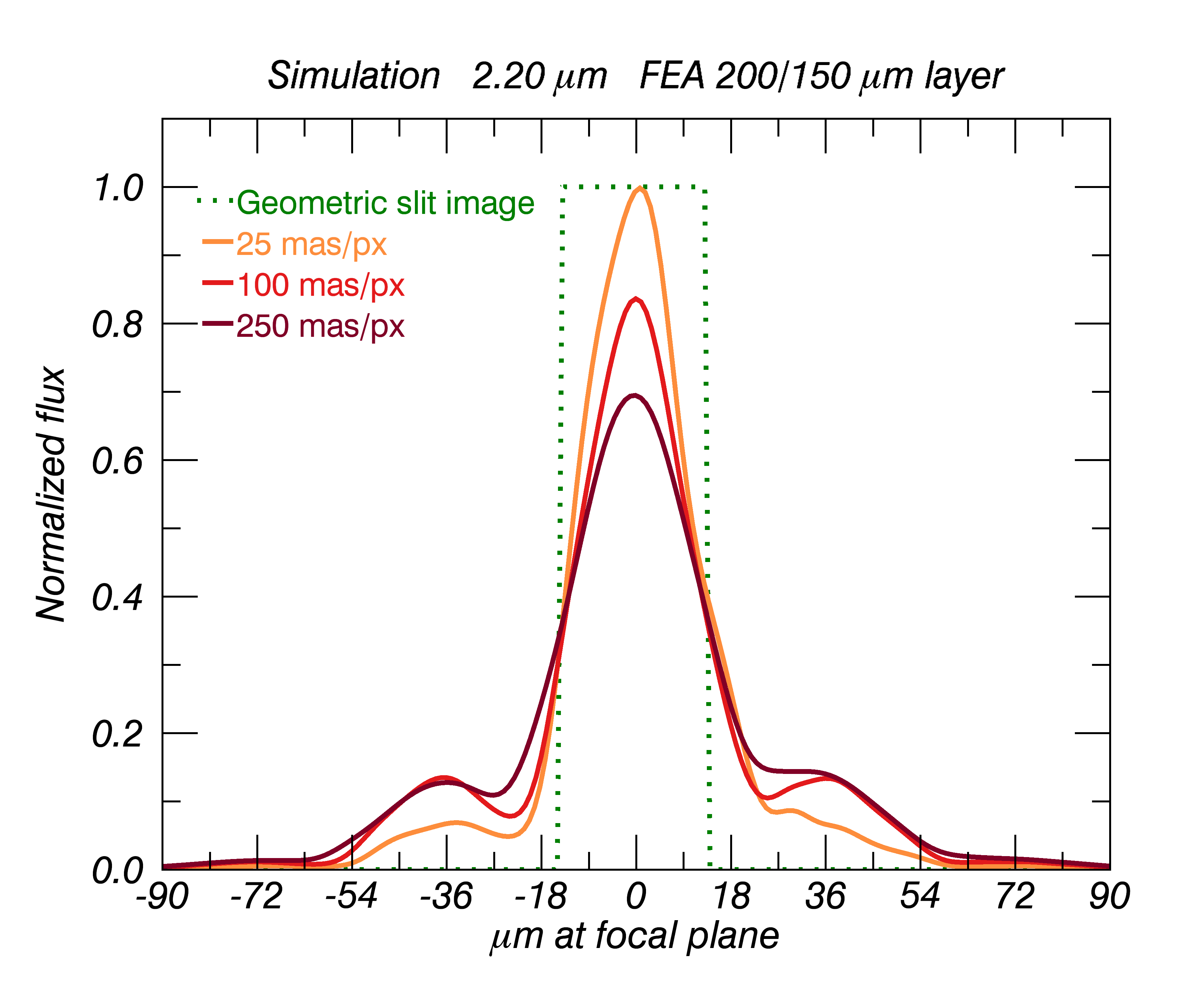}
}
\caption[K-band data vs. simulation]{K-band data ({\textbf{left}}) vs. simulation ({\textbf{right}}) using the grating deformation from a 200 \SI{}{\micro\meter} thick NiP layer polished to 75\% (150 \SI{}{\micro\meter}) thickness in all three pixel scales.}
\label{fig:Kcompare}
\end{center}
\end{figure}

In K-band, the simulation using the grating deformation from a 200 \SI{}{\micro\meter} thick NiP layer polished to 75\% (150 \SI{}{\micro\meter}) thickness matches the K-band data quite well, shown in figure \ref{fig:Kcompare}. The simulation with pupil centered on the grating is also adequate to describe the data in the 25 mas pixel scale, though the line profile variation in K-band with a shifted pupil is rather low (see right plot in figure \ref{fig:pupilshift}), so it would be difficult to tell if the pupil were shifted on the grating. The measured line profiles also show a low variation along the slitlet (see Gr\"{a}ff (2016)\cite{graeff16} for an example of this small variation).

\subsection{H-band}
\label{sec:Hband}

\begin{figure}[htbp!]
\begin{center}
\resizebox{1.0\textwidth}{!}{
\includegraphics[width=1.0\textwidth]{h_variation_pxscale_smoothnew.png}
\includegraphics[width=1.0\textwidth]{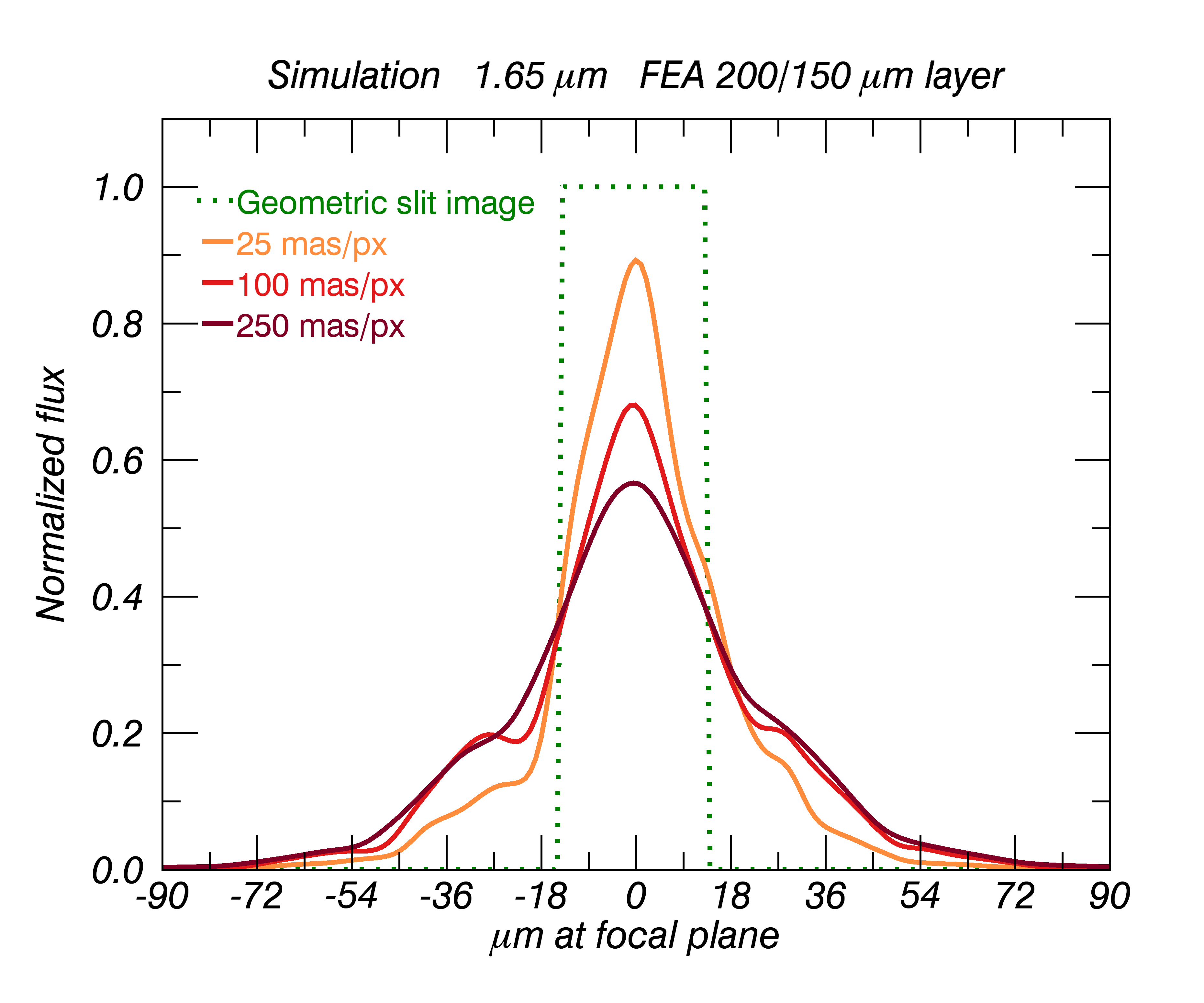}
}
\caption[H-band data vs. simulation]{H-band data ({\textbf{left}}) vs. simulation ({\textbf{right}}) using the grating deformation from a 200 \SI{}{\micro\meter} thick NiP layer polished to 75\% (150 \SI{}{\micro\meter}) thickness in all three pixel scales.}
\label{fig:Hcompare}
\end{center}
\end{figure}

In H-band the comparison is also straightforward. As in the K-band case, the simulation using the grating deformation from a 200 \SI{}{\micro\meter} thick NiP layer polished to 75\% (150 \SI{}{\micro\meter}) thickness matches the H-band data, shown in figure \ref{fig:Hcompare}. As in K-band, having the pupil centered on the grating is also adequate to describe the data in the 25 mas pixel scale, though a shift left or right of a few percent would give similar results, as the profile variation with pupil shift is low in H-band with this deformation amplitude. The measured line profiles also show a low variation of the line profiles with slitlet position. (see Gr\"{a}ff (2016)\cite{graeff16} for an example of this small variation).

\clearpage

\subsection{J-band}
\label{sec:Jband}

\begin{figure}[htbp!]
\begin{center}
\resizebox{1.0\textwidth}{!}{
\includegraphics[width=1.0\textwidth]{j_variation_pxscale_smoothnew.png}
\includegraphics[width=1.0\textwidth]{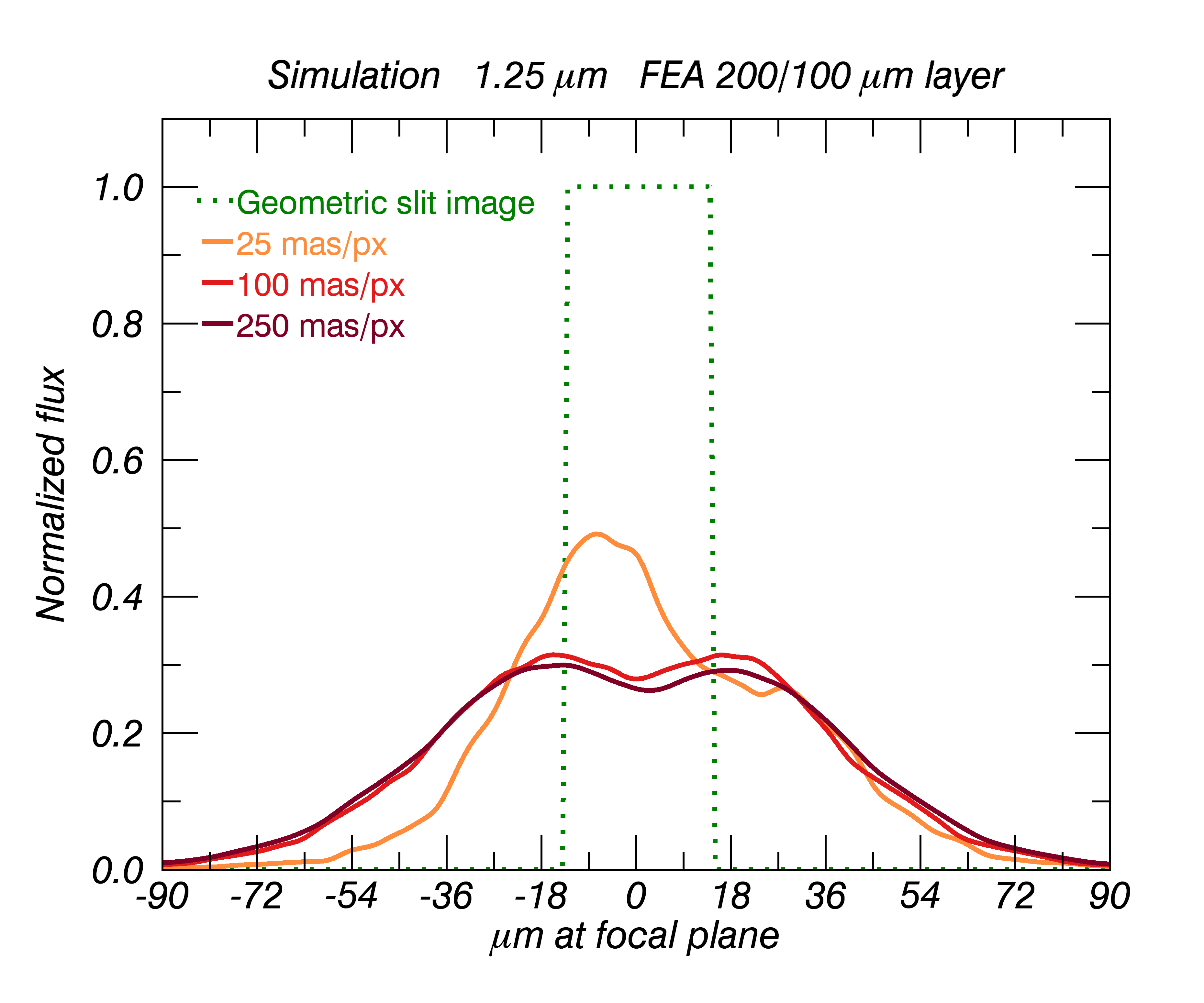}
}
\caption[J-band data vs. simulation]{J-band data ({\textbf{left}}) vs. simulation ({\textbf{right}}) using the grating deformation from a 200 \SI{}{\micro\meter} thick NiP layer polished to 50\% (100 \SI{}{\micro\meter}) thickness in all three pixel scales. In the simulated line profiles, the pupil position has been shifted approximately 4 mm in the dispersion direction to reproduce the asymmetric peak heights in the 25 mas pixel scale.}
\label{fig:Jcompare}
\end{center}
\end{figure}

As already mentioned in section \ref{sec:lineProfiles} the J-band line profiles appear different from the other two bands. The most obvious difference is the double-peaked structure, most visible in the 100 and 250 mas pixel scales. Additionally, there is an obvious asymmetry in the peak heights in the 25 mas pixel scale. Given the results of the simulations, it became clear that to reproduce the J-band line profiles, a larger amplitude grating deformation was required to get the central order suppression necessary to result in a double-peaked line profile. Of the layer thickness/polishing simulated, we found that the 200 \SI{}{\micro\meter} thick layer with 50\% (100 \SI{}{\micro\meter}) thickness remaining after polishing gave results that matched the simulations the best. Figure \ref{fig:Jcompare} shows the measured line profiles and the best-matching simulation from the 200/100 NiP layer deformation. 

In addition, the J-band line profiles in the 25 mas pixel scale vary strongly across a single slitlet. We could reproduce this behaviour by varying the pupil position on the grating. One possible source of movement of the pupil position is the image slicer, which is described in detail in Tecza et al. (2000)\cite{tecza00}. A torsional deformation of a slicer mirror with a measured amplitude of 2 HeNe fringes (close to what could be expected based on measurements of the slicer mirrors; see Eisenhauer et al. (2003)\cite{eisenhauer03} for an example of a slicer interferogram) would result in the pupil shifting on the grating by approximately $\pm$ 10 mm, which is nearly an entire period of the grating deformation function. Both the asymmetry in the 25 mas pixel scale line profiles and the variation in line profile along a slitlet can be explained with a shift in the pupil (due to, for example, a torsion in the slicer mirrors) in combination with the grating deformation. 

\begin{figure}[htbp!]
\begin{center}
\resizebox{1.0\textwidth}{!}{
\includegraphics[height=7cm, trim={0.5cm 0.1cm 0.3cm 0.0cm}, clip]{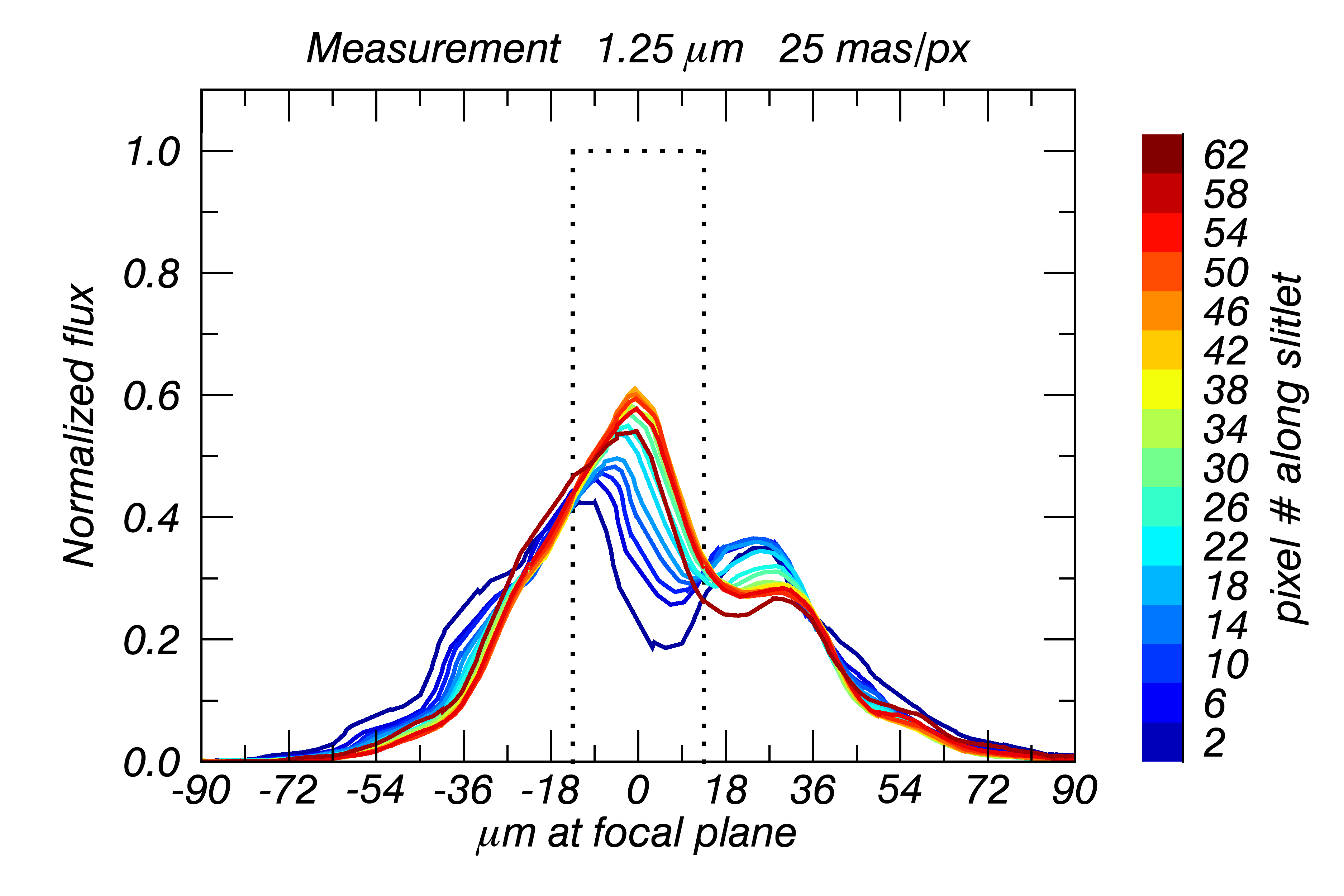}
\includegraphics[height=7cm, trim={1.2cm 0.1cm 0.3cm 0.0cm}, clip]{revX_withdet_CodeV_2dPPC_withcol_LSF_ypupilshift_fea200100_1p25um25mas.png}
}
\caption[J-band variation data vs. simulation]{J-band data ({\textbf{left}}) showing the variation in line profile over a single slitet, vs. simulation ({\textbf{right}}) showing the line profile variation with a pupil shift of $\pm$10\%. The geometric slit image is shown as a dotted line.}
\label{fig:Jvariation}
\end{center}
\end{figure}

Figure \ref{fig:Jvariation} compares the simulated and measured line profile variation in the smallest pixel scale. A shift in the central pupil position of approximately 4 mm in the dispersion direction on the grating can provide the necessary asymmetry to reproduce the behavior of the central 25 mas/px line profile (shown in figure \ref{fig:Jcompare}), however, to fully explain the variation in the line profiles over the full length of the slitlets, a continuously varying pupil position is required. If a torsion in the image slicer is the reason for a continuously varying pupil position, then we expect the same pupil position variation on all diffraction gratings. This variation is only noticeable in the measurements on the J-band grating. As noted in the previous two sections, both measurements and simulations of the H- and K-band line profiles showed very low variation with pupil position on the grating.

In J-band it is clear that the simulation is not a perfect reproduction of the line profiles--the 100 and 250 mas pixel scale simulations are slightly wider and the central dip is less pronounced, while in the 25 mas pixel scale the right shoulder is less obvious. To explore this, we also tested purely sinusoidal grating deformations with a period equal to the spacing of the lightweighting holes and  variety of P/V values. A sinusoidal deformation with a P/V  of 0.58 \SI{}{\micro\meter} also gives results that match the measured line profiles reasonably well, but with narrower line profiles. This indicated to us that while the primary component of the grating deformation that affects the line profiles is the amplitude of the Fourier component corresponding to the period of the lightweighting holes, the exact shape of the deformation (i.e. other Fourier components of the deformation) play a role in the width of the profiles. Without a cryogenic measurement of the gratings that are currently in use in SPIFFI, we cannot know exactly what the deformation amplitude and shape of each grating is. However, we can say with confidence that to reproduce the behaviour seen, the J-band grating needs to have a different, and higher, deformation amplitude than the other two gratings. This is discussed more in the following section.

\subsection{Single Wavelength Measurements on Multiple Gratings}
\label{sec:multiwave}
To cement our result, we attempted to reproduce the measurements that led us to suspecting the grating blank was causing problems in the first place--namely, producing simulations of a single wavelength on the three diffraction gratings. We choose a wavelength at the center of J-band (1.25 \SI{}{\micro\meter}), as shorter wavelengths are more sensitive to the details of the grating deformation. The simulation setup uses the K-, H-, and J- band gratings in the 4th, 3rd, and 2nd diffraction orders respectively. The grating deformations used in the simulation are the ones needed to reproduce the in-band results of the previous sections, namely the deformation from NiP layer thicknesses of 200/150 \SI{}{\micro\meter} for the H- and K- band gratings, and NiP layer thicknesses of 200/100 \SI{}{\micro\meter} for the J-band grating. 

\begin{figure}[htbp!]
\begin{center}
\resizebox{1.0\textwidth}{!}{
\includegraphics[width=1.0\textwidth]{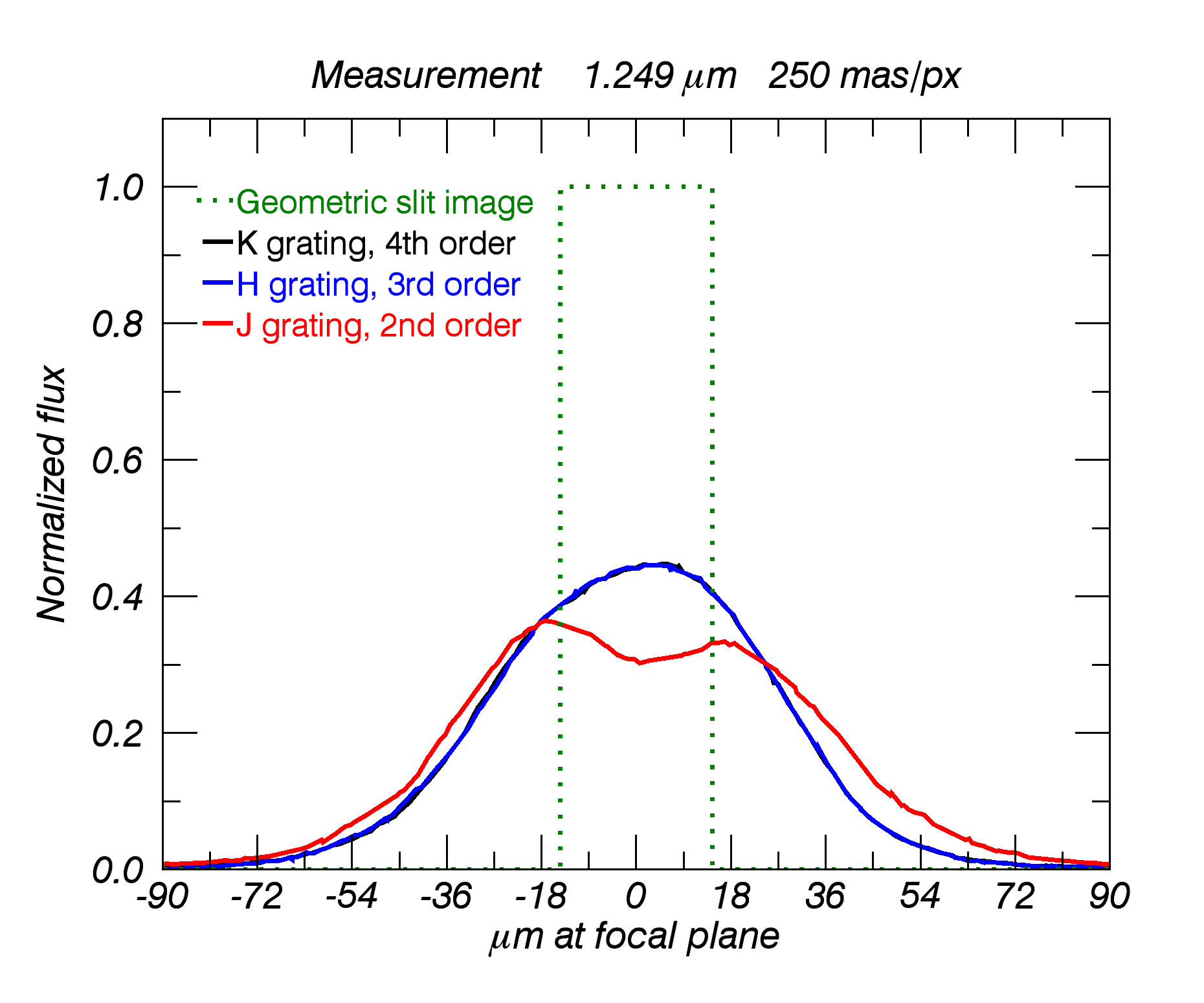}
\includegraphics[width=1.0\textwidth]{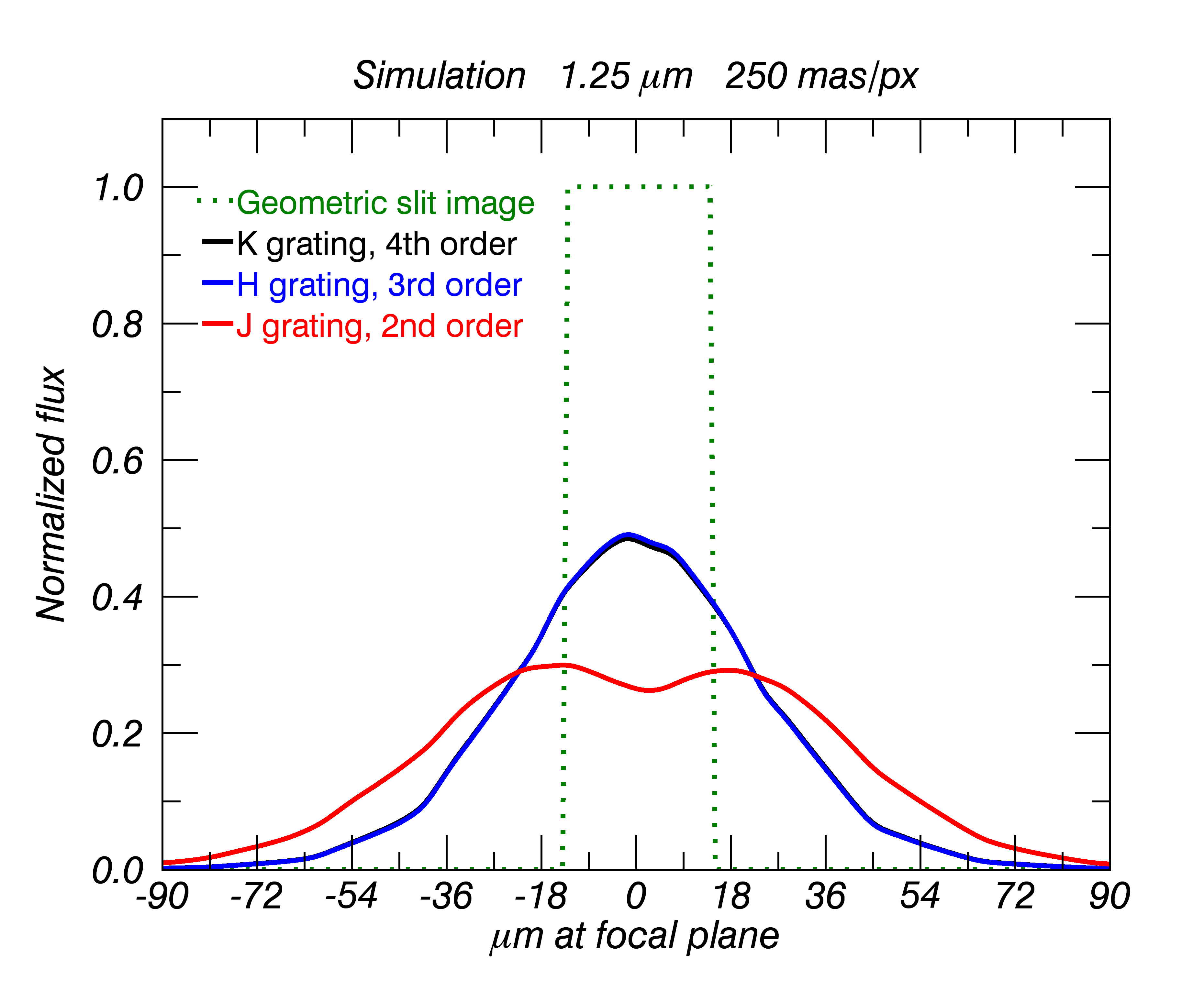}
}
\caption[J-band data vs. simulation on multiple gratings]{J-band (1.25 \SI{}{\micro\meter}) wavelength line on all three diffraction gratings ({\textbf{left}}) vs. simulation ({\textbf{right}}) using the grating deformations found in the previous three sections. A slightly higher deformation amplitude on the H- and K- band gratings results in a better match to the data.}
\label{fig:multigrat}
\end{center}
\end{figure}

Figure \ref{fig:multigrat} shows the results for 1.25 \SI{}{\micro\meter} in the 250 mas pixel scale. The agreement between measurement and simulation is quite good, though the line profile is clearly higher and narrower in the H- and K- band grating simulations than in the measurements. 
In the smaller pixel scales, there is a similar slight disagreement between the simulation and measurements. (Figure \ref{fig:multigratdata} in the appendix shows these measurements in all pixel scales and wavelengths). We were able to more closely reproduce the measured line profiles by increasing the amplitude of deformation on the H- and K- band gratings by approximately 20\%, which was achieved by scaling the amplitude of the 200/150 \SI{}{\micro\meter} simulation by a factor of 1.2. However, here we do not attempt to fine-tune the deformation amplitudes to match the measurements exactly--we simply note that a slightly higher deformation amplitudes on the H- and K- band gratings (though still smaller than the deformation of the J-band grating)  reproduces the measurements better. 

The results of this test leave us to conclude that in order to reproduce the entire data set we have obtained from 1.0-2.5 \SI{}{\micro\meter} on three diffraction gratings, the H- and K- gratings need to be very similar to each other, while the J-band grating needs to be different. The primary difference required is that the deformation amplitude induced by the lightweighting structure needs to be larger on the J-band grating. We found in our notes on the production of the gratings that the J-band blank was re-polished at least once to achieve higher surface flatness. The results of the FEA of the grating blanks show that more polishing results in higher deformation amplitudes, so this result is consistent with the information we have on the grating manufacturing.

\section{Discussion and Conclusion}
\label{sec:discussionconclusion} 

SPIFFI has complex line profile shapes that vary with wavelength and pixel scale, the origins of which have been sought since the instrument construction. Because SPIFFI is still in use at the telescope as part of SINFONI, we investigated the line profiles based on measurements we could take with the instrument calibration unit, as well as laboratory measurements of a spare SPIFFI diffraction grating. Cryogenic measurements of the spare SPIFFI diffraction grating showed significant periodic deformation due to the lightweighted structure and the bimetallic bending effect between the aluminum blank material and the NiP polishing layer. We performed a finite element analysis of the grating blank, and found that the amplitude of the deformation depends on the initial layer thicknesses and amount of surface polishing. The deformation from cryogenic measurements of the spare grating falls in the range of values expected based on NiP layer thicknesses and polishing depths provided by the manufacturer.

We found that inserting the grating deformation into an optical simulation gives rise to satellite peaks in the diffraction pattern of the grating, and reproduces the behavior of the SPIFFI line profiles with both wavelength and pixel scale as measured with the instrument calibration unit. We determined that to reproduce the measured data, the deformation amplitude on the J-band grating must be higher than the other two gratings, likely due to re-polishing of the J-band grating blank. The result of our study is that we have proven that cryogenic deformation of the diffraction gratings due to the bimetallic bending effect on the lightweighted blank is responsible for the non-ideal line profiles of SPIFFI, and that the diffraction gratings should be replaced for optimal instrument performance. The plan is for all diffraction gratings to be replaced in the upgrade of SPIFFI for use in ERIS.

When building a cryogenic instrument, simulations of the deformations of optical elements when cooling to cryogenic temperatures can provide valuable information on the effect of lightweighted structures, especially in the presence of components with mismatched CTE. Cryogenic wavefront measurements provide powerful diagnostics for the as-built performance of a cooled optical system, and can be included in an instrument optical model. A full instrument optical model and simulation including IFS slit diffraction effects\cite{raab03, antichi09} is important to assist in instrument design, debugging, and performance analysis work.  

Some practical information for instrument design came out of this work. Our analyses showed that materials with mismatched CTEs combined with lightweighted structure results in significant deformation. Mismatched NiP layer thicknesses between the inside of the lightweighting holes and the grating surfaces resulted in larger deformation, however, deformation was still present even with matched NiP layer thicknesses. Careful attention should be paid to the design of any complex structure that could be subject to bimetallic stresses. Second, a test optical simulation we carried out with the period of the lightweighting structure doubled resulted in the first few diffraction orders falling inside of the slit in the focal plane, significantly improving the spectral line profiles. In general, the spatial frequencies of lightweighting structures should be designed to minimize the effects of potential diffraction on instrument performance.

The design of the new grating blanks for ERIS (currently undergoing FDR) takes into account what we have learned from this analysis. The new blanks will not have any lightweighting structure. Instead, optically unused area around the edges of the blanks has been removed to reduce weight. This entirely removes from the optics the spatial frequencies of the lightweighting structure that resulted in the observed diffraction effects in the SPIFFI line profiles. It has been determined that an NiP polishing layer is still required for a surface finish with high enough quality for grating ruling. However, the new blank material will be a high silicon content aluminum alloy such as AlSi40 that has a CTE that is matched to NiP to within 0.5 x 10$^{-6}$. This significantly reduces the bimetallic bending effect.\cite{kinast14} Finally, finite element analysis will be performed on the final grating blank/coating design to ensure that it is within specifications, and any deformation found can be incorporated into an optical model. Given the simulation results of figure \ref{fig:CodeV_2DPPC_LSF_nograt} (the ``perfect grating" case), if the new gratings are within specifications, ERIS/SPIFFIER should have a spectral resolution close to the design value of R $\sim$4k in J-, H-, and K-bands.

\acknowledgments 
We thank Johannes Hartwig and Kurt Dittrich from MPE for the preparation of the cryogenic test facility and Christian Rau from MPE for preparing the cryogenic drive motors, both used for the cryogenic grating wavefront measurements. We thank ESO and the Paranal staff for providing day-time access and support for the SINFONI instrument to complete super-sampled line profile measurements using the instrument calibration unit. We also thank the two anonymous referees for their helpful comments on this manuscript.


\bibliography{../../BIBTEX/mpe.bib}   
\bibliographystyle{spiejour}   

\pagebreak
\appendix{}

\section{Plots of line profiles on different gratings}
\label{sec:appendixa} 

This appendix contains an example of the super-sampled line profiles of the instrument as a function of wavelength. The wavelengths shown are roughly at the band centers, and are selected from the bright arc-lamp calibration lines. The arc lamps used are Argon for J-band measurements, Xenon and Argon together for H-band measurements, and Neon and Argon together for K-band measurements. The lines plotted come from the same location on the image slicer. This means that the light producing each line passes through the same optical path in the spectrometer collimator and camera. 

Each wavelength was measured on as many diffraction gratings as possible by additionally utilising the non-blaze orders of the gratings. For example, for lines falling in the J-band wavelengths, we measured each line in 2nd order on the J-band grating, 3rd order on the H-band grating, and 4th order on the K-band grating. The limiting factor was that we did not wish to turn the gratings by more than a few degrees during a measurement, in order to keep the grating as uniformly illuminated as possible. This means that for lines falling in the H-band and K-band wavelengths, we were only able to obtain line profiles produced from 2 of the 3 gratings. 

The resulting line profiles are similar between the three diffraction gratings, however, they are not identical. In particular, one can see that the H- and K-band gratings produce nearly identical line profiles, while the J-band grating produces line profiles with more pronounced shoulders, and in some cases, double peaks. These results, when taking the simulations into account, indicate that the deformation on the J-band grating is slightly larger than on the H or K band gratings.

\begin{figure}[htbp!]
\begin{center}
\resizebox{1.0\textwidth}{!}{
\includegraphics[height=7cm, trim={0.4cm 0.1cm 0.9cm 0.3cm}, clip]{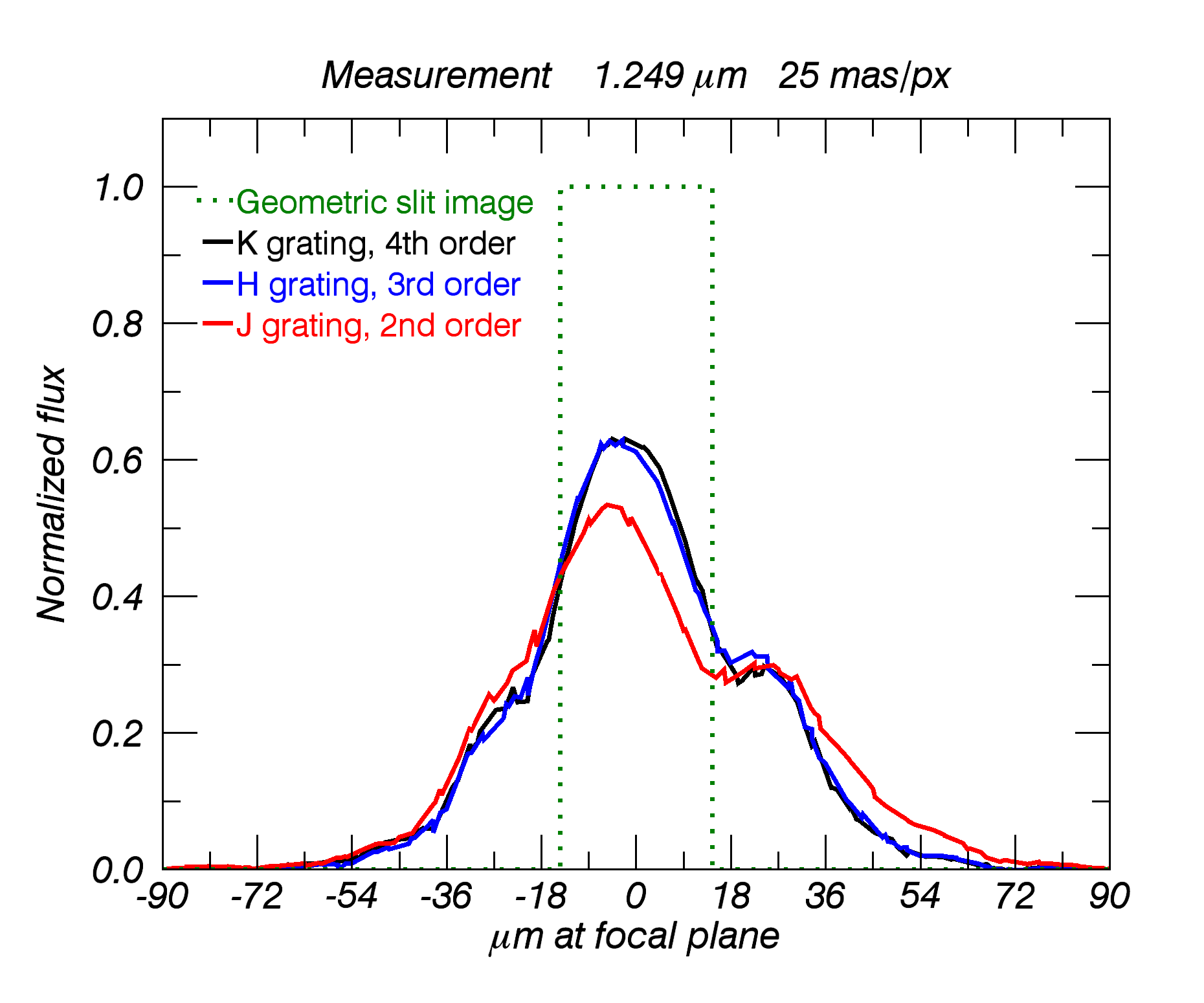}
\includegraphics[height=7cm, trim={1.0cm 0.1cm 0.9cm 0.3cm}, clip]{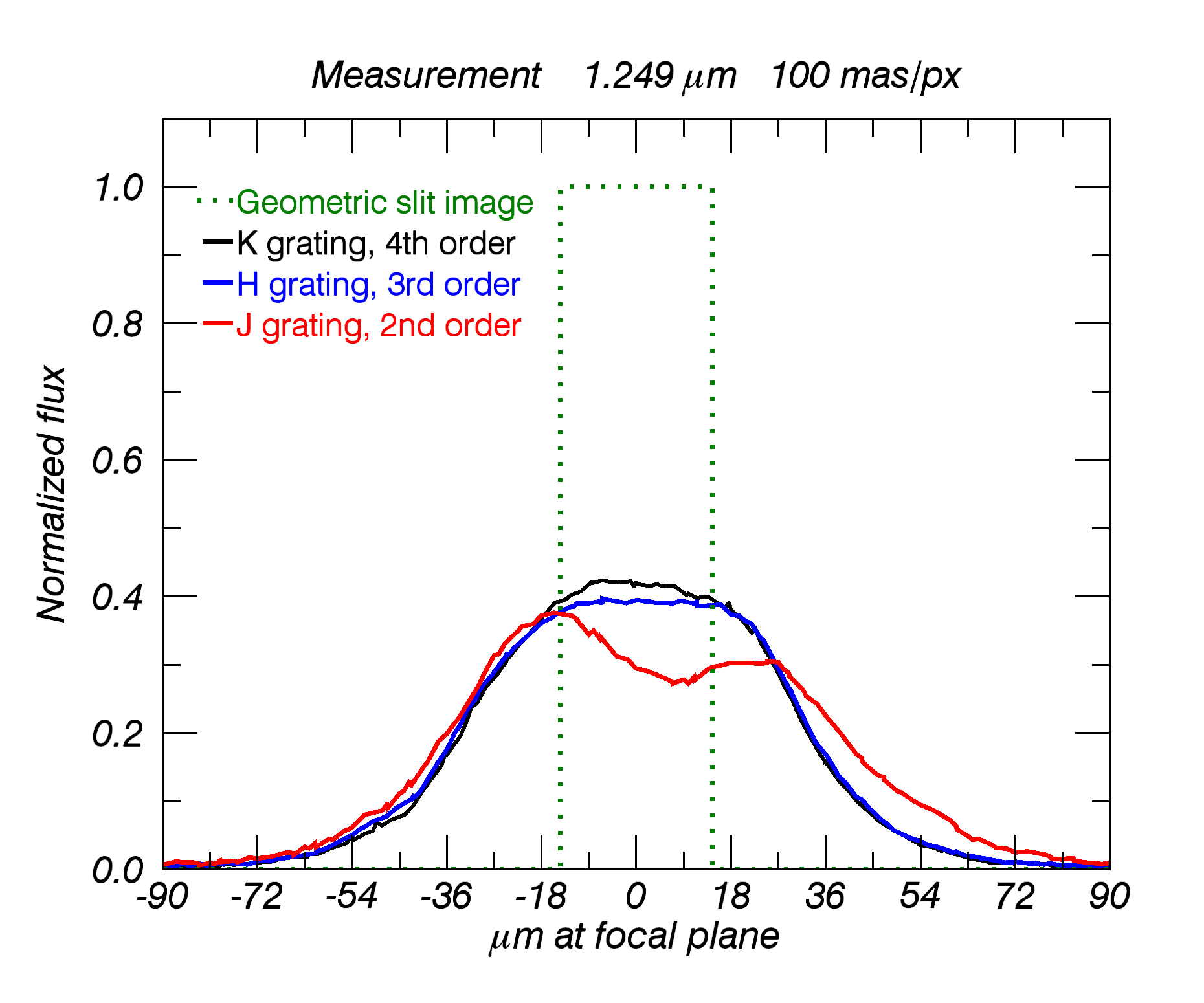}
\includegraphics[height=7cm, trim={1.0cm 0.1cm 0.9cm 0.3cm}, clip]{revX_opt_Jband_250mas_norm1.png}
}
\resizebox{1.0\textwidth}{!}{
\includegraphics[height=7cm, trim={0.4cm 0.1cm 0.9cm 0.3cm}, clip]{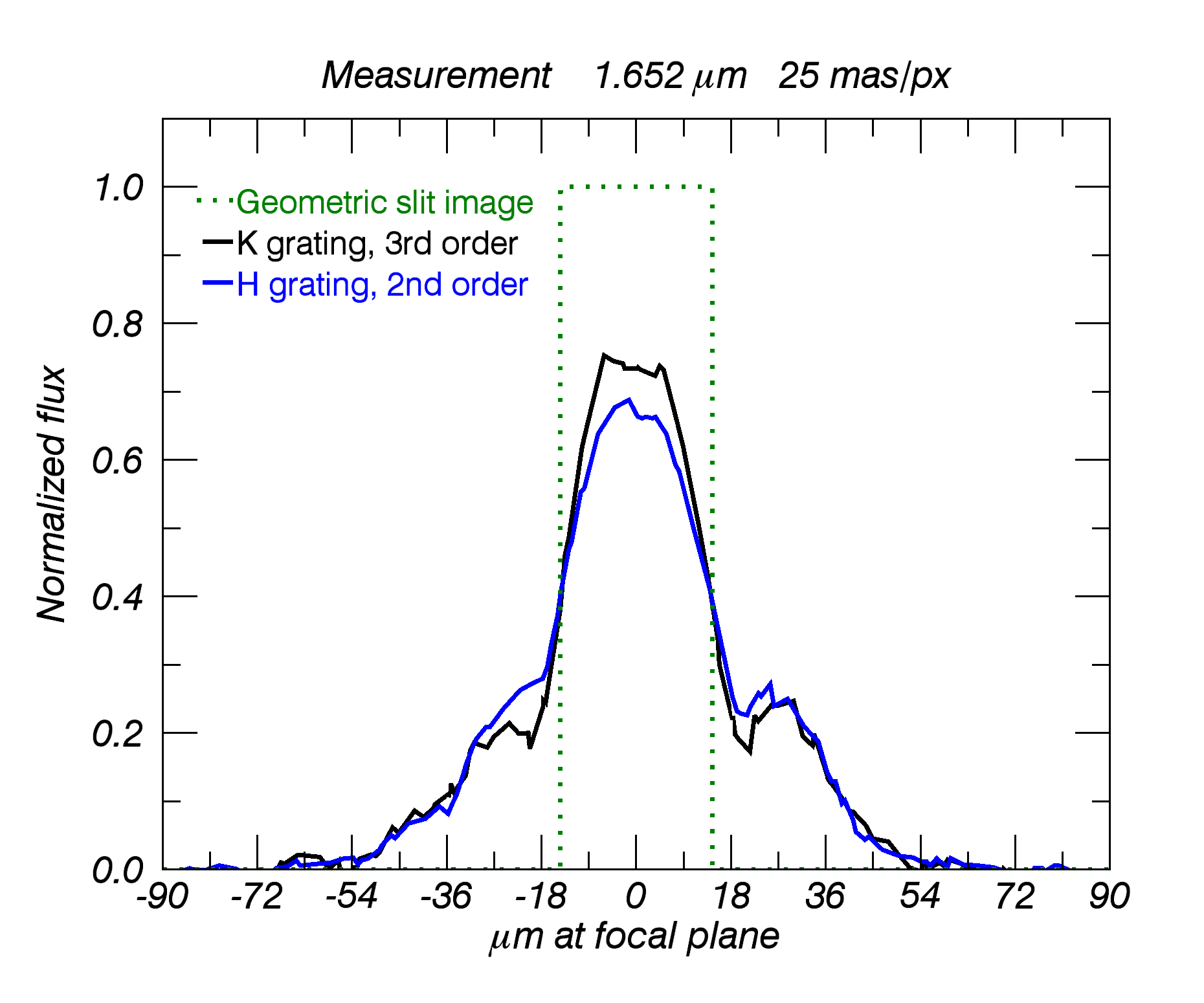}
\includegraphics[height=7cm, trim={1.0cm 0.1cm 0.9cm 0.3cm}, clip]{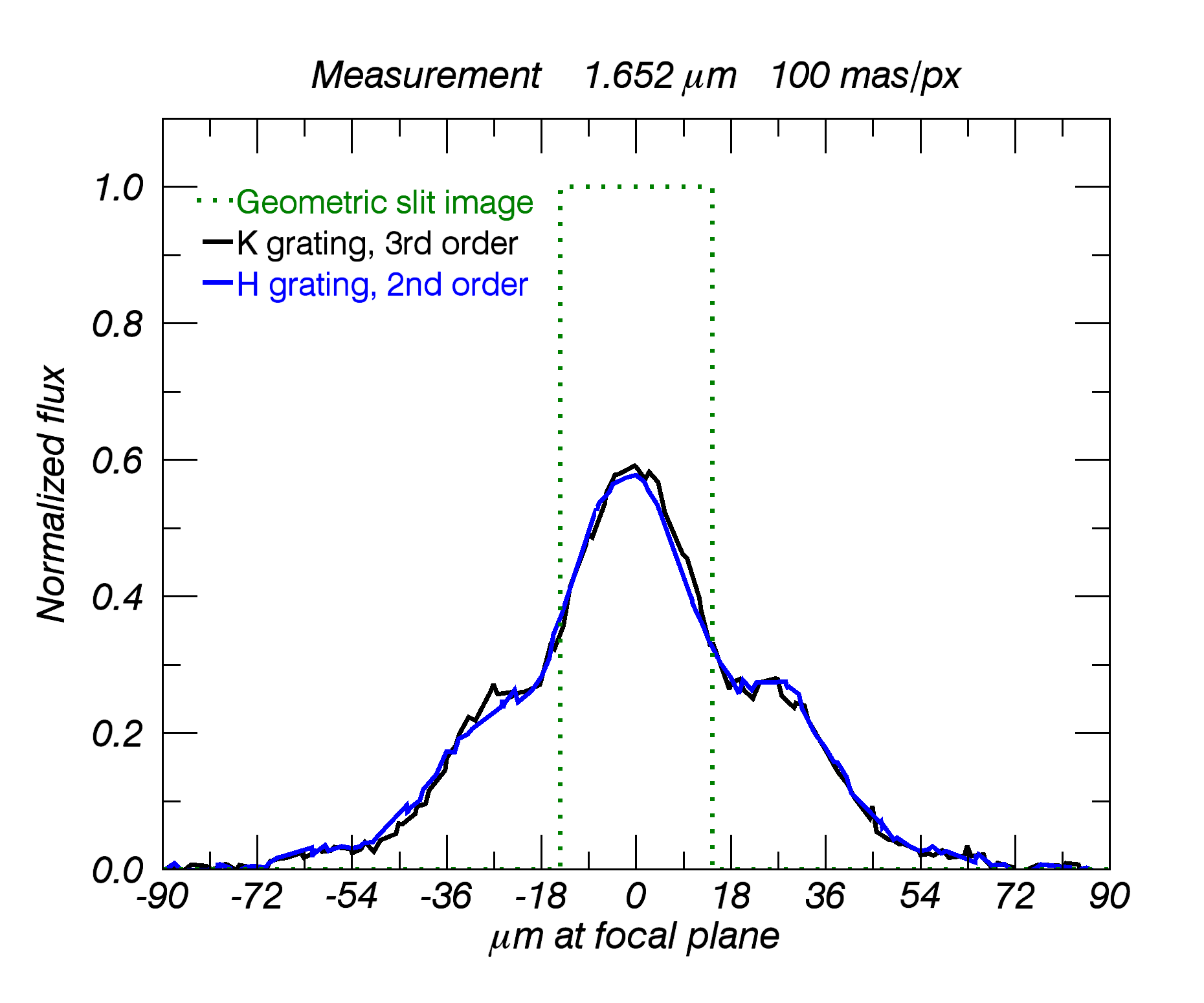}
\includegraphics[height=7cm, trim={1.0cm 0.1cm 0.9cm 0.3cm}, clip]{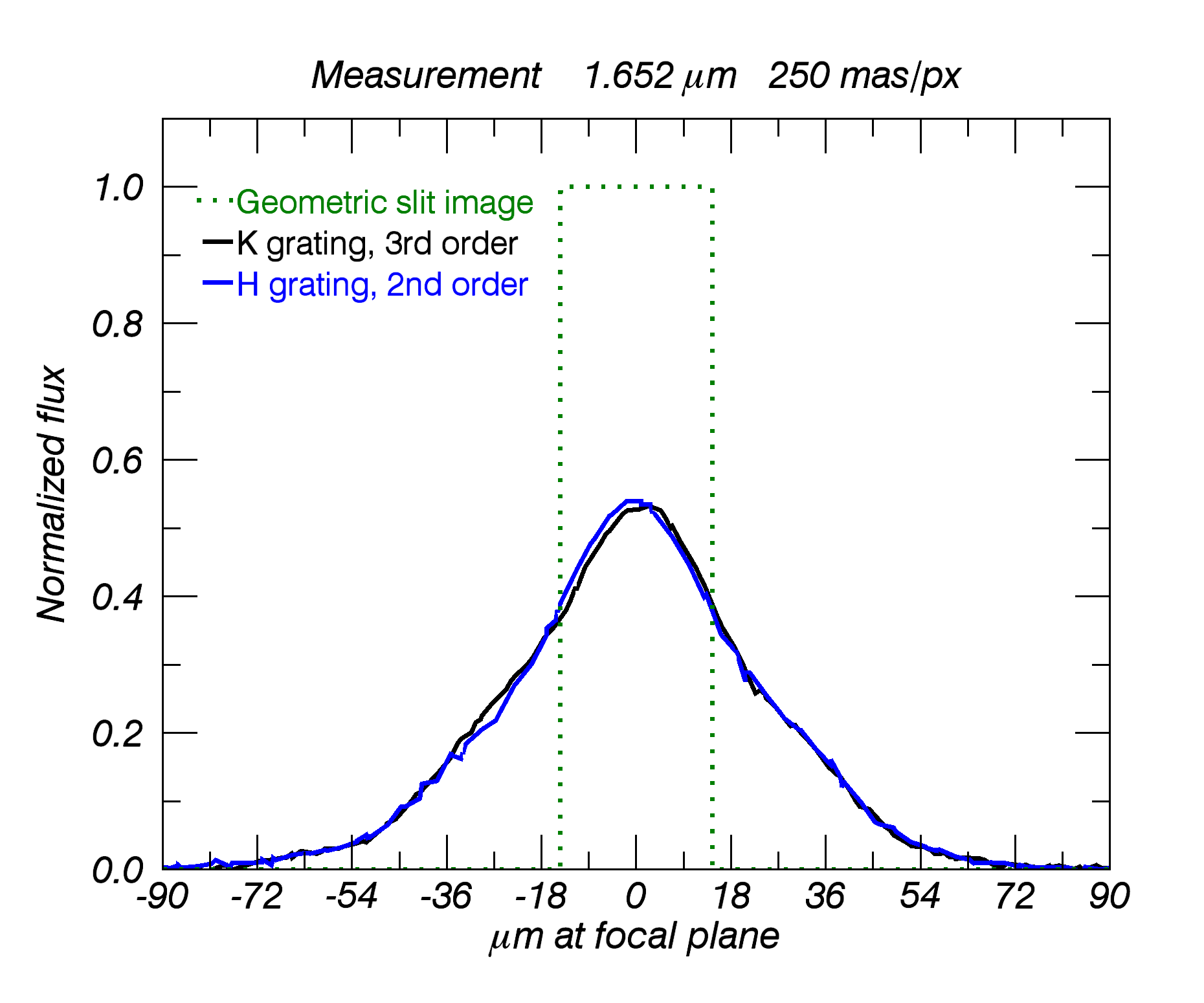}
}
\resizebox{1.0\textwidth}{!}{
\includegraphics[height=7cm, trim={0.4cm 0.1cm 0.9cm 0.3cm}, clip]{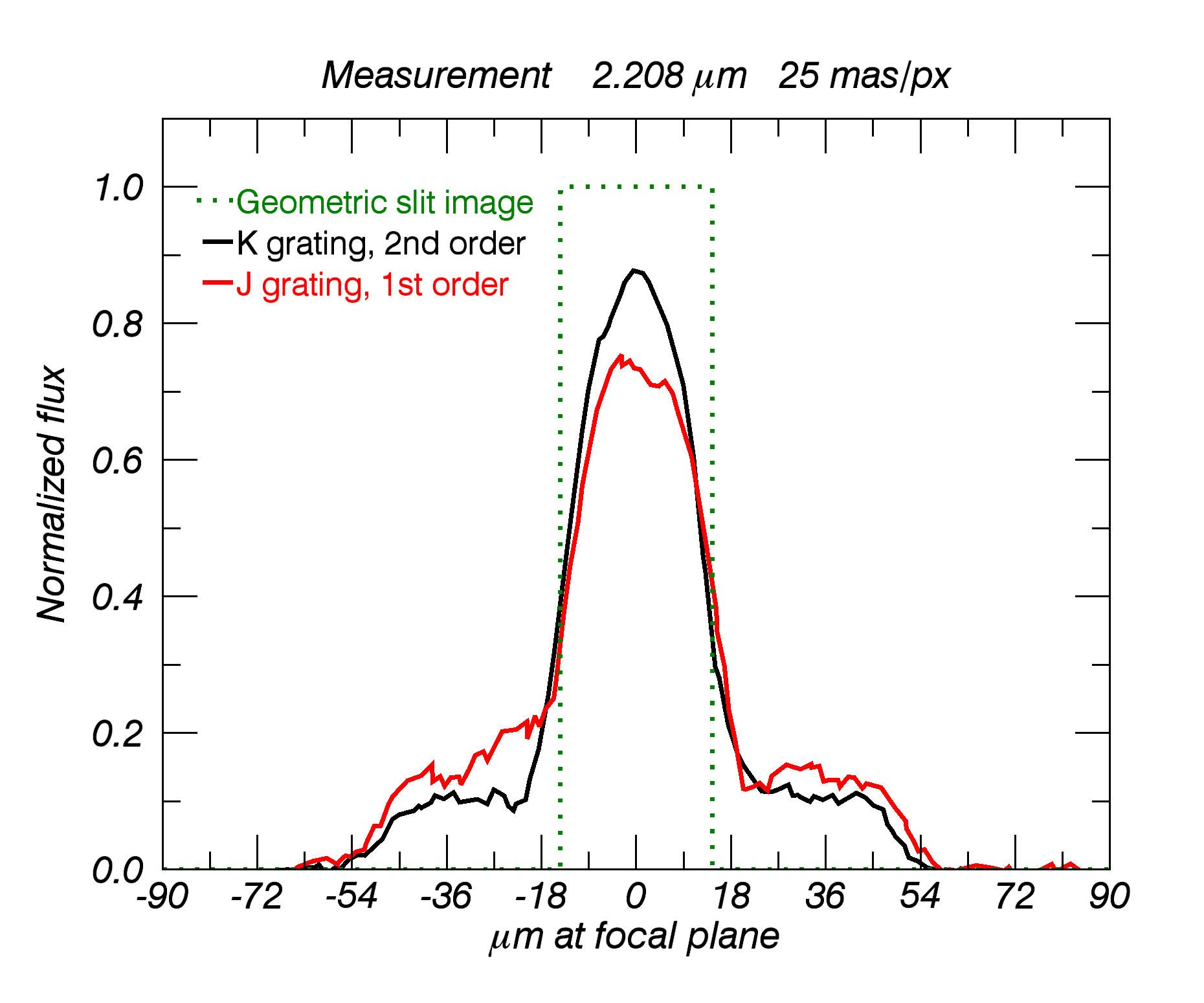}
\includegraphics[height=7cm, trim={1.0cm 0.1cm 0.9cm 0.3cm}, clip]{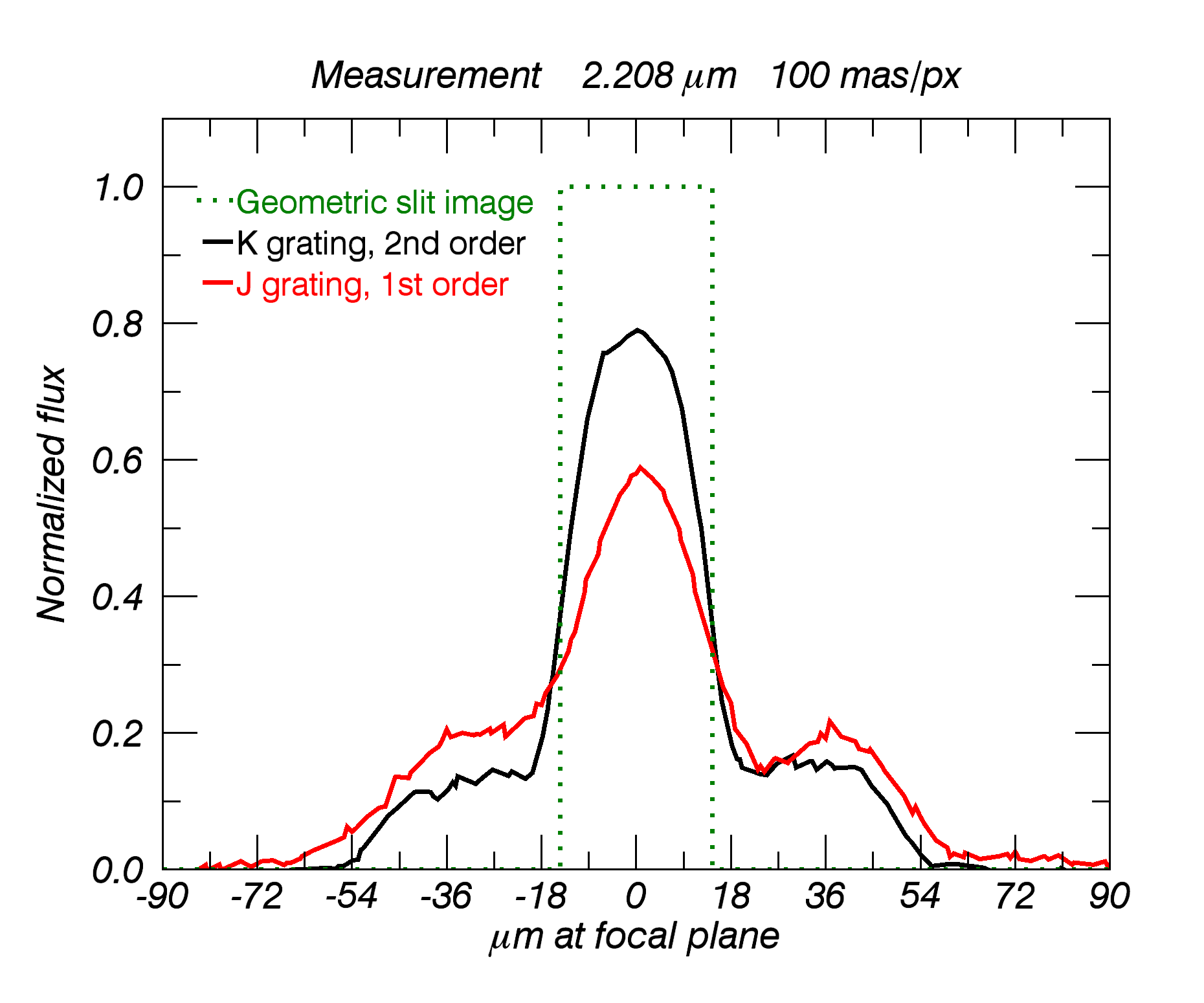}
\includegraphics[height=7cm, trim={1.0cm 0.1cm 0.9cm 0.3cm}, clip]{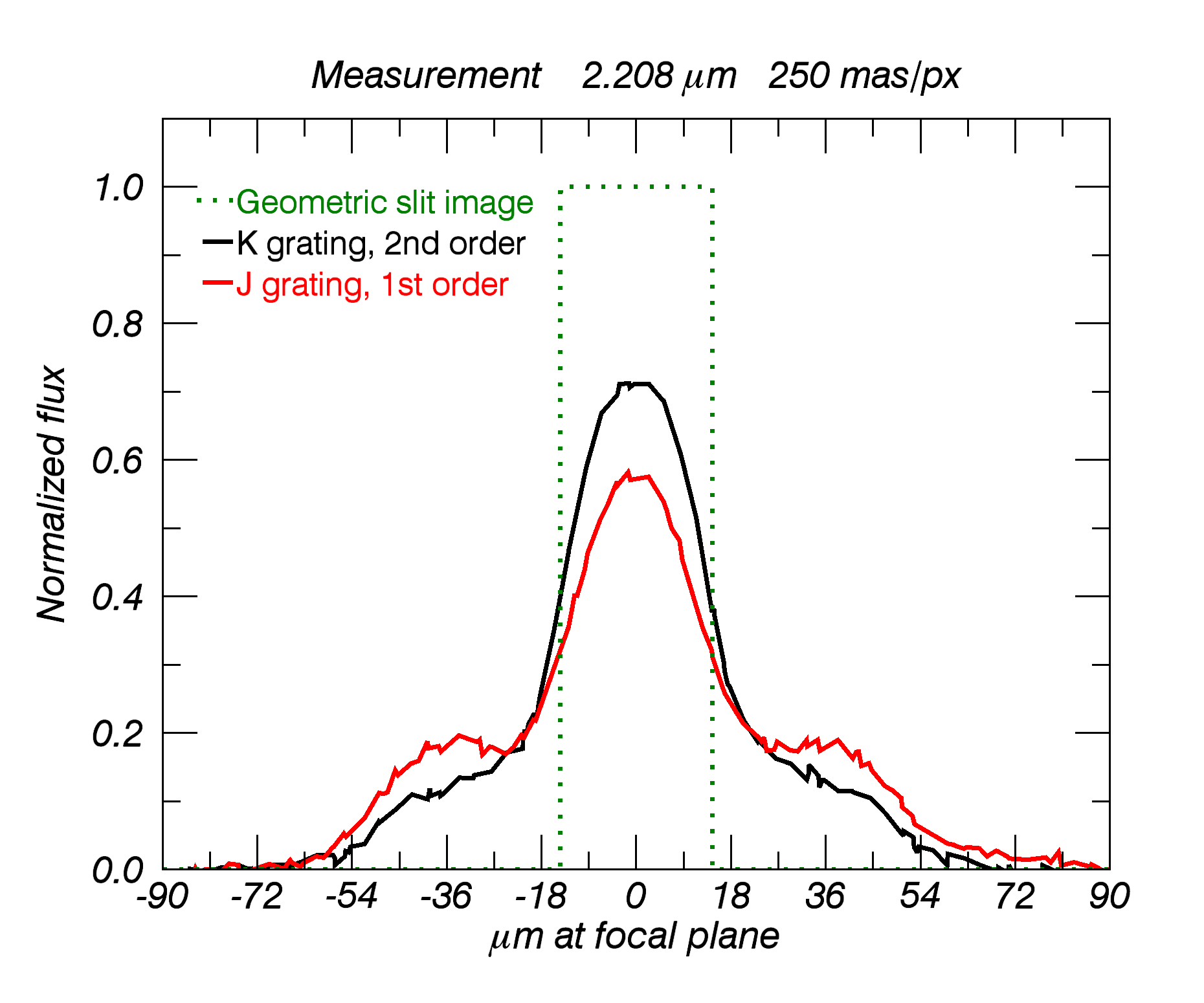}
}

\caption[Measured LSF at single wavelengths on multiple diffraction gratings]{Measured LSF at single wavelengths on multiple diffraction gratings in all three pixel scales ({\textbf{left column:}} 25 mas/px, {\textbf{middle column:}} 100 mas/px, {\textbf{right column:}} 250 mas/px). The wavelengths shown are at roughly the center wavelength of each band, and are selected from the bright arc-lamp calibration lines ({\textbf{top row:}} 1.249 \SI{}{\micro\meter}, {\textbf{middle row:}} 1.652 \SI{}{\micro\meter}, {\textbf{bottom row:}} 2.208  \SI{}{\micro\meter}). Each line was measured on as many gratings as possible.}
\label{fig:multigratdata}
\end{center}
\end{figure}
\pagebreak

\section{Further Finite Element Analyses of Grating Blanks}
\label{sec:appendixB} 

\begin{figure}[htbp!]
	\begin{center}
		\resizebox{1.0\textwidth}{!}{
			\includegraphics[height=7cm, trim={3.2cm 0.1cm 30.85cm 1.1cm}, clip=true]{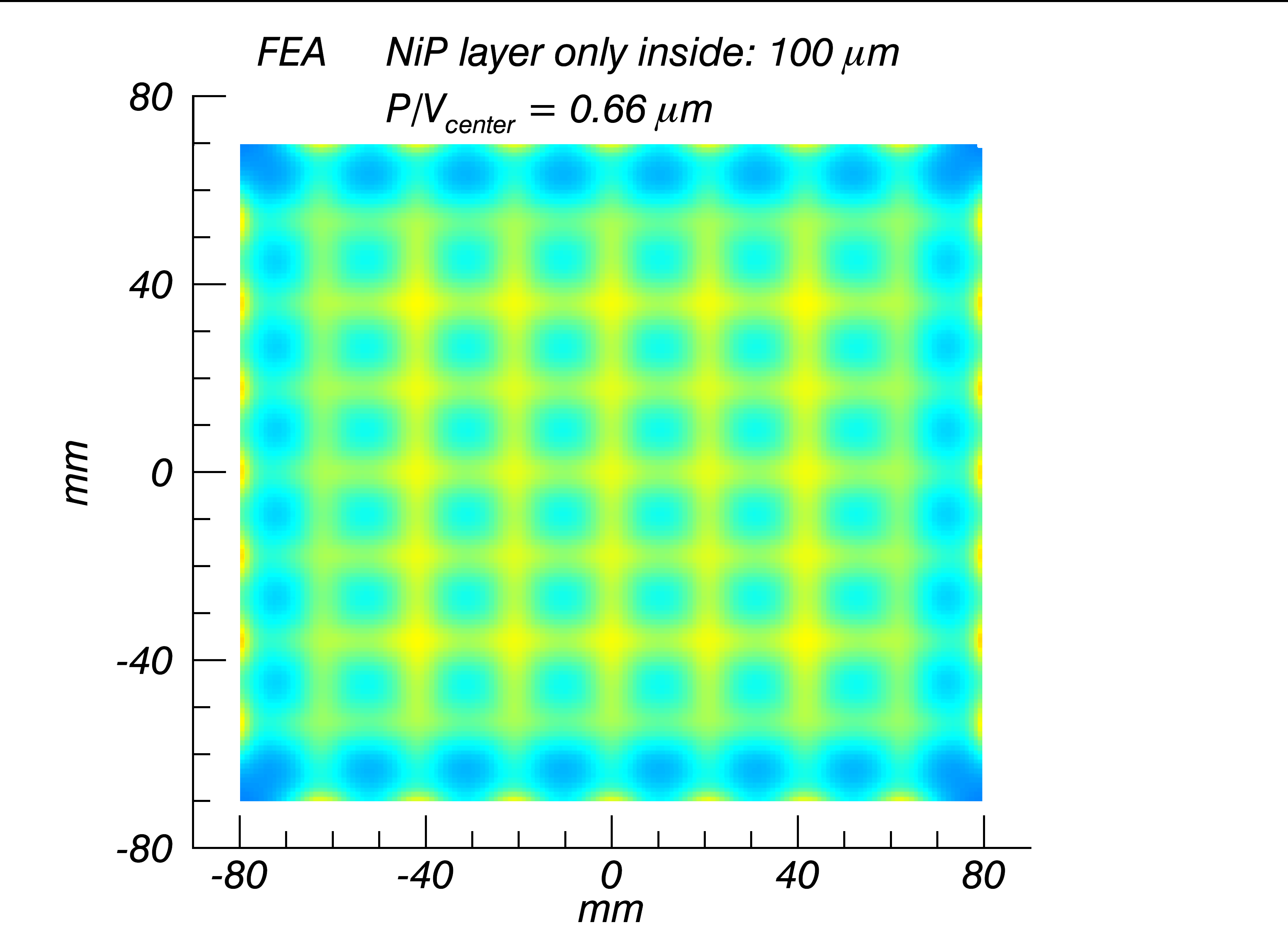}
			\includegraphics[height=7cm, trim={10.5cm 0.1cm 30.85cm 1.1cm}, clip=true]{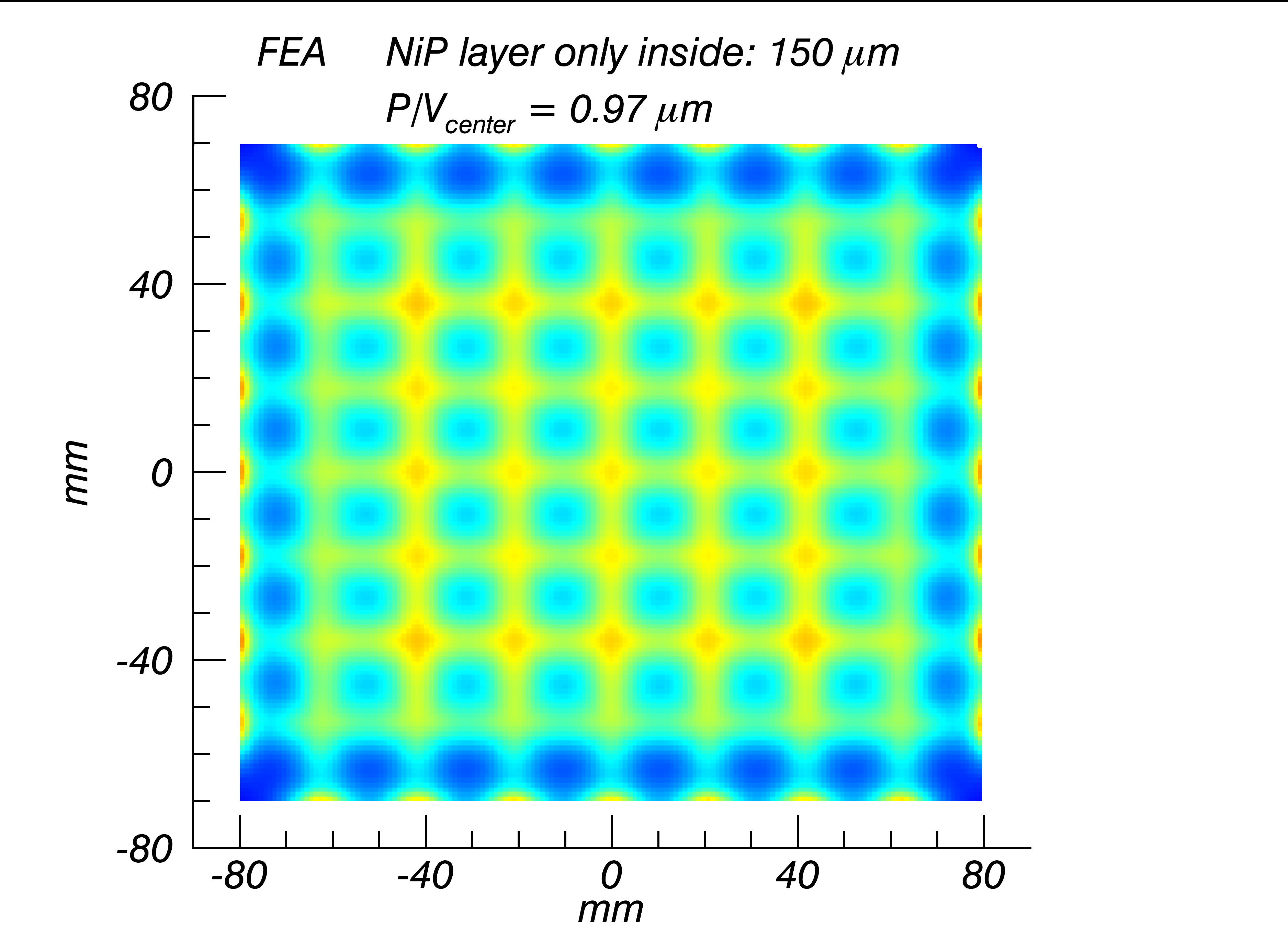}
			\includegraphics[height=7cm, trim={10.5cm 0.2cm 0 1.2cm}, clip=true]{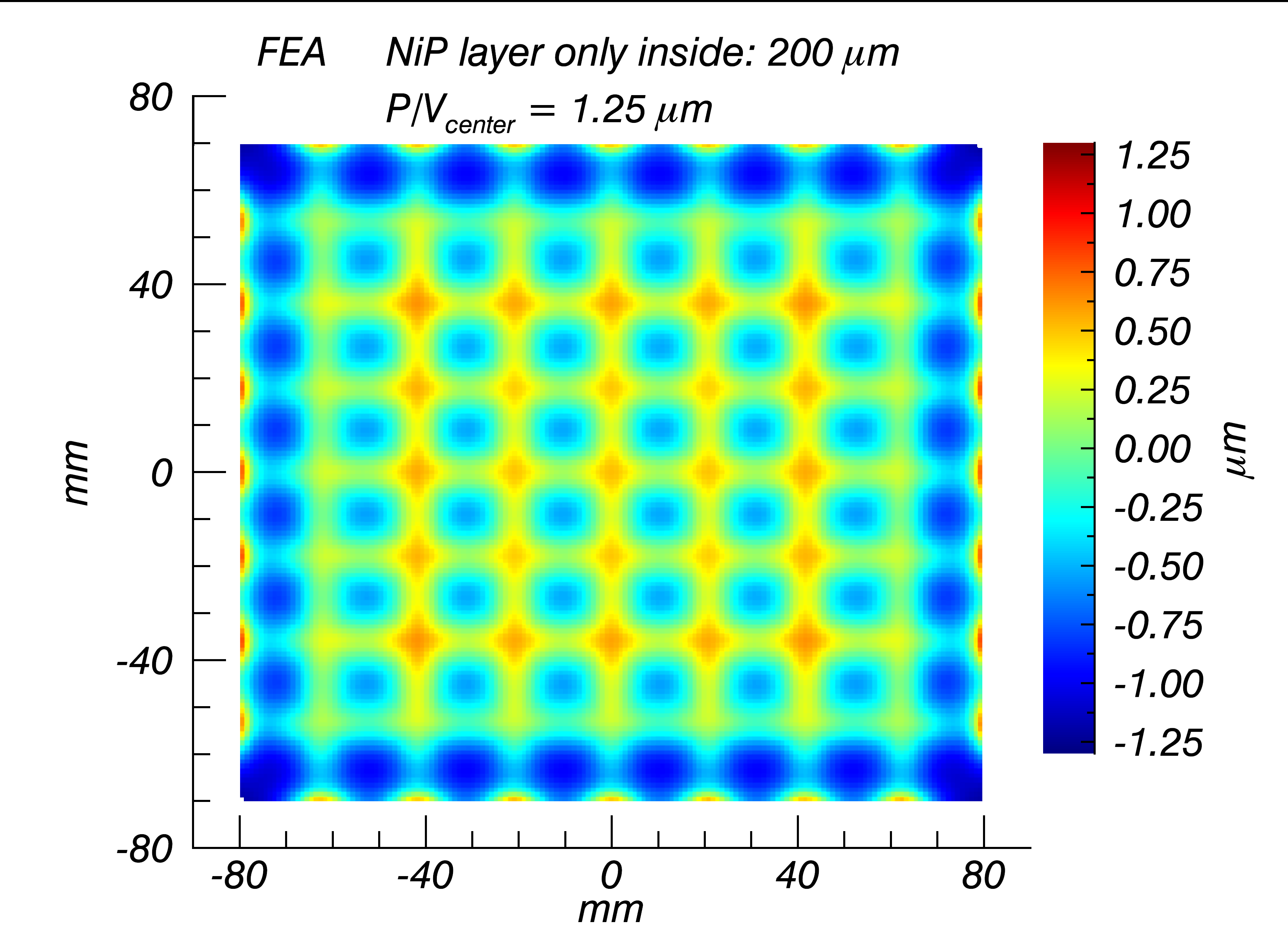}
		}
		\resizebox{1.0\textwidth}{!}{
			\includegraphics[height=7cm, trim={3.2cm 0.1cm 30.85cm 1.1cm}, clip=true]{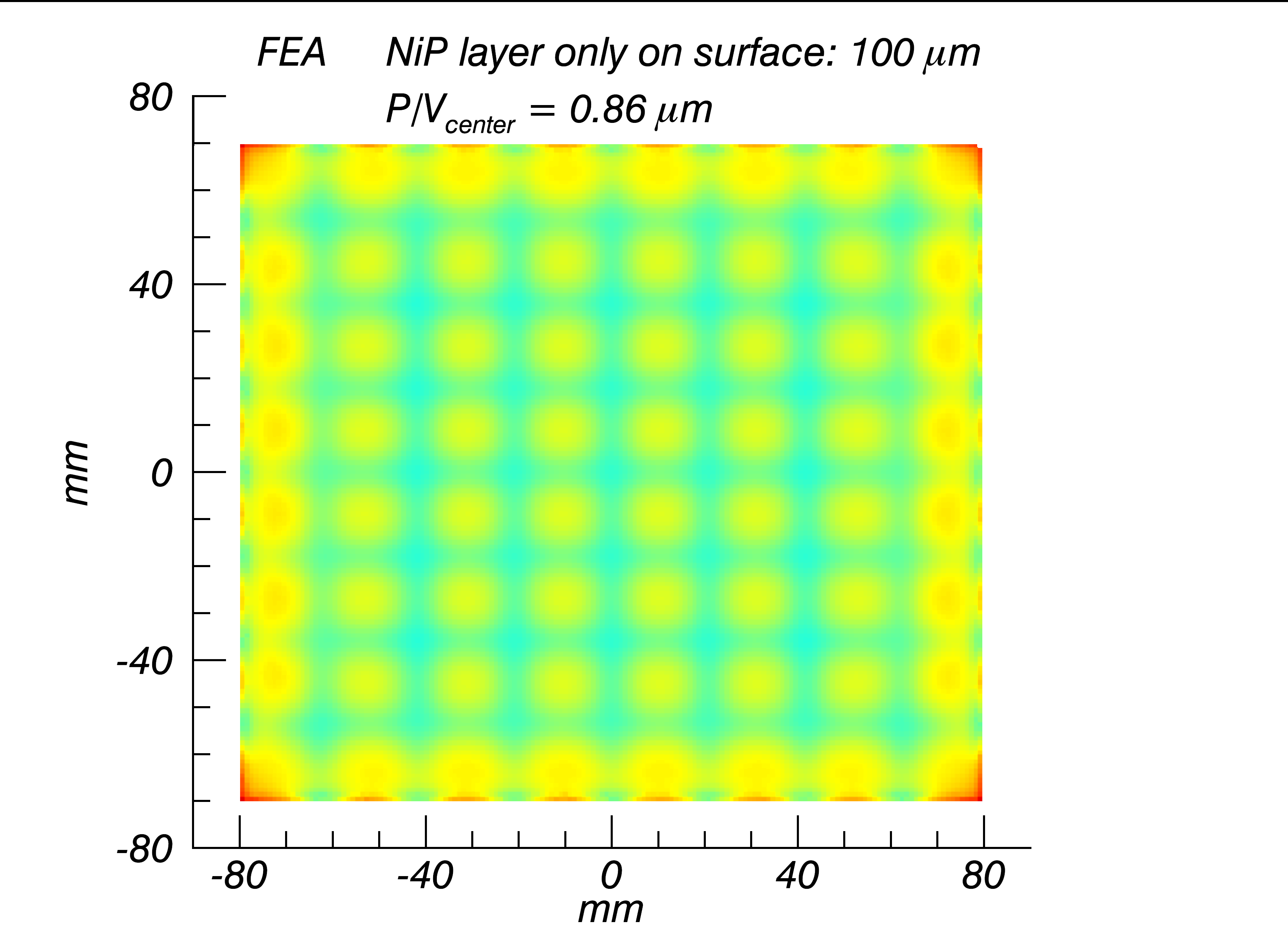}
			\includegraphics[height=7cm, trim={10.5cm 0.1cm 30.85cm 1.1cm}, clip=true]{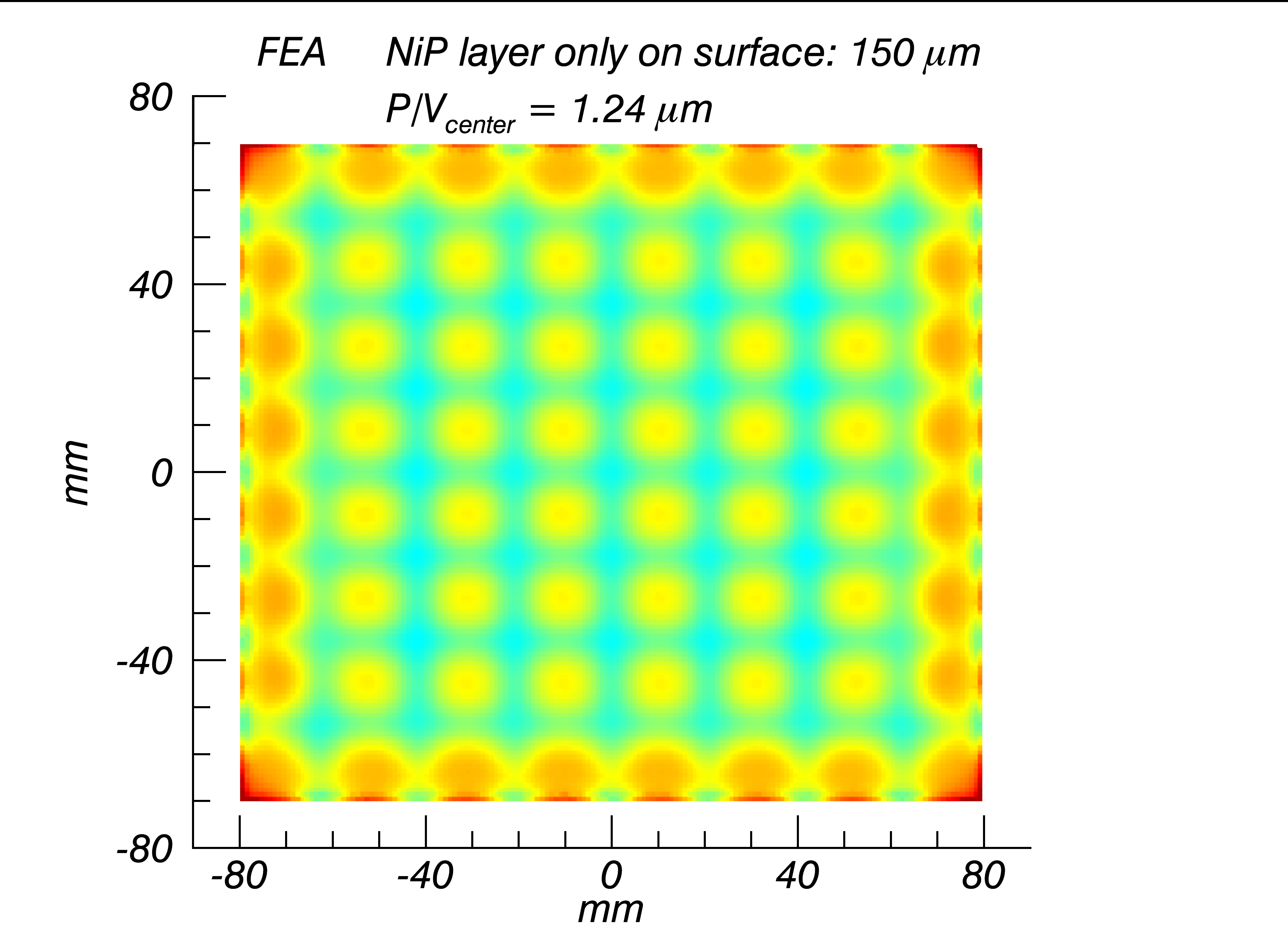}
			\includegraphics[height=7cm, trim={10.5cm 0.2cm 0 1.2cm}, clip=true]{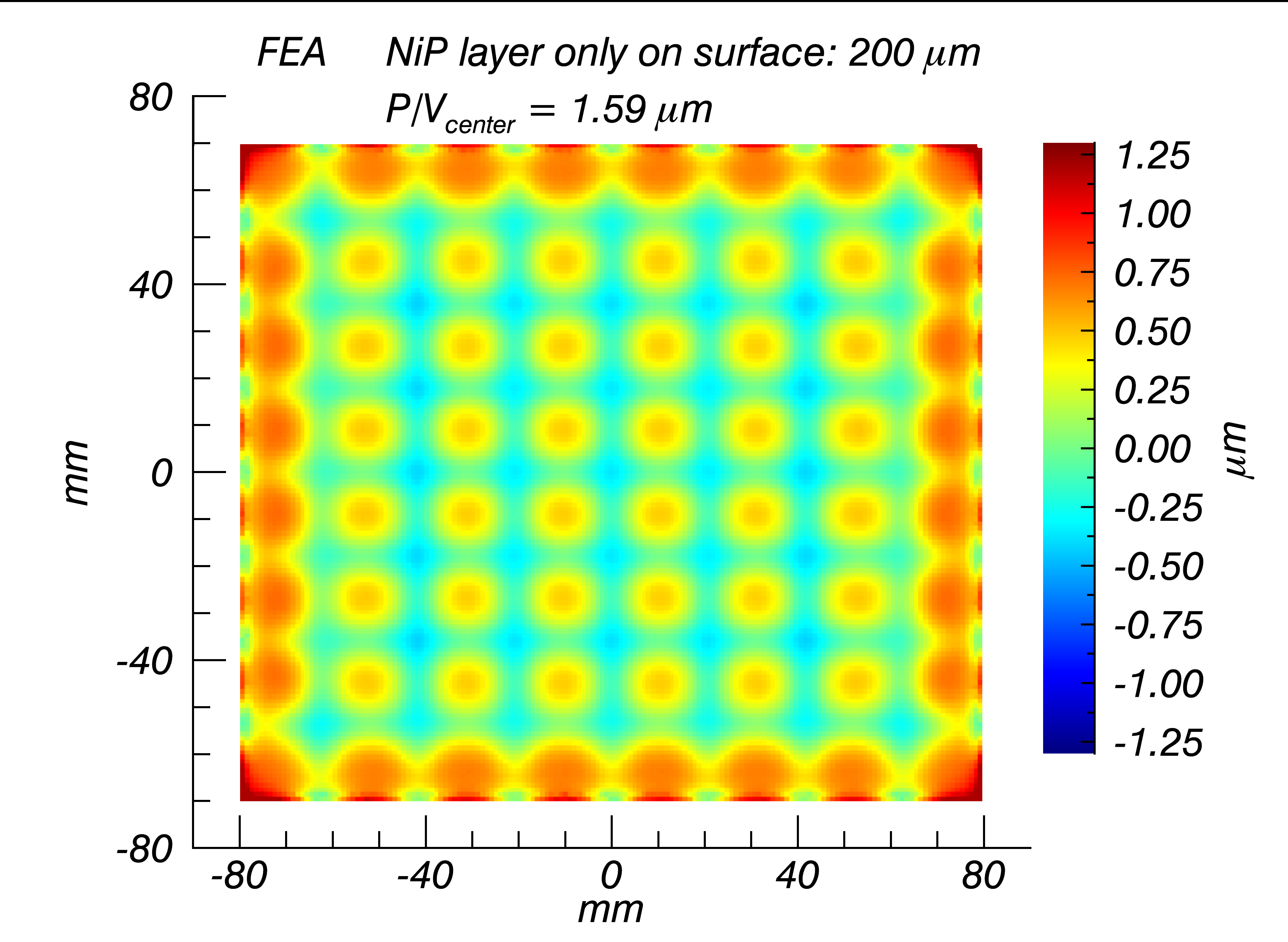}
		}
		\caption[Further FEA simulations of surface deformation]{FEA simulation of the grating surface deformation when cooling a grating blank with  NiP layer thickness of 100, 150, and 200 \SI{}{\micro\meter} (left to right) from room temperature to 80K. The top row shows the surface deformation with a layer of NiP only inside of the lightweighting holes, while the bottom row shows the surface deformation with NiP only on the outer surfaces of the grating blank. The peak-to-valley in the centre region (P/V$_{center}$ values) are calculated for rectangle of 82 x 72 mm in the center of the grating corresponding to a grid of four lightweighting holes, and corresponds to opposite directions of deformation for the two cases. The deviations at the edges are larger. All images are on the same color scale.}
		\label{fig:FEAdeformation2}
	\end{center}
\end{figure}

In section \ref{sec:FEA} we showed that more polishing (e.g. mismatched NiP layer thicknesses) results in more deformation. Here we present the edge cases: We simulate a grating blank with either NiP only inside of the lightweighting holes, or only on the outside surfaces of the blank, for 100, 150, and 200 \SI{}{\micro\meter} thick NiP layers. In general, thicker layers result in more deformation. Gratings with NiP only inside of the holes show a deformation in the same direction as the ``realistic" grating blanks presented in section \ref{sec:FEA}, but with a larger amplitude. Gratings with NiP only on the surfaces of the blank show a deformation in the opposite direction. This shows visually how the two NiP layers oppose each other. The P/V deformation values are tabulated in table \ref{tab:FEAresults2}, and plots of the surface deformations are shown in figure \ref{fig:FEAdeformation2}. A simulation with no NiP layer anywhere resulted in no deformation, as expected.

\begin{table}[htbp!]
\begin{center}
\begin{tabular}{c|cc}
\hline
 NiP Layer & Inside Holes & On Blank surfaces \\
 thickness [\SI{}{\micro\meter}] & P/V$_{center}$ [\SI{}{\micro\meter}] & P/V$_{center}$ [\SI{}{\micro\meter}]  \\
\hline
\hline
200 & 1.20 & 1.55* \\ 
150 & 0.93 & 1.21* \\
100 & 0.63 & 0.84* \\
 \hline
\end{tabular}
\vspace{0.05in}
\caption[Table of further FEA grating deformations]{FEA simulated P/V deformation of the grating in the central region of 82 mm x 72 mm (orientated on the lightweighting holes) for different NiP thicknesses either inside the lightweighting holes only, or only on the surfaces of the grating blank. {\textbf{}}* The deformation is in opposite directions the two cases of NiP location. The P/V deformation at the edges is larger by up to a factor of two due to edge effects.}
\label{tab:FEAresults2}
\end{center}
\end{table}

\clearpage
\pagebreak







\end{document}